\documentclass[aps,prd,twocolumn,superscriptaddress,nofootinbib]{revtex4-2}

\usepackage{soul}
\usepackage{xcolor}
\usepackage{amsmath}
\usepackage{graphicx}

\newcommand{\mathst}[1]{\text{\st{$#1$}}}
\newcommand{\threedots}[1]{\overset{\makebox[0pt]{\raisebox{-1.5ex}[0.1ex][0.1ex]{\ensuremath{\dot{}\,\dot{}\,\dot{}}}}}{#1}}
\newcommand{\orcid}[1]{\href{https://orcid.org/#1}{\includegraphics[width=9pt]{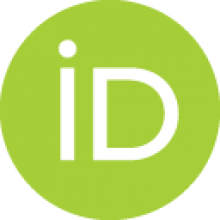}}}

\begin{document}

\title{Collapsar Disk Outflows III: Detectable Neutrino and Gravitational Wave Signatures}

\author{Rodrigo Fern\'andez\orcid{0000-0003-4619-339X}}
\email[]{rafernan@ualberta.ca}
\affiliation{Department of Physics, University of Alberta, Edmonton, AB T6G 2E1, Canada}

\author{Silas Janke\orcid{0009-0000-4192-5390}}
\affiliation{Department of Physics, University of Alberta, Edmonton, AB T6G 2E1, Canada}
\affiliation{Department of Physics and Astronomy, Heidelberg University, Im Neuenheimer Feld 226, 69120 Heidelberg, Germany}

\author{Coleman Dean\orcid{0000-0001-9364-4785}}
\affiliation{Department of Physics, University of Alberta, Edmonton, AB T6G 2E1, Canada}

\author{Irene Tamborra\orcid{0000-0001-7449-104X}}
\affiliation{Niels Bohr International Academy and DARK, Niels Bohr Institute, University of Copenhagen, Blegdamsvej 17, 2100 Copenhagen, Denmark}

\date{\today}

\begin{abstract}
We investigate the neutrino and gravitational wave (GW) signals from
accretion disks formed during the failed collapse of a rotating massive star (a collapsar).
Following black hole formation, a neutrino-cooled, shocked accretion disk forms, 
which displays non-spherical oscillations for a period of seconds before becoming advective
and exploding the star.
We compute the neutrino and GW signals (matter quadrupole, 
frequencies $\lesssim 100$\,Hz) 
from collapsar disks using
global axisymmetric, viscous hydrodynamic simulations.
The neutrino signal with typical energies of $\mathcal{O}(10)$\,MeV is maximal during 
the neutrino-cooled (NDAF) phase that follows shock formation. This phase lasts for a few seconds
and is easily detectable
within $\mathcal{O}(10$--$100)$\,kpc by the IceCube Neutrino Telescope.
Additional neutrino signatures 
from a precursor equatorial shock and by stochastic accretion plumes during the advective phase are
detectable within the galaxy. The GW signal during the NDAF phase is detectable
in the galaxy by current and next-generation ground-based observatories. The 
explosion (memory) GW signal is similar to that of standard core-collapse 
supernovae and can be probed with
a deci-Hertz space-based detector. Shock oscillations during the NDAF phase impart
time variations with frequency $\mathcal{O}(10-100)$~Hz to the neutrino and GW signals, 
encoding information
about the shock dynamics and inner disk. These time variations can be detectable  in  neutrinos
 by IceCube within
$\mathcal{O}(1$--$10)$\,kpc depending on progenitor model, flavor transformation scenario, and  
detailed properties of the angular momentum transport mechanism. 
\end{abstract}

\maketitle

\section{Introduction}

Core-collapse supernovae (CCSNe) are intrinsic multi-messenger sources (e.g., \cite{adams_2013,tamborra_2025}). 
SN 1987A showed that both the electromagnetic
and neutrino signals from these events are observable when occurring 
within our galaxy or nearby satellites \citep{hirata_1987,bionta_1987,alekseev_1987,arnett_1989}. 
With current facilities, 
about a million neutrinos will be detected from the next CCSN in the
galaxy or its vicinity, 
providing information on the protoneutron star (PNS), multi-dimensional dynamics 
of the SN engine, stellar progenitor, fundamental neutrino physics, and
potential signatures of new physics
(e.g., \cite{mirizzi_2016,janka_2017,horiuchi_2018,scholberg_2018,mueller_2019}). 
Current ground-based gravitational wave (GW) observatories can
also detect a CCSN in the galaxy, providing independent information
about the explosion dynamics and protoneutron star properties
(e.g., \cite{abdikamalov_2022,christensen_2022,powell_2022}).

Collapsars \cite{woosley_1993} are a subset of CCSNe, corresponding to  
rapidly-rotating progenitors that fail to explode via standard channels like
the delayed-neutrino mechanism, resulting in the formation of
a black hole (BH) accretion disk fed by the infalling stellar layers. 
Initially, the disk undergoes 
significant neutrino cooling (e.g., \cite{popham1999,chen_2007}), and its subsequent evolution is thought
to power long-duration
gamma-ray bursts (LGRBs) and the associated broad-line type Ic (Ic-BL) 
SNe \cite{macfadyen_1999,macfadyen_2003,woosley_2006}. 
The neutrino-cooled phase (NDAF) of the disk involves significant neutronization, which could in
principle provide a site for the rapid neutron capture process ($r$-process; \cite{pruet_2003,kohri_2005,siegel_2019,DF24b}). 
As the disk accretes and the density drops, neutrino production becomes dynamically unimportant,
and the disk transitions to an advection-dominated (ADAF) phase, powering an outflow that is able
to explode the star even when a relativistic jet is not produced.

The NDAF phase in collapsar disks results in neutrino emission at levels comparable
to the PNS accretion phase of CCSNe prior to BH formation (e.g., \cite{lopezcamara_2009,sekiguchi_2011,just_2022}). 
The shock that bounds the accretion disk after BH formation is known to also undergo
oscillatory motion (e.g., \cite{taylor_2011,gottlieb_2022}), 
analogous to the Standing Accretion Shock Instability (SASI) in CCSNe \cite{blondin_2003,foglizzo_2007},
but with a different underlying instability mechanism \cite{molteni_1999,gu_foglizzo_2003,gu_lu_2006,
nagakura_yamada_2008,nagakura_yamada_2009}. 
SASI oscillations in CCSNe are known to modulate
accretion onto the PNS and involve non-spherical matter motions, imparting
$\sim 100$\,Hz modulations in the neutrino signal, which are detectable for
sufficiently nearby galactic CCSNe 
\cite{lund_2010,lund_2012,tamborra_2013,tamborra_2014,mueller_2014_neutrinos,walk_2018,
walk_2019,vartanyan_2019,walk_2020,lin_2020,nagakura_2021,lin_2023,beise_2024,beise_2025}.
The SASI also provides a distinct imprint on the GW signal from CCSNe 
(e.g., \cite{kotake_2009,murphy_2009,cerda-duran_2013,yakunin_2015,andresen_2017,pajkos_2019}), which is correlated with 
the neutrino signal modulation \cite{kuroda_2017,takiwaki_2018,shibagaki_2021,drago_2023}.
 
Neutrino production in collapsar disks has been studied extensively
through global, time-dependent simulations (e.g., \cite{lee_2006,nagataki_2007,sekiguchi_2011,harikae_2009a,obergaulinger_2017,fujibayashi_2022,just_2022,issa_2025,shibata_2025}),
although (to our knowledge) direct detectability of MeV neutrinos has only been computed semi-analytically \cite{liu_2016}.
The contribution of collapsars to the high-energy tail of the diffuse supernova 
neutrino background has also been explored \cite{schilbach_2019,wei_2024,martinez-mirave_2024}, showing that 
this signal can potentially provide crucial input on the properties of the population of collapsing massive stars.
GW emission from collapsar accretion disks has been modeled semi-analytically, considering the
contribution of matter motions (e.g., \cite{vanputten_2001,fryer_2002,kobayashi_2003,piro_2007}),
disk/jet precession \cite{romero_2010,sun_2012},
and anisotropic-neutrino emission \cite{hiramatsu_2005,suwa_2009}. Waveform
extraction and detectability estimates from global collapsar simulations has been carried out 
in hydrodynamics \cite{ott_2011,kotake_2012,cerda-duran_2013}
and more recently in magnetohydrodynamics (MHD) \cite{gottlieb_2024}, over timescales of up to
$25$\,s.

Here, we examine the diagnostic potential of the neutrino and GW signals from collapsar
accretion disks and their outflows, in the limiting case where a relativistic jet is not present.
For the first time, we assess the detectability
of the MeV neutrino signal using global time-dependent simulations that include the relevant microphysics.
We post-process the axisymmetric, viscous hydrodynamic simulations presented
in \cite{DF24a} (hereafter Paper~I). 
These models include gray, 3-species neutrino transport through a 
leakage scheme for emission and annular light bulb-type absorption, a 19-isotope nuclear reaction network, and
Newtonian self-gravity with a pseudo-Newtonian BH. For each simulation, we extract the neutrino luminosities 
and mean energies for all flavors and compute the GW emission due to the matter quadrupole moment.
The resulting waveforms span durations $\sim 100$\,s, allowing for predictions at lower frequencies
than most CCSN simulations, which typically last $\lesssim 10$\,s.

The paper is structured as follows. Section~\ref{s:methods} provides a brief
summary of the computational models employed in this work 
and describes the extraction of the neutrino and GW signal, along with
the characterization of the shock that bounds the accretion disk.
Our results are presented in Section~\ref{s:results}, divided
into neutrino signal detectability, GW signal detectability,
and analysis of time variations in all observables.
Section~\ref{s:summary} provides a summary
and discussion of our results. Appendix~\ref{app:noise_power} presents our
calculation of the spectral power of Poisson noise. 

\section{Methods \label{s:methods}}

\subsection{Collapsar Models \label{s:models}}

We extract and post-process the neutrino and GW signals from the
simulations presented in Paper~I. Each simulation starts 
from a pre-collapse progenitor and is evolved in spherical symmetry
to BH formation using {\tt GR1D} version~1 \cite{oconnor_2010}. 
The resulting profile is then mapped onto {\tt FLASH} version~3.2 
\cite{fryxell00,dubey2009} for
two-dimensional axisymmetric evolution with rotation (``2.5D''). 
The computational domain is discretized in 
spherical coordinates $(r,\theta)$, with a grid that is logarithmic in radius and equispaced in $\cos\theta$. 
The hydrodynamic equations are solved with the split version of the Piecewise Parabolic Method \cite{colella84}.
Source terms include a viscous stress with non-zero $r\theta$ and $r\phi$ components, with magnitude
parameterized as in \cite{FM12}, multipole Newtonian self-gravity 
\cite{MuellerSteinmetz1995,FMM19}, a pseudo-Newtonian potential for the BH \cite{artemova1996,FKMQ14},
neutrino emission through a gray, 3-species leakage scheme with lightbulb-type
absorption \cite{FM13,fernandez_2022}, a 19-isotope nuclear reaction network \cite{weaver1978,Timmes1999} with 
a nuclear statistical equilibrium solver \cite{seitenzahl_2008}, and the 
equation of state (EOS) of \cite{timmes2000}.
Simulations are run to times beyond shock breakout from 
the outer edge of the stellar progenitor, which takes
$\Delta T_{\rm sim}\sim 100$\,s from BH formation. 

The original models from Paper~I produced spatial output at time intervals of $10$\,ms
(sampling frequency $f_{\rm s}=100$\,Hz). This interval
was chosen to limit the total amount of data produced, considering that the entire
post-BH evolution lasts for $\Delta T_{\rm sim }= 100-400$\,s. Since we compute the GW signal 
in post-processing, and in order to obtain a better temporal
sampling during the NDAF phase, we re-run all models from Paper~I with spatial output at $1$\,ms 
intervals ($f_{\rm s}=1$\,kHz),
over a time period of
$10$\,s that covers shocked disk formation, the NDAF phase, and the onset of runaway expansion.
Given the stochastic character of the system, the evolution was not identical to the original models,
but quantitatively the same for time-integrated quantities.
We focus the discussion on detectability using the models with finer time sampling, 
and for clarity append {\tt -dt} to their names.
Neutrino information is output at every time step for all
models ($\Delta t\sim 10^{-7}$\,s).

Table~\ref{tab:models} lists all models studied in this paper, showing the simulations with high temporal
sampling frequency ({\tt -dt}) as well as the original models from Paper I.
The presupernova progenitors employed are models 16TI and 35OC from \cite{woosley_2006b}.
Paper~I presents evolution of these progenitors using the nuclear EOS 
SFHo \cite{steiner_2013} in {\tt GR1D} (our baseline model {\tt 16TI-SFHo}, and {\tt 35OC-SFHo}, respectively), 
as well as a model that uses the DD2 EOS
\cite{hempel_2012} with the 16TI progenitor 
(model {\tt 16TI-DD2}).
Also, the strength of the viscosity parameter is varied around the baseline model 
({\tt 16TI-SFHo-{lo$\alpha$}} and {\tt 16TI-SFHo-{hi$\alpha$}}; we have renamed these two models
relative to Paper I for clarity). 
The neutrino quantities, GW strains, and shock time series information are available at
\cite{zenodo_repo}.

\begin{table}
\caption{Properties of the collapsar models. Columns from left to right show model name, progenitor star, 
nuclear EOS used in pre-BH evolution, post-BH viscosity parameter $\alpha$, simulation duration, 
and spatial output frequency. The bottom half of the table shows models originally
presented in Paper I, while the top half shows new models which are re-runs of their original
counterparts but at higher spatial output frequency (suffix {\tt -dt}). Note that we have renamed the models
that vary the viscosity parameter $\alpha$, the suffix correspondence relative to Paper I is 
$\alpha01\to{\rm hi}\alpha$
and 
$\alpha001\to{\rm lo}\alpha$.
\label{tab:models}} 
\begin{ruledtabular}
\begin{tabular}{lccccc}
Model & Prog. & EOS & $\alpha$ & $\Delta T_{\rm sim}$ & $f_{\rm s}$ \\
      &       &     &          &     (s)              &   (Hz)       \\
\noalign{\smallskip}
\hline
{\tt 16TI-SFHo-dt}              & 16TI & SFHo & 0.03 & 10 & $10^3$ \\
{\tt 16TI-SFHo-lo$\alpha$-dt}   & 16TI & SFHo & 0.01 & 10 & $10^3$ \\
{\tt 16TI-SFHo-hi$\alpha$-dt}   & 16TI & SFHo & 0.10 & 10 & $10^3$ \\
{\tt 16TI-DD2-dt}               & 16TI & DD2  & 0.03 & 10 & $10^3$ \\
{\tt 35OC-SFHo-dt}              & 35OC & SFHo & 0.03 & 10 & $10^3$ \\
\noalign{\smallskip}
\hline
\noalign{\smallskip}
{\tt 16TI-SFHo}                 & 16TI & SFHo & 0.03 & 217 & 100 \\
{\tt 16TI-SFHo-lo$\alpha$}      & 16TI & SFHo & 0.01 & 293 & 100 \\
{\tt 16TI-SFHo-hi$\alpha$}      & 16TI & SFHo & 0.10 & 424 & 100 \\
{\tt 16TI-DD2}                  & 16TI & DD2  & 0.03 & 297 & 100 \\
{\tt 35OC-SFHo}                 & 35OC & SFHo & 0.03 & 102 & 100 \\
\end{tabular}
\end{ruledtabular}
\end{table}

\subsection{Neutrino Signal Extraction \label{s:neutrinos}}

Our gray leakage scheme outputs the total energy- and number luminosities for $\nu_e$, $\bar{\nu}_e$, and 
the combined heavy lepton neutrinos and antineutrinos ($\nu_\mu$, $\bar{\nu}_\mu$, $\nu_\tau$, and $\bar{\nu}_\tau$).
Here, we use $\nu_x$ and  $\bar{\nu}_x$ to denote any one species of heavy lepton neutrino and antineutrino, respectively. 
Since leakage schemes only track energy production, the luminosities have no directional information. Previous studies
of time-variations in the neutrino signal from multi-dimensional CCSNe simulations with more advanced neutrino
transport have computed projected luminosities
as a function of viewing angle \cite{tamborra_2013,tamborra_2014,mueller_2014_neutrinos,walk_2018,walk_2019,walk_2020}. 
In our calculations, we use an average flux obtained by assuming that the total luminosity 
is radiated equally in all directions. For anisotropic emission, this is equivalent to
averaging projected luminosities over all viewing angles, which results in the smearing out of 
stronger or weaker emission features that may arise along specific directions.
The net outgoing luminosities are obtained by subtracting the total absorbed power from the emitted one, as in \cite{fernandez_2022}.

Following \cite{ruffert_1996}, we obtain the mean energies $\langle \epsilon_{\nu_i}\rangle$ of each species 
$\nu_i$ as the ratio of energy luminosity $L_{\nu_i}$ to number luminosity $N_{\nu_i}$,
\begin{equation}
\label{eq:mean_energy}
\langle \epsilon_{\nu_i}\rangle = \frac{L_{\nu_i}}{N_{\nu_i}}.
\end{equation}
To compute the detection rate, we need to compute the number flux spectrum at Earth.
This requires (1) reconstructing the output spectrum at the source from our gray
leakage scheme and (2) accounting for neutrino flavor transformation. 

We first estimate the number flux of neutrinos of type $\nu_i$ at the source in a given energy bin
\begin{equation}
\Delta F_{\nu_i}(\epsilon) = \frac{dF_{\nu_i}}{d\epsilon}\Delta \epsilon,
\end{equation}
where $F_{\nu_i}$ is the number flux of neutrinos 
and $\Delta \epsilon$ is the width of the energy bin.
For a source at a distance $D$, the (average) number flux in terms
of the total number luminosity is given by
\begin{equation}
\label{eq:number_flux_total}
F_{\nu_i} = \frac{1}{4\pi D^2}N_{\nu_i}.
\end{equation}
The leakage scheme assumes that the spectrum follows a Fermi-Dirac distribution
with zero chemical potential, but with a normalization given by the integral of the leakage emissivity \cite{fernandez_2013}. 
The total number flux is related to the Fermi-Dirac distribution as
\begin{eqnarray}
\label{eq:fermi_dirac}
F_{\nu_i} & \propto & \int_0^\infty \frac{\epsilon^2\,d\epsilon}{\exp({\epsilon/kT_{\nu_i}}) + 1}\nonumber\\
      & \propto & (kT_{\nu_i})^3 \int_0^\infty \frac{x^2\,dx}{\exp(x) + 1} = \mathcal{F}_2(0)\, (kT_{\nu_i})^3,
\end{eqnarray}
where $\mathcal{F}_2(0) = 3\zeta(3)/2 \simeq 1.803085$ is a 
Fermi-Dirac integral for zero chemical potential 
\cite{bludman_1978}. Combining Eqs.~(\ref{eq:number_flux_total}) and (\ref{eq:fermi_dirac}), we can write a function
with the desired energy dependence that also integrates to $F_{\nu_i}$:
\begin{equation}
\frac{dF_{\nu_i}}{d\epsilon} = \frac{N_{\nu_i}}{4\pi D^2}\frac{1}{\mathcal{F}_2(0)\,kT_{\nu_i}}
                            \frac{(\epsilon/kT_{\nu_i})^2}{[\exp(\epsilon/kT_{\nu_i}) + 1]}.
\end{equation}
The neutrino temperature is defined (consistent with Eq.~\ref{eq:mean_energy}) from the 
mean energy as 
\begin{equation}
\label{eq:mean_energy_temp}
kT_{\nu_i} = \frac{\langle \epsilon_{\nu_i}\rangle}{\mathcal{F}_3(0) / \mathcal{F}_2(0)} 
\simeq \frac{1}{3.1}\langle \epsilon_{\nu_i}\rangle.
\end{equation}
The number flux spectrum in terms of leakage quantities is thus
\begin{equation}
\label{eq:nu_spectrum}
\frac{dF_{\nu_i}}{d\epsilon} = \frac{3.1}{4\pi D^2\,\mathcal{F}_2(0)}\frac{N_{\nu_i}}{\langle \epsilon_{\nu_i}\rangle} 
                            \frac{(3.1\epsilon/\langle \epsilon_{\nu_i}\rangle)^2}{[\exp(3.1\epsilon/\langle \epsilon_{\nu_i}\rangle) + 1]}.
\end{equation}
In the CCSN literature, it is conventional to take 
$\langle \epsilon_{\nu_i}\rangle = [\mathcal{F}_4(0) / \mathcal{F}_3(0)] kT_{\nu_i}\simeq 4.1kT_{\nu_i}$, which corresponds
to an energy-spectrum weighted mean energy. We estimate the uncertainty in our detection rate calculation by 
using both the coefficient $3.1$ and $4.1$ in Eq.~(\ref{eq:nu_spectrum}). The latter coefficient
results in a factor $\sim 2$ decrease in the event rate,
which is generally larger than the shot noise in the signal. 

Flavor transformation  while neutrinos propagate in the source envelope and on their way to Earth  
has significant implications for detectability \cite{mirizzi_2016,janka_2017,horiuchi_2018,mueller_2019}. 
This is particularly important for collapsar accretion disks, because the electron flavor neutrino
and antineutrino luminosities are significantly higher than the sum of all heavy lepton
flavor luminosities (e.g., Paper~I). 
Because of the uncertainties linked to the impact of neutrino-neutrino interactions on flavor conversion physics in the source
\cite{mirizzi_2016,tamborra_2021,volpe_2024},
we  bracket detectability (as in \cite{tamborra_2013}) 
by comparing a case of no flavor transformation (most optimistic), full flavor conversion 
(most pessimistic and extreme),
and an intermediate case assuming adiabatic Mikheyev–Smirnov–Wolfenstein (MSW) 
neutrino flavor conversion in matter, for comparison with previous work.
For the intermediate case, we assume normal mass ordering and ignore neutrino self interaction,
expressing the number fluxes of neutrinos observed on Earth as (\cite{dighe_2000}) 
\begin{eqnarray}
\label{eq:oscillated_luminosities}
\tilde{F}_{{\nu}_e}     &  = & {p}_{ee}F_{{\nu}_e} + (1 - {p}_{ee})F_{{\nu}_x}\\
\tilde{F}_{\bar{\nu}_e} &  = & \bar{p}_{ee}F_{\bar{\nu}_e} + (1 - \bar{p}_{ee})F_{\bar{\nu}_x}\\
\tilde{F}_{{\nu}_x}     &  = & \frac{1}{2}\left(1-{p}_{ee}\right)F_{{\nu}_e} + \frac{1}{2}(1 + {p}_{ee})F_{{\nu}_x}\\
\tilde{F}_{\bar{\nu}_x} &  = & \frac{1}{2}\left(1-\bar{p}_{ee}\right)F_{\bar{\nu}_e} + \frac{1}{2}(1 + \bar{p}_{ee})F_{\bar{\nu}_x}
\end{eqnarray}
where the tilde denotes oscillated quantities, untilded quantities are those from the simulation,
the heavy lepton quantities correspond to one species ($\bar{\nu}_x = \nu_x$), and the survival 
probabilities are
\begin{eqnarray}
p_{ee}       & = & \sin^2\theta_{13}\\
\bar{p}_{ee} & = & \cos^2\theta_{12}\cos^2\theta_{13},
\end{eqnarray}
where we adopt
mixing angles
$\sin^2\theta_{12}=0.303$ and $\sin^2\theta_{13}\simeq 0.022$ \cite{esteban_2020},
hence $\bar{p}_{ee} \simeq 0.682$. 
The case with no flavor transformation (``unoscillated'') 
corresponds to setting tilded quantities
equal to their untilded counterparts, and full flavor conversion (``full swap'') corresponds to 
\begin{eqnarray}
\tilde{F}_{{\nu}_e} = F_{{\nu}_x}; & \tilde{F}_{\bar{\nu}_e} = F_{\bar{\nu}_x}\\
\tilde{F}_{{\nu}_x} = F_{{\nu}_e}; & \tilde{F}_{\bar{\nu}_x} = F_{\bar{\nu}_e}.
\end{eqnarray}

We focus on neutrino detection prospects for the IceCube Neutrino Telescope \cite{Abbasi_2011}. IceCube, 
followed by  Super-Kamiokande, is the largest operating Cherenkov detector sensitive to the signal from collapsing massive stars~\cite{mirizzi_2016,scholberg_2018}. 
Cherenkov neutrino detectors are mainly sensitive to electron antineutrinos through inverse beta decay 
($\bar\nu_e p \rightarrow n e^+$). 

We employ the IceCube rate calculator in
{\tt SNOwGLoBES} \cite{Malmenbeck_2019} through
the {\tt SNEWPY} software package \cite{baxter_2022}. 
A data cube with an energy spectrum 
$dF_{\nu_i}/d\epsilon$ 
for every neutrino flavor and time bin is used as input for the \texttt{RateCalculator} class of \texttt{SNEWPY}, which calculates the count rate of IceCube for possible interaction channels \cite{Malmenbeck_2019}. The sum of all channels is used as a mean value to draw the total rate 
from a Poissonian distribution, to account for statistical fluctuations.
Since the \texttt{SNEWPY} implementation does not incorporate background noise \cite{Malmenbeck_2019}, we 
consider a reference Poissonian background  with a mean of $1.5\times 10^{6}$ counts per second, 
arising from a background rate of $286\,\text{s}^{-1}$ for the
4,800 standard digital optical modules (DOMs) and $1.25\times 286\,\text{s}^{-1}$ for the 360 DeepCore DOMs, in both
cases accounting for $250\mu$s dead time after every count \cite{Abbasi_2011}. As a cross check, we verify that
our detection rates are consistent with the analytic estimate for a CCSN signal in IceCube by \cite{halzen_2009}.

\subsection{GW Signal Extraction \label{s:gw}}

We estimate the GW signal from the time-varying mass quadrupole of the system, 
following previous work on Newtonian simulations (e.g., \cite{oohara_1989,rasio_1992,ruffert_1996,murphy_2009}).
While the contribution from anisotropic neutrino emission to the GW strain in collapsars is estimated to
be larger than that from matter motions \cite{kotake_2012}, the absence of directional
information in our neutrino scheme prevents us from carrying out  a detailed calculation.
The matter quadrupole formula is valid when the GW wavelength
is larger than the source size \cite{gottlieb_2024}. This restriction means that given the size of the disk during the 
NDAF phase ($\sim 10^7-10^8$\,cm), predictions from this approximation are valid up to $\sim 50-500$\,Hz. We thus focus our
analysis on frequencies below this limit.

For an axisymmetric source, only the + mode is non-zero, with a strain amplitude $h_+$ at a distance $D$ given by
(e.g., \cite{finn_1990})
\begin{equation}
\label{eq:Dhplus}
D h_{+} = \frac{3G}{2c^4}\,\sin^2\theta\, \ddot{\mathst{I}}_{\rm zz},
\end{equation}
where $\theta$ is the angle measured from the symmetry axis ($z$), and 
$\ddot{\mathst{I}}_{\rm zz}$ is the second time derivative of 
the $zz$ component of the reduced quadrupole moment (e.g., \cite{rasio_1992})
\begin{eqnarray}
\label{eq:ddIzz_reduced}
\ddot{\mathst{I}}_{\rm zz} & = & 2\int \rho d^3 x\left[v_z^2-\frac{1}{3}\mathbf{v}^2 
   -z\frac{\partial \Phi}{\partial z} + \frac{1}{3}\mathbf{x}\cdot\nabla \Phi\right]\\
   & = & 2\int \rho d^3 x\left[(v_r\cos\theta -v_\theta\sin\theta)^2 - \frac{1}{3}(v_r^2 + v_\theta^2 + v_\phi^2)\right. \nonumber\\ 
   && \qquad\qquad\left. + \frac{2}{3}P_2(\cos\theta) r g_r + \frac{1}{3}\frac{\partial P_2(\cos\theta)}{\partial\theta} r g_\theta\right].
\end{eqnarray}
Here,  $\rho$ is the mass density, $v_z$ is the $z$-velocity, and $\Phi$ is the
gravitational potential. The second equation makes use of quantities obtained directly
from the spherical coordinate grid, including the rotational velocity $v_\phi$, the components of the 
acceleration of gravity $\mathbf{g}=-\nabla\Phi$ and the $\ell=2$ Legendre polynomial $P_2(x) = (3x^2-1)/2$. 

For our axisymmetric source, the luminosity in GWs is given by
\begin{equation}
L_{\rm GW} = \frac{3}{10}\frac{G}{c^5} \left(\threedots{\mathst{I}}_{\rm zz}\right)^2,
\end{equation}
where $\threedots{\mathst{I}}_{\rm zz}$ is the third time derivative of the $zz$ component of
the reduced quadrupole moment (in axisymmetry, $\mathst{I}_{\rm xx}=\mathst{I}_{\rm yy}=-\frac{1}{2}\mathst{I}_{\rm zz}$,
with all other components vanishing, e.g. \cite{finn_1990}). Following \cite{rasio_1992} or \cite{murphy_2009},
we compute $\threedots{\mathst{I}}_{\rm zz}$ by numerically differentiating $\ddot{\mathst{I}}_{\rm zz}$ in time, otherwise
we have to solve a Poisson equation for the time derivative of $\Phi$ to obtain a closed expression for 
$\threedots{\mathst{I}}_{\rm zz}$ (e.g., \cite{blanchet_1990,ruffert_1996}). 

The spectral energy density is obtained from Parseval's theorem. The energy emitted in GWs is:
\begin{eqnarray}
\label{eq:Egw_definition}
E_{\rm GW} & = &\int_{-\infty}^\infty L_{\rm GW}\,dt 
            =    \frac{3}{10}\frac{G}{c^5} \int_{-\infty}^\infty \left(\threedots{\mathst{I}}_{\rm zz}\right)^2\,dt\\
           & = & \frac{3}{10}\frac{G}{c^5} \int_{-\infty}^\infty \big|\widetilde{\threedots{\mathst{I}}}(f)_{\rm zz}\big|^2 df,
\end{eqnarray}
where the Fourier transform of a function $\Upsilon$ is defined as
\begin{equation}
\widetilde{\Upsilon}(f) = \int_{-\infty}^\infty \Upsilon(t) e^{-i2\pi ft}\,dt.
\end{equation}
Using the fact that $\widetilde{\threedots{\mathst{I}}}_{\rm zz}(f) = i2\pi f \widetilde{\ddot{\mathst{I}}}_{\rm zz}(f)$, and that
the Fourier transform of a real function in time is an even function in frequency, we can write
\begin{equation}
\label{eq:Egw_def}
E_{\rm GW} = \frac{3}{5}\frac{G}{c^5} \int_0^\infty (2\pi f)^2\,\big|\widetilde{\ddot{\mathst{I}}}(f)_{\rm zz}\big|^2 df.
\end{equation}
The spectral energy density for an equatorial observer ($\sin\theta = 1$) is then
\begin{eqnarray}
\frac{dE_{\rm GW}}{df} & = & \frac{3}{5}\frac{G}{c^5}(2\pi f)^2\,\big|\widetilde{\ddot{\mathst{I}}}(f)_{\rm zz}\big|^2\\
                       & = & \frac{4}{15}\frac{c^3}{G} D^2 (2\pi f)^2\, \big|\widetilde{h_+}\big|^2.
\end{eqnarray}

The standard approach to estimate detectability is to compute a ``characteristic strain'' \cite{flanagan_1998,ott_2012}:
\begin{eqnarray}
\label{eq:hchar_definition}
h^2_{\rm char}(f) & = &\frac{2}{\pi^2}\frac{G}{c^3}\frac{1}{D^2}\frac{dE_{\rm GW}}{df}\\
                  & = & \frac{32}{15}f^2 \big|\widetilde{h_+}\big|^2\simeq 2 f^2 \big|\widetilde{h_+}\big|^2,
\end{eqnarray}
where zero redshift is assumed for a galactic source. This is to be compared with the root-mean-square 
strain associated with the detector noise
\begin{equation}
h^2_{\rm rms}(f) = f S_h(f),
\end{equation}
where $S_h(f)$ is the spectral density of the strain noise in the detector (units of Hz$^{-1}$).
The signal to noise ratio SNR is then given by
\begin{equation}
\label{eq:snr_def}
(\textrm{SNR})^2 = \int_0^\infty d\ln f\, \frac{h^2_{\rm char}}{h^2_{\rm rms}}.
\end{equation}
For a given source, Eq.~(\ref{eq:snr_def}) implies that the SNR scales inversely to the product $\sim D\sqrt{S_h(f)}$.

In the remainder of the paper, we present results for an equatorial observer ($\sin\theta=1$), given that the
angular dependence {of the signal computed with this method} is straightforward 
(i.e., no GW signal when the system is viewed along the rotation axis). 
{However, allowing for deviations from axisymmetry, anisotropic neutrino 
emission \cite{kotake_2012},  and higher frequency matter components 
can result in significant GW emission along the rotation axis ($\sin\theta=0$).}

\subsection{Shock Analysis \label{s:shock}}

As in Paper I, we characterize the evolution of the shock surface $R_{\rm s}(\theta,t)$ that encloses the accretion disk
in terms of an expansion in Legendre polynomials $P_\ell$
\begin{equation}
R_{\rm s}(\theta,t) = \sum_\ell a_\ell(t) P_\ell(\cos\theta),
\end{equation}
where the coefficients are defined by
\begin{equation}
\label{eq:legendre_coeff}
a_\ell(t) = \frac{2\ell + 1}{2}\int_{-1}^1 R_{\rm s}(\theta,t) P_\ell(\cos\theta) d(\cos\theta).
\end{equation}
The position of the shock surface is found following the method of Paper I, using
discontinuities in the pressure, velocity, and/or ${}^{56}$Ni mass fraction depending on position
and model.

Since we are only interested in large-scale deformations of the shock, we focus the discussion on the average
shock radius $a_0$, and the normalized dipole $a_1/a_0$ and quadrupole $a_2/a_0$ coefficients. 

\begin{figure*}
\includegraphics*[width=\textwidth]{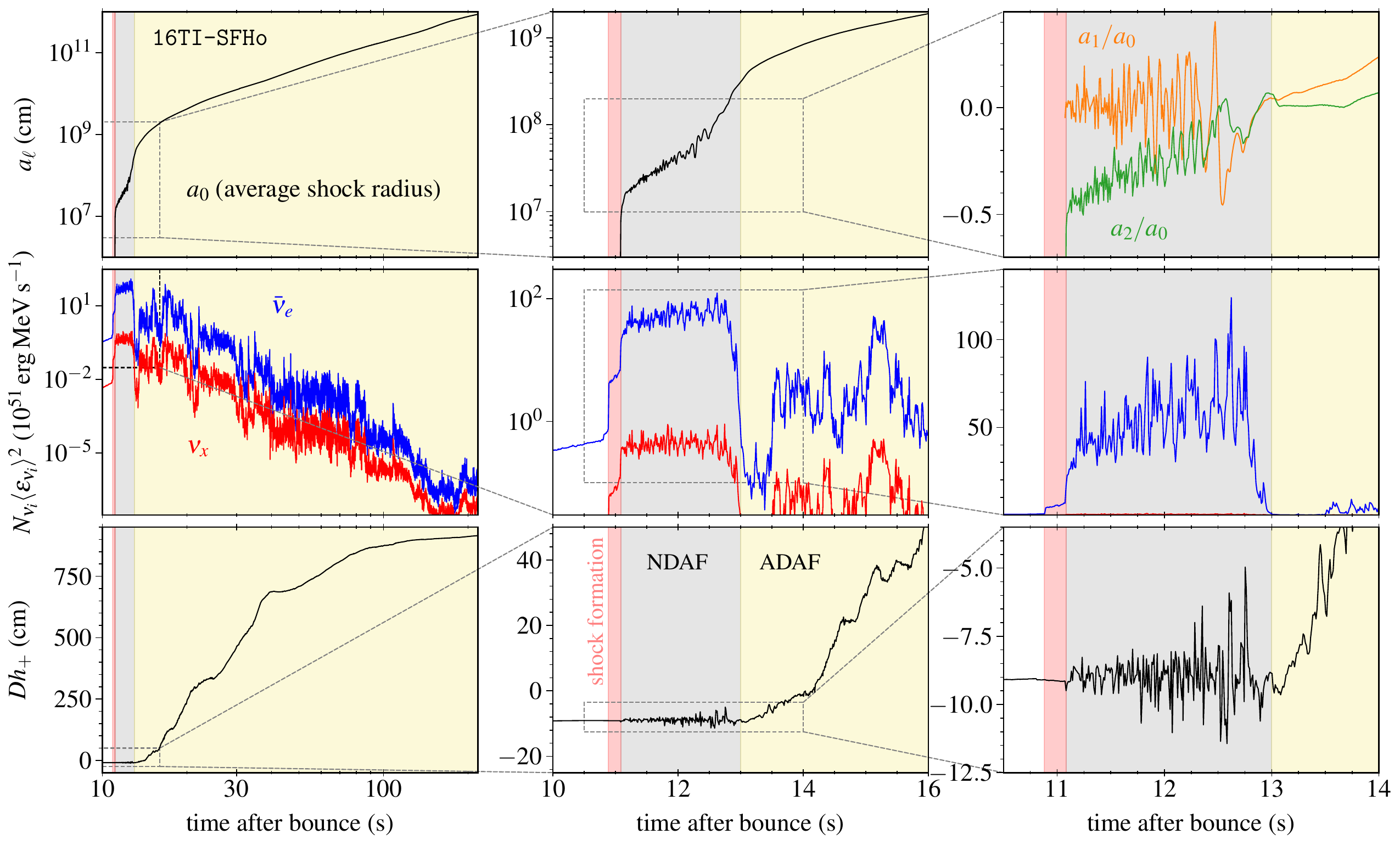}
\caption{Key physical quantities and relevant phases in the evolution of the baseline
model {\tt 16TI-SFHo}. Shown are the shock radius coefficients (top row, Eq.~\ref{eq:legendre_coeff}),
product of number luminosity times mean energy for electron antineutrinos $\bar{\nu}_e$ and one species
of heavy lepton neutrinos $\nu_x$ (middle row), and GW displacement 
for an equatorial observer (bottom row, Eq.~\ref{eq:Dhplus} with $\sin\theta=1$).
Panels on the right are zoom-ins of panels  to their left, as shown by dashed lines.
The pink, gray, and yellow shaded regions encompass shock formation, NDAF phase, and
ADAF phase, respectively. The average shock radius (monopole) $a_0$ 
illustrates the timing of formation, oscillations, and expansion of the shock, while the normalized dipole ($a_1/a_0$) 
and quadrupole ($a_2/a_0$) coefficients show the amplitude of large-scale oscillations during the NDAF phase.
The product $N_{\nu_i}\langle \epsilon_{\nu_i}\rangle^2$ is proportional to the neutrino absorption rate via inverse beta decay, 
and thus assesses detectability in IceCube. The large difference between $\bar{\nu}_e$ and $\nu_x$
illustrates the sensitivity of neutrino detection to flavor transformation.
}
\label{fig:shock_strain_rate}
\end{figure*}

\section{Results \label{s:results}}

\subsection{Overall Signal}

We first illustrate the general evolution of observables relying on the
baseline model {\tt 16TI-SFHo}. Fig.~\ref{fig:shock_strain_rate} 
shows the concurrent evolution of the shock surface, neutrino emission, and GW displacement
over various timescales. The shaded areas indicate distinct phases in the evolution.
Note that for the period shown, BH formation has already occurred (at $2.72$\,s post-bounce for this model), prior to which
there would be neutrino and GW emission associated to the PNS accretion phase. We do not
show the latter here as it was obtained using 1D evolution; see Paper I for neutrino luminosities
pre- and post-BH formation. 

The NDAF phase takes place over the interval $11-13$\,s post-bounce, where neutrino
emission is maximal and both the shock surface and GW strain display oscillatory behavior.
The average shock radius $a_0$ expands gradually during this phase, and the amplitude
of the GW displacement ($Dh_+\sim 5$\,cm) is comparable to the
GW signal induced by SASI motions in the PNS accretion phase of CCSNe (e.g., \cite{shibagaki_2021}).
This phase is the most promising for the detection of time variations with multiple cosmic messengers.

The neutrino detectability  is assessed in
Fig.~\ref{fig:shock_strain_rate} by the product of number
luminosity times squared mean energy
at the source. This product is proportional to the number flux times cross section for inverse
beta decay, which is the dominant detection process for IceCube \cite{Abbasi_2011}.
We show electron-type antineutrinos as well as one species of heavy lepton neutrinos
to illustrate why flavor transformation can have a significant impact on detectability
(the two curves differ by a factor $100$).

Just prior to the NDAF phase, there is a step-like increase in neutrino emission indicated
by the pink shaded area. This corresponds to the process of shock formation, which
involves an equatorial buildup of material, increasing the density and thus neutrino 
emission (cf. Paper I). The average shock radius $a_0$ is only computed 
after the full shocked bubble forms, once 
this transient shock formation stage ends. 
The GW displacement has little sensitivity to the shock formation process.

The end of the NDAF phase is marked by a rapid drop in neutrino emission, at which point
the disk enters the ADAF phase. Shock oscillations freeze out and the average shock
radius starts to expand at a faster rate. Neutrino emission during the ADAF phase
can be non-negligible albeit stochastic in nature, due to the highly
turbulent state of the disk. As the disk accretes and decreases in density, neutrino
emission gradually decreases on a timescale of $\sim 30$\,s.
The GW signal in this phase displays low frequency
changes while steadily increasing, reaching  $Dh_+\sim 800$\,cm over a timescale of $\sim 200$\,s.
This is the characteristic ``memory signal'' associated with stellar explosions (e.g., \cite{choi_2024}).

\subsection{Neutrino Detectability}

\begin{figure*}
\includegraphics*[width=\columnwidth]{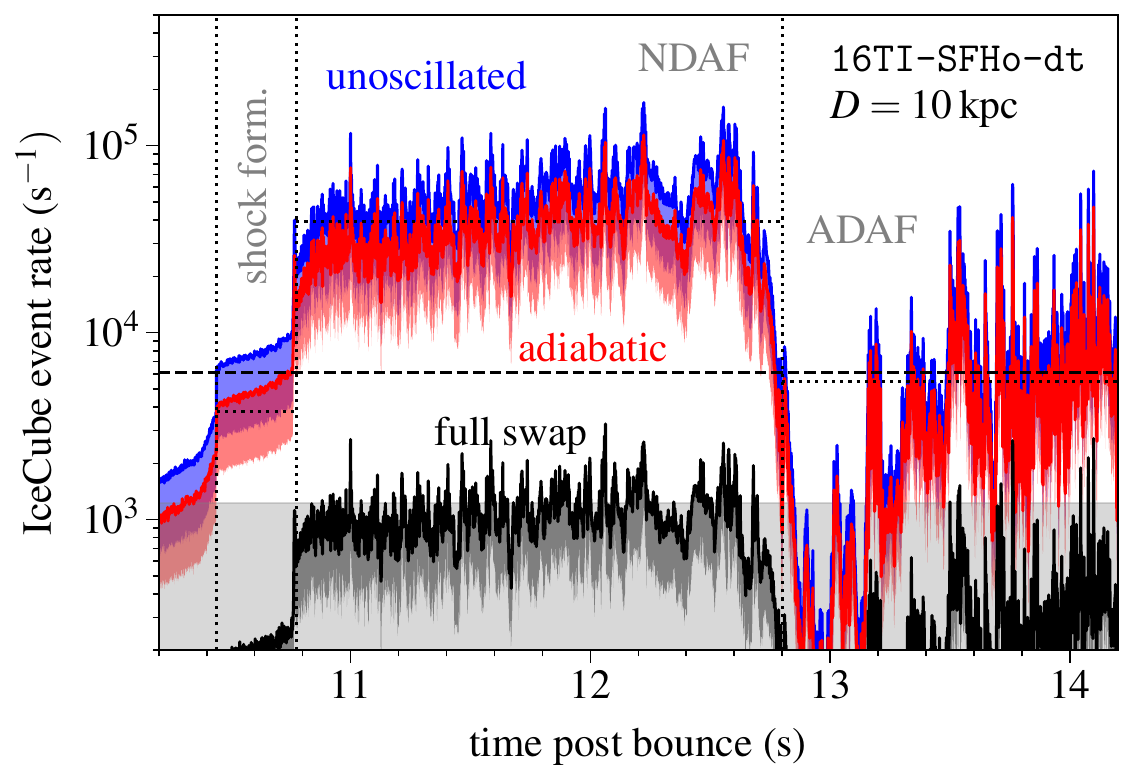}
\includegraphics*[width=\columnwidth]{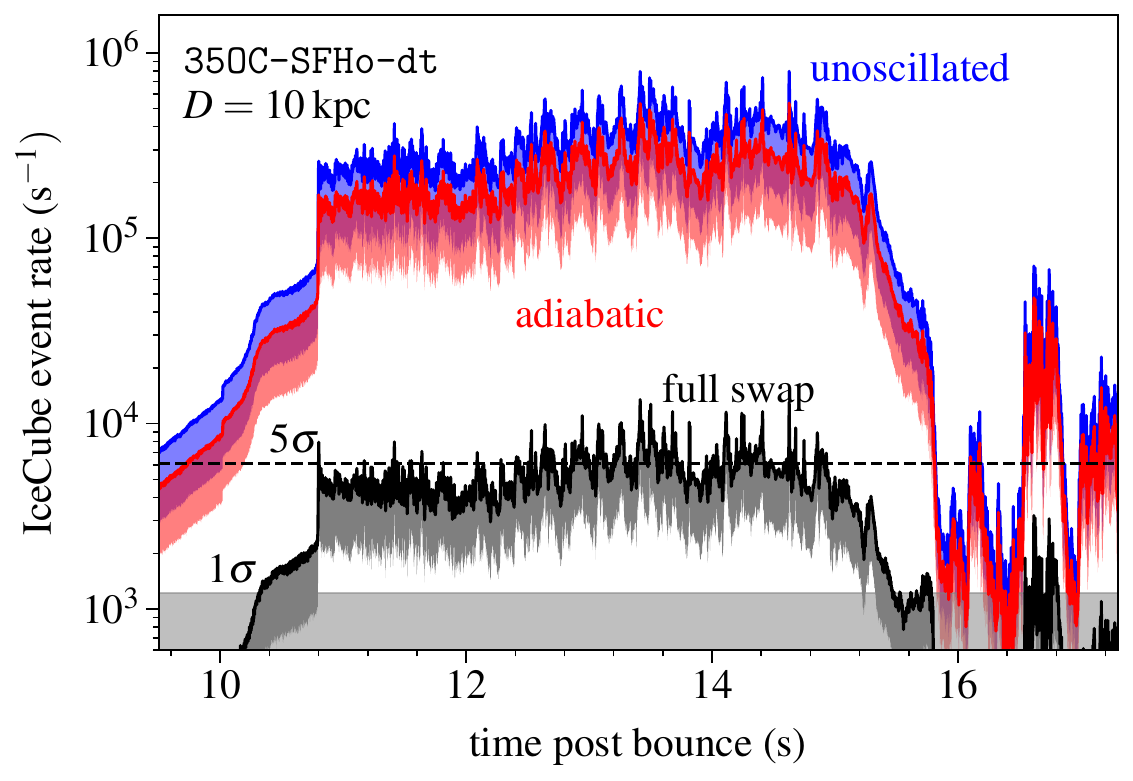}
\includegraphics*[width=\columnwidth]{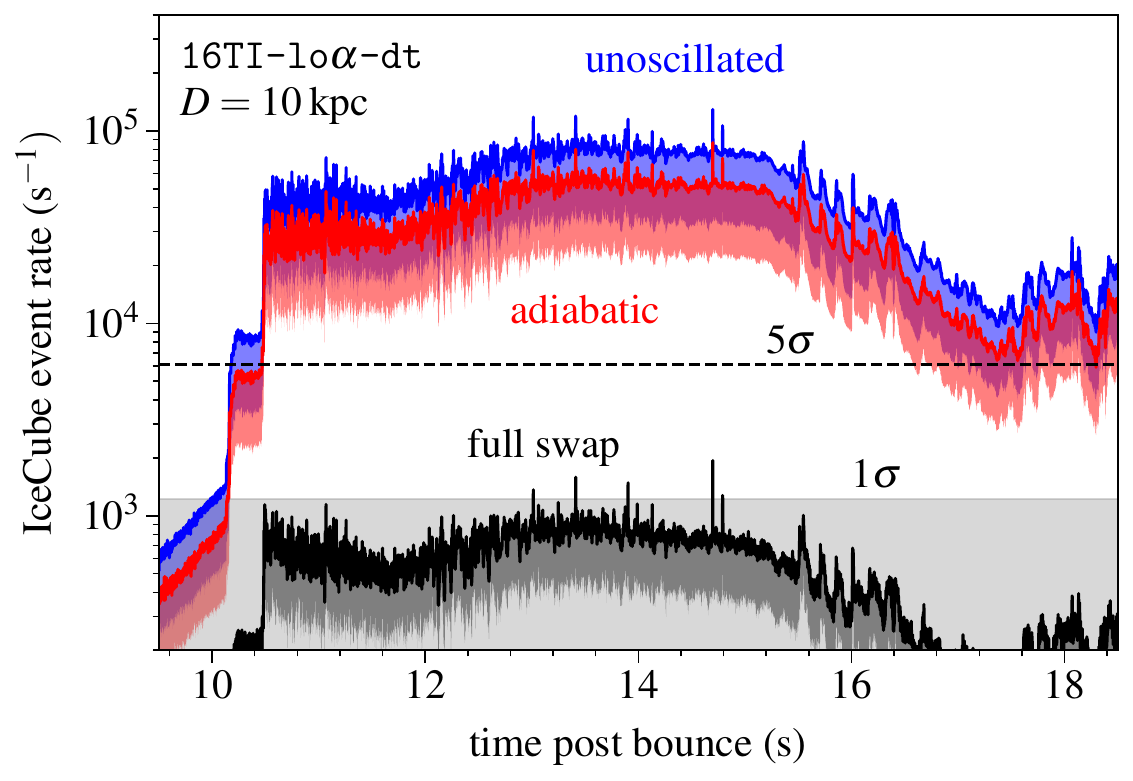}
\includegraphics*[width=\columnwidth]{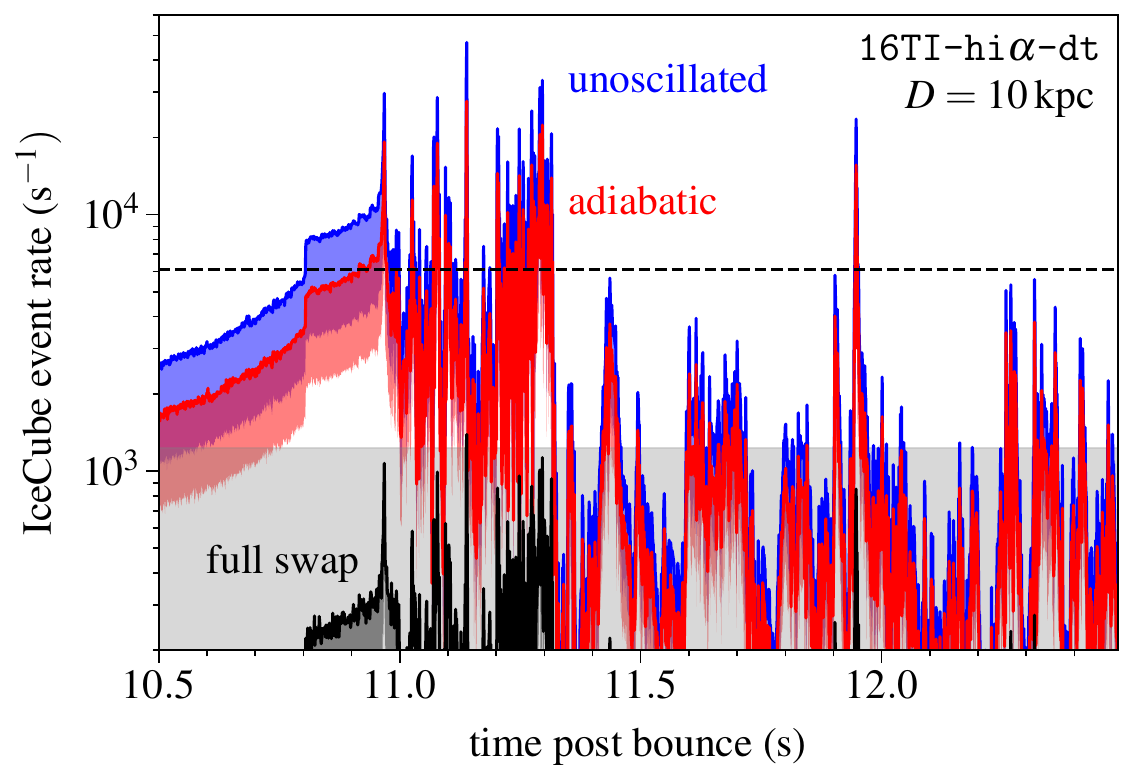}
\caption{IceCube event rate over the period of shock formation, NDAF phase, and onset of 
explosion, for models {\tt 16TI-SFHo-dt} (upper left) and {\tt 35OC-SFHo-dt} (upper right),
{\tt 16TI-lo$\alpha$-dt} (lower left), and {\tt 16TI-hi$\alpha$-dt} (lower right),
assuming a fiducial distance of $D = 10$\,kpc. Blue, red, and black curves correspond to unoscillated, adiabatic,
and full swap flavor transformation scenarios, respectively (\S\ref{s:neutrinos}). 
The semi-transparent region below each curve indicates
the uncertainty range obtained by using $\langle \epsilon_\nu\rangle = 4.1kT_\nu$ instead of Eq.~(\ref{eq:mean_energy_temp}) 
when computing the neutrino spectrum for input in {\tt SNOwGLoBES} (\S\ref{s:neutrinos}). 
This uncertainty is larger than the shot noise in the signal.
The light gray horizontal band at the bottom indicates the $1\sigma$ 
detection threshold, and the dashed line the $5\sigma$ threshold. 
The dotted lines in the upper left panel indicate the time ranges and event rate levels 
(adiabatic case with default mean energy coefficient) 
used to compute the detection horizons in Table~\ref{tab:neutrino}, see text for details.
At this distance, the NDAF phase is detectable at high significance 
for likely flavor transformation scenarios.
The high viscosity model
{\tt 16TI-hi$\alpha$-dt} does not experience an NDAF phase.  
}
\label{fig:neutrino_rate}
\end{figure*}

Figure~\ref{fig:neutrino_rate} shows the IceCube event rate at $10$\,kpc
for  four of the models in Table~\ref{tab:models}, over a period of time around the NDAF phase:
the baseline model {\tt 16TI-SFHo-dt},
varying the stellar progenitor ({\tt 35OC-SFHo-dt}), 
and varying the viscosity parameter
({\tt 16TI-lo$\alpha$-dt} and {\tt 16TI-hi$\alpha$-dt}).  
The 16TI and 35OC progenitors
differ in their core compactness and hence have different overall luminosities in the NDAF phase.
The strength of viscosity determines the existence and duration of the NDAF phase, with the high-viscosity
model experiencing dynamically negligible neutrino cooling except for the lead up to shock formation.
The low-viscosity model has an  NDAF phase longer than the baseline model, extending to the
end of the period shown.

The NDAF phase, when present, dominates the signal, as a plateau in the event rate with rapid time
variations superimposed. We study these time variations in detail in \S\ref{sec:time_variations}, for now
we focus on the overall detectability of the various components of the neutrino signal.
The IceCube event rates for the unoscillated and adiabatic flavor transformation cases 
are significantly higher than for the full swap case, as expected, given the lower heavy lepton luminosity relative to
electron-flavor neutrinos. For all models shown in Fig.~\ref{fig:neutrino_rate}, 
the unoscillated and adiabatic cases result in detection
of the NDAF phase
at $10$\,kpc with $5\sigma$ significance, whereas the full swap case is only detectable for the 
35OC progenitor. The compactness of the stellar model and the flavor conversion scenario 
are thus crucial for determining the detectability of the neutrino signal.

\begin{table*}
\caption{
IceCube $5\sigma $ detection horizon $D_{\rm max}$ (kpc) for the three 
neutrino signatures from collapsar disks identified here, assuming different flavor transformation scenarios, 
for all {\tt -dt} models (cf.~Table~\ref{tab:models}).
The lower and upper distance range 
corresponds to using $\langle \epsilon_\nu\rangle = 4.1kT_\nu$ and $3.1kT_\nu$ in in Eq.~(\ref{eq:mean_energy_temp}),
respectively, when constructing the spectral flux for {\tt SNOwGLoBES} (\S\ref{s:neutrinos}). 
The shock formation detection horizon is obtained by setting the onset of the
neutrino precursor to disk formation equal to the $5\sigma$ detection threshold. 
The NDAF horizon is obtained by computing the root-mean-square (rms) of the event rate over the duration of this
phase and setting it equal to the $5\sigma$ threshold. Likewise, the ADAF horizon is obtained by computing the 
rms event rate over an interval of $2$\,s starting at the transition from NDAF to ADAF. 
The dotted lines in the upper-left panel of Fig.~\ref{fig:neutrino_rate}
show these times and event rate levels for model {\tt 16TI-SFHo-dt}. Values for the corresponding low output frequency models 
(without {\tt -dt} in their names) are very similar and not shown for conciseness. 
The high-viscosity model does not experience an NDAF phase, and the low-viscosity
{\tt -dt} model does not reach the ADAF phase over the time simulated.
\label{tab:neutrino}} 
\begin{ruledtabular}
\begin{tabular}{l|ccc|ccc|ccc}
Model &  \multicolumn{3}{c}{Shock Formation precursor} & \multicolumn{3}{c}{NDAF rms}  & \multicolumn{3}{c}{ADAF rms.}\\
      & unosc. & adiabatic & swap & unosc. & adiabatic & swap & unosc. & adiabatic & swap\\
\hline
{\tt 16TI-SFHo-dt}        & 6.5--10  & 5.3--7.9  & 1.1--1.8  & 20--31  & 17--25 & 2.9--4.2 & 7.6--12  & 6.3--9.5 & 1.5--2.2\\
{\tt 16TI-lo$\alpha$-dt}  & 6.0--9.5 & 5.0--7.3  & 0.95--1.6 & 20--31  & 17--25 & 2.3--3.2 & --       &    --    &     --  \\
{\tt 16TI-hi$\alpha$-dt}  & 7.2--11  & 5.9--8.9  & 1.3--1.9  &   --    &   --   &    --    & 5.1--7.8 & 4.2--6.3 & 1.0--1.5\\ 
{\tt 16TI-DD2-dt}         & 5.5--8.6 & 4.5--6.8  & 0.88--1.5 & 20--30  & 16--24 & 2.7--3.9 & 2.4--3.6 & 2.0--3.0 & 0.48--0.72\\
{\tt 35OC-SFHo-dt}        & 17--27   & 14--21    & 3.1--4.7  & 48--73  & 40--60 & 6.4--9.3 & 11--17   & 9.2--14  & 2.4--3.6\\
\end{tabular}
\end{ruledtabular}
\end{table*}

Formation of the shock wave that bounds the accretion disk  
is not instantaneous, and has a clear neutrino signature in the form of a step-like increase
in event rate prior to the onset of the NDAF phase (pink shaded region in Fig.~\ref{fig:shock_strain_rate}).
As the rotating star collapses, a 
\emph{dwarf disk} forms (e.g., Paper I), increasing gradually in density and accounting for a steady
rise in neutrino emission following BH formation. The transition from \emph{dwarf disk} to fully shocked
disk involves an initially equatorial shock that emerges from the inner edge of the 
dwarf disk, due to pileup of material with angular momentum above the local Keplerian value outside the ISCO.
This equatorial shock generates a step-like increase in neutrino emission, visible at $t=10.4$\,s and
$10.0$\,s for models {\tt 16TI-SFHo-dt} and {\tt 35OC-SFHo-dt}, respectively (Fig.~\ref{fig:neutrino_rate}).
This precursor shock expands for $\sim 0.3$\,s until a more spherical shocked bubble forms, which
grows quickly, 
resulting in a second rapid increase in neutrino emission
at $t=10.8$\,s and $10.4$\,s for models {\tt 16TI-SFHo-dt} and {\tt 35OC-SFHo-dt}, respectively. 
This second increase marks the beginning of the NDAF phase.

The shock formation signature in the neutrino light curve is detectable with $5\sigma$
signficance at a distance of $10$\,kpc
for model {\tt 35OC-SFHo-dt} in the adiabatic and unoscillated cases, but barely in model 
{\tt 16TI-SFHo-dt} in the optimistic unoscillated case.
While the shock formation signature is present in all of our models, 
the detailed temporal shape  
can be dependent on the physics used to model it (EOS, neutrino scheme, or MHD), as well as 
on the angular momentum distribution of the progenitor star.
An in-depth study of all the dependencies of this
precursor shock signal is left for future work.

The transition from NDAF to ADAF phase is marked by a rapid drop in the event rates from the 
NDAF plateau, as shown in 
Fig.~\ref{fig:neutrino_rate}. Thereafter, stochastic
enhancements in the neutrino emission 
persist over an extended period of time ($\sim 10$\,s). 
These bursts of emission are associated
with turbulent fluctuations in the inner disk, which bring the maximum density above 
$\sim 10^9$\,g\,cm$^{-3}$, temporarily enhancing neutrino emission and accretion.
This behavior can be described in terms of the ``ignition accretion rate'' \cite{chen_2007,de_2021}:
turbulent flutuations in the ADAF phase bring the accretion rate episodically
above this threshold. Fig.~\ref{fig:shock_strain_rate} shows that this 
turbulent behavior continues for the rest of the ADAF phase, along with a gradual
decrease of the overall level of emission over a timescale of tens of seconds.

Figure~\ref{fig:neutrino_rate} shows that these ADAF
spikes in neutrino emission are detectable with $5\sigma$ 
significance 
at $10$\,kpc for all models shown in the
unoscillated and adiabatic flavor transformation cases.
These episodic spikes in neutrino emission
should be a robust feature of the ADAF phase, although
their detailed temporal spectrum should depend of the nature of the underlying turbulence.
In our simulations, this turbulence is driven thermally by viscous heating, whereas in
realistic collapsars it should be driven by the magnetorotational instability (e.g., \cite{balbus_1998}).

Table~\ref{tab:neutrino} summarizes the $5\sigma$ detection horizon in IceCube for the
three neutrino signatures of collapsar disks discussed thus far. 
The detection horizon for the NDAF phase is obtained by 
setting the root-mean-square (rms) event rate over the
duration of this phase
equal to the $5\sigma$ threshold. The time interval spanning the NDAF phase as well
as the rms
event rate obtained is shown with dotted lines in the upper left panel of Fig.~\ref{fig:neutrino_rate}.
This is a conservative estimate of the position of the 
plateau in neutrino emission; if we take the maximum of the spikes during the NDAF phase we
obtain a detection horizon $\sim 50\%$ larger.
IceCube would be able to detect this signature for
a galactic collapsar that resembled any of our models  
undergoing an NDAF phase, 
as long as flavor transformation is minimal.  
A more pessimistic full swap case reduces the detection distance to a few kpc. 
 
Since neutrino emission prior to BH formation is expected to be 
stronger than in the NDAF phase (e.g., Paper I), the rapid drop in the event
rate at BH formation will be followed by a period of a few seconds without detection, until
the neutrino-cooled disk signature emerges. This characteristic temporal pattern encodes
information about the rotation
rate and density profile of the progenitor star, which determine disk formation. The extent of the
NDAF phase would provide additional information about the disk thermodynamics, as well as the angular momentum
transport mechanism and/or the presence of a large-scale magnetic field 
(cf. the general relativistic MHD simulations of \cite{issa_2025}).

Also shown in Table~\ref{tab:neutrino} is the detection horizon for the precursor to disk formation, 
which is obtained by setting the event rate at the onset of the equatorial shock precursor equal 
to the $5\sigma$ threshold (cf. upper left panel of Fig.~\ref{fig:neutrino_rate}). 
This signature is, by definition, fainter than the NDAF plateau, and thus 
has a smaller detection horizion (few kpc or less for all models except {\tt 35OC-SFHo-dt}). 
Finally, the detecton horizon for ADAF neutrino bursts
is also obtained as the rms event rate over an interval of 2\,s starting at the
transition to the ADAF phase (cf. Fig.~\ref{fig:neutrino_rate}).
Like our estimate for the NDAF phase, this ADAF detection horizon
is also conservative, yielding values similar to those of the shock formation precursor. 
Fig.~\ref{fig:neutrino_rate} shows that the spikes in emission can easily reach the NDAF plateau value, hence
the actual detection horizon is likely larger.

\begin{figure*}
\includegraphics*[width=0.49\textwidth]{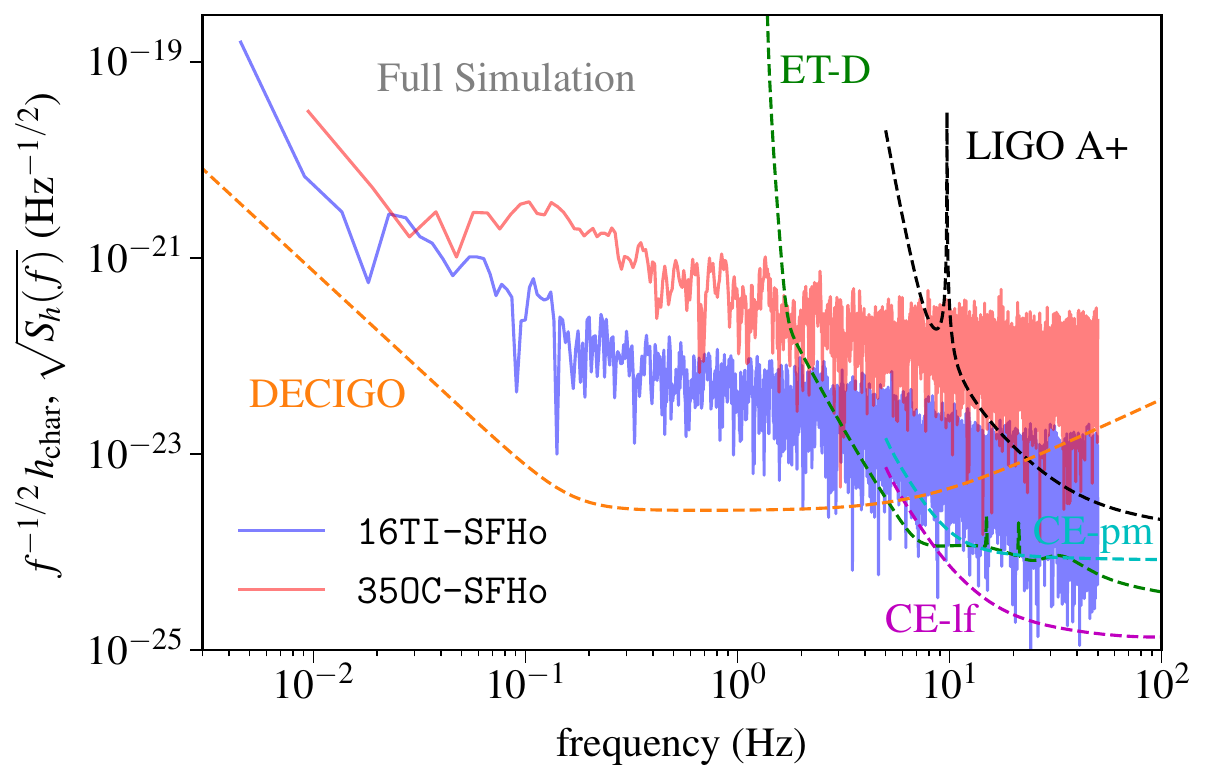}
\includegraphics*[width=0.49\textwidth]{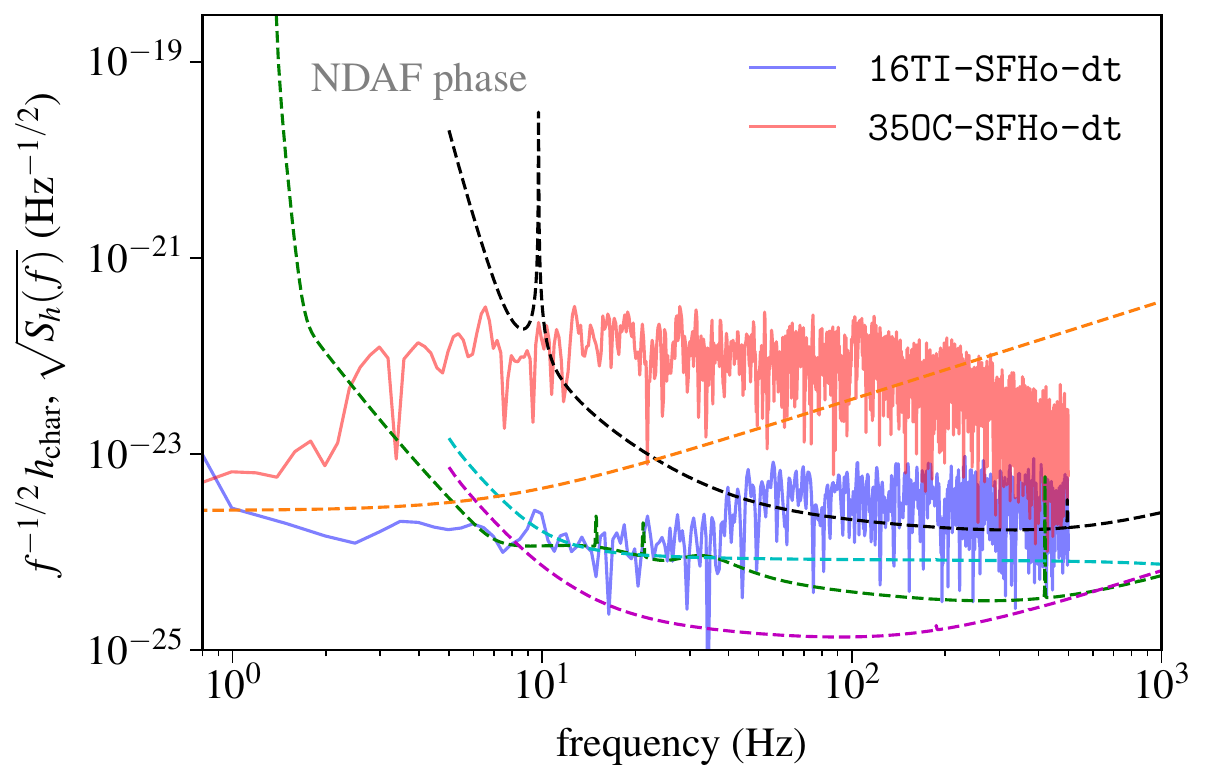}
\caption{\emph{Left:} Characteristic strain $h_{\rm char}$ (Eq.~\ref{eq:hchar_definition}) at $D=10$\,kpc 
for an equatorial observer, for models {\tt 16TI-SFHo} and {\tt 35OC-SFHo}, 
calculated from the waveform over the entire simulation (cf. Table~\ref{tab:models}),
and applying a Hann window (i.e., a ``short Fourier transform'').
Also shown with dashed lines are the strain noise sensitivities for LIGO A+ \cite{ligo_lrr_2020}, 
Cosmic Explorer for 20\,km post-merger (CE-pm) and 40\,km low-frequency (CE-lf) configurations \cite{srivastava_2022}, 
Einstein Telescope (ET-D \cite{hild_2011}), and Deci-Herz Interferometer 
Gravitational Wave Observatory (DECIGO, \cite{yagi_2017}), as labeled on each curve.
\emph{Right:} Same as the left panel, but now using the {\tt -dt} models, and restricting the
time period to the NDAF phase (cf. Fig.~\ref{fig:neutrino_rate}), over which the Hann window is applied.
}
\label{fig:gw_spec}
\end{figure*}

\subsection{Gravitational Wave Detectability}

The characteristic GW strain ($h_{\rm char}$, Eq.~\ref{eq:hchar_definition}) 
for models {\tt 16TI-SFHo} and {\tt 35OC-SFHo} (as a representative subset) is shown in
Fig.~\ref{fig:gw_spec}. 
Given that the frequency range is dependent on the rate of spatial output of the simulations ($f_s$),
we show the spectrum for the original runs from Paper I (Table~\ref{tab:models}), 
which cover low frequencies but sample at $f_{\rm s}=100$\,Hz. Separately, we show the
re-runs ({\tt -dt} suffix) of these models, which extend to higher frequencies 
but have a limited duration in time, so we restrict them to the NDAF phase.

As with the neutrino signal, the higher compactness of the 35OC
progenitor results in more power emitted in GWs than for model {\tt 16T-SFHo} at all frequencies.
The long-duration models {\tt 35OC-SFHo} and {\tt 16TI-SFHo} have similar overall spectral shapes,
with model-specific differences manifesting around frequencies $\sim 0.1$\,Hz, reflecting the fact that
the path to explosion is different for each model on timescales of $10-100$\,s.
Over the NDAF phase, the {\tt -dt} models show more distinct spectral shapes, in addition to
an overall normalization difference. Model {\tt 16TI-SFHo-dt} has a rising spectrum
in the frequency range $10-100$\,Hz, whereas the spectrum is essentially flat in that same 
range for model {\tt 35OC-SFHo-dt}. {Note that the amplitude of the signal above 
$\sim 100$\,Hz is likely overestimated by the quadrupole formula, since it is beyond
its applicability limit; Ref.~\cite{gottlieb_2024} employs 
an alternative approach suitable for such high-frequency regime based on Green's function.}
As we elaborate in the next subsection, the GW strain {at frequencies $10-100$\,Hz} 
is a sensitive measure of the global shocked cavity that contains the disk, and it is thus influenced
by turbulence.

\begin{table*}
\caption{GW detection horizon $D_{\rm max}$ and energy emitted $E_{\rm GW}$ (Eq.~\ref{eq:Egw_definition}). 
The top group of rows shows results for {\tt -dt} models, restricting
the waveform to the time period of the NDAF phase (high viscosity model not shown: no NDAF phase). 
The detection horizon is computed assuming $\textrm{SNR} = 8$ (Eq.~\ref{eq:snr_def}) 
for an equatorial observer $(\sin\theta = 1)$, and using the sensitivity function $S_h(f)$
from LIGO A+ \cite{ligo_lrr_2020},  Cosmic Explorer for 20\,km post-merger (CE-pm) 
and 40\,km low-frequency (CE-lf) configurations \cite{srivastava_2022}, and the
Einstein Telescope (ET-D \cite{hild_2011}). The bottom group of rows shows results
for models from Paper I over the entire simulation ($\sim 100$\,s), using
the sensitivity function of the Deci-Herz Interferometer Gravitational Wave Observatory (DECIGO, \cite{yagi_2017})
to compute the detection horizon.
\label{tab:gw}} 
\begin{ruledtabular}
\begin{tabular}{lcccccc}
Model &  \multicolumn{5}{c}{$D_{\rm max}$ (in kpc, $\textrm{SNR}=8$, $\sin\theta=1$)} & $E_{\rm GW}$\\
      & LIGO A+ & CE-pm & CE-lf & ET-D & DECIGO &  ($10^{-12}\,M_\odot c^2$)   \\
\noalign{\smallskip}
\hline
\noalign{\smallskip}                                     
NDAF phase ($f \sim 1-500$\,Hz)  &     &     &     &     &      &     \\ 
\noalign{\smallskip}                                     
{\tt 16TI-SFHo-dt}       & 4.1  & 10  & 53  & 22  & & 1.9  \\
{\tt 16TI-lo$\alpha$-dt} & 13  & 56  & 270  & 78  & & 11  \\
{\tt 16TI-DD2-dt}        & 6.4 & 21  & 110  & 35  & & 2.9 \\
{\tt 35OC-SFHo-dt}       & 84 & 320 & 1600  & 500 & & 420  \\
\noalign{\smallskip}                                     
\hline 
\noalign{\smallskip}                                     
Full Simulation ($f \sim 0.01-50$\,Hz) &     &     &     &     &      &     \\ 
\noalign{\smallskip}                                     
{\tt 16TI-SFHo}          &  & & & & 110   & 2.6 \\
{\tt 16TI-lo$\alpha$}    &  & & & & 81    & 2.4 \\
{\tt 16TI-hi$\alpha$}    &  & & & & 28    & 0.8 \\
{\tt 16TI-DD2}           &  & & & & 53    & 0.8 \\
{\tt 35OC-SFHo}          &  & & & & 910   & 56  \\
\end{tabular}
\end{ruledtabular}
\end{table*}

Also shown in Fig.~\ref{fig:gw_spec} are the strain noise sensitivity curves
for LIGO A+ \cite{ligo_lrr_2020} as well as third-generation detectors. Table~\ref{tab:gw} shows
the corresponding detection horizon $D_{\rm max}$, assuming $\text{SNR}=8$ (Eq.~\ref{eq:snr_def}) for each detector. 
The NDAF phase alone can be detected out to several kpc by LIGO A+ in the case of the baseline {\tt -dt}
model. The low-viscosity model has a longer NDAF phase (by a factor 3), with a corresponding
increase in the detection horizon. For the 35OC progenitor, the NDAF phase can be detected
within the entire galaxy and its satellites. Thus, detailed properties of the progenitor and physics 
involved (e.g., magnetic fields that regulate angular momentum transport) can greatly influence
detectability of the NDAF phase in GWs. {We caveat} 
that the detection horizons shown {include contributions from frequencies for which} the quadrupole formula becomes invalid, 
{hence} a full general-relativistic
treatment {is required} to 
{quantify high-frequency} components {and
obtain a more reliable estimate in the LIGO A+ band} (e.g., \cite{gottlieb_2024,yuan_2025}).

Regarding third generation GW detectors, the Einstein Telescope (ET-D, \cite{hild_2011})
and Cosmic Explorer \cite{srivastava_2022} in 20\,km post-merger (CE-pm) and 40\,km low-frequency (CE-lf)
configurations extend coverage to lower frequencies (in addition to more sensitivity)
relative to LIGO A+. The resulting detection horizons are significantly higher,
with CE-lf extending the detectability of the NDAF phase up to $1.6$\,Mpc for the 35OC progenitor.

\begin{figure}
\includegraphics*[width=\columnwidth]{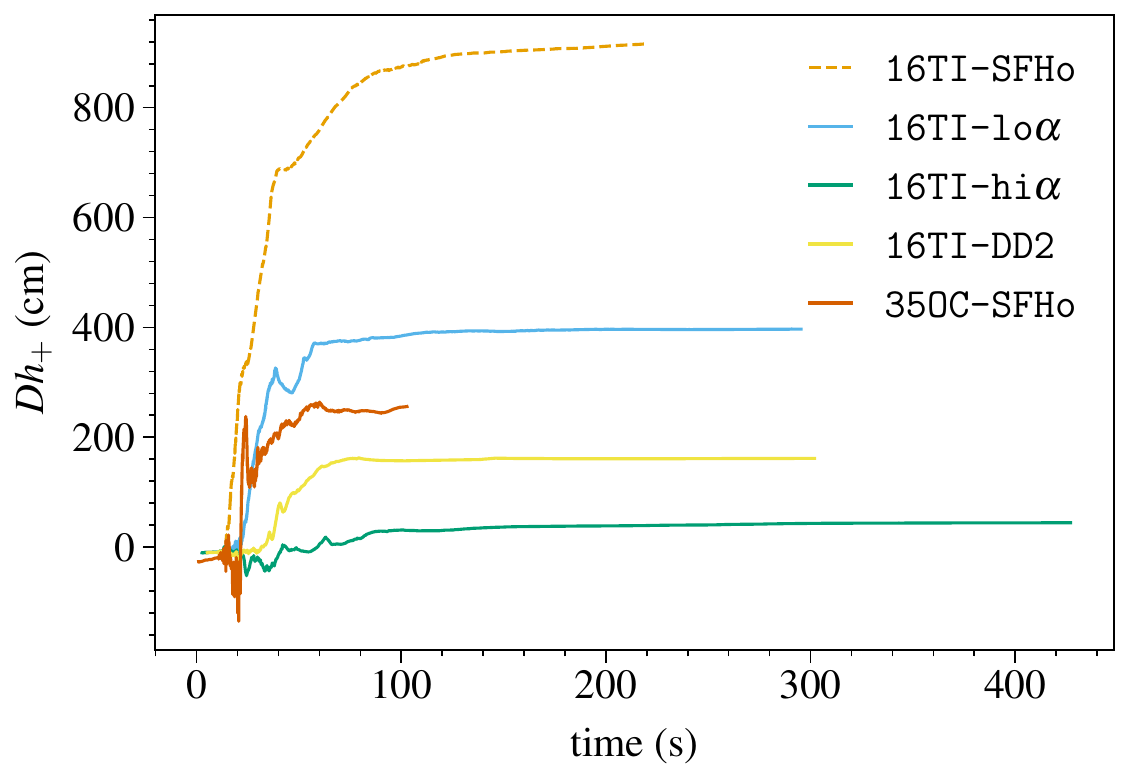}
\caption{Gravitational wave displacement (Eq.~\ref{eq:Dhplus}) as a function of time, for 
the models originally presented in Paper~I, as labeled. }
\label{fig:gw_memory}
\end{figure}

\begin{figure*}
\includegraphics*[width=0.985\columnwidth]{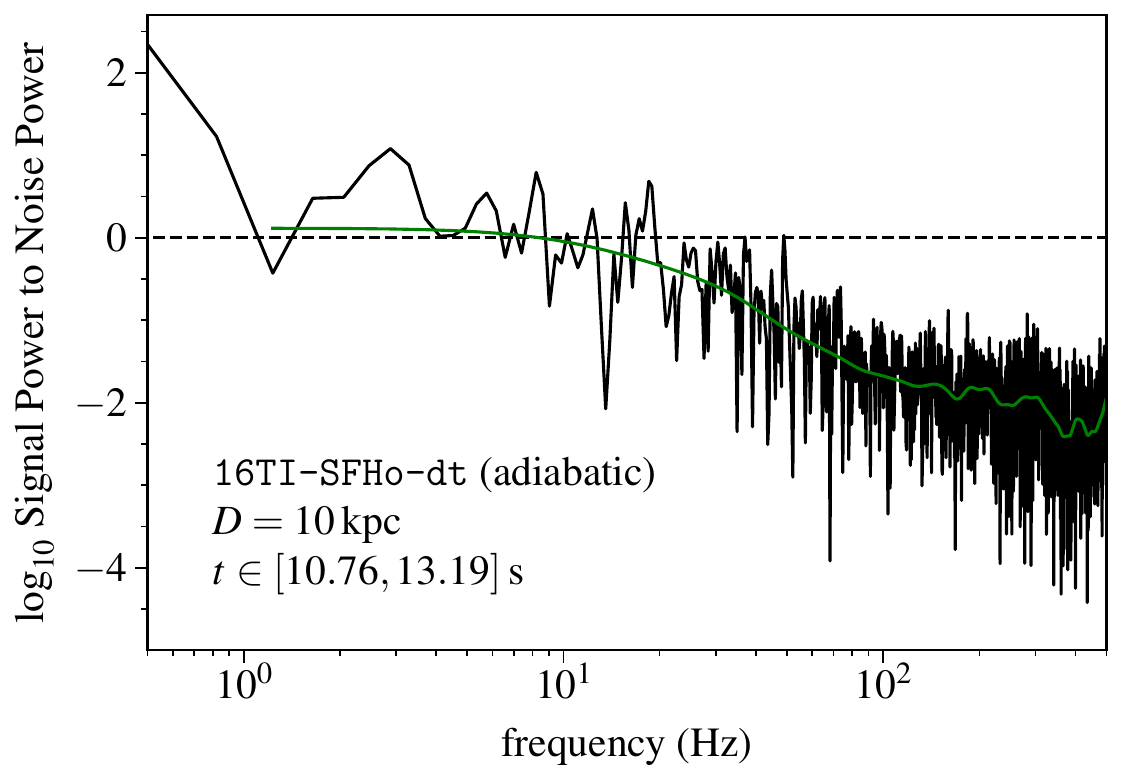}
\includegraphics*[width=\columnwidth]{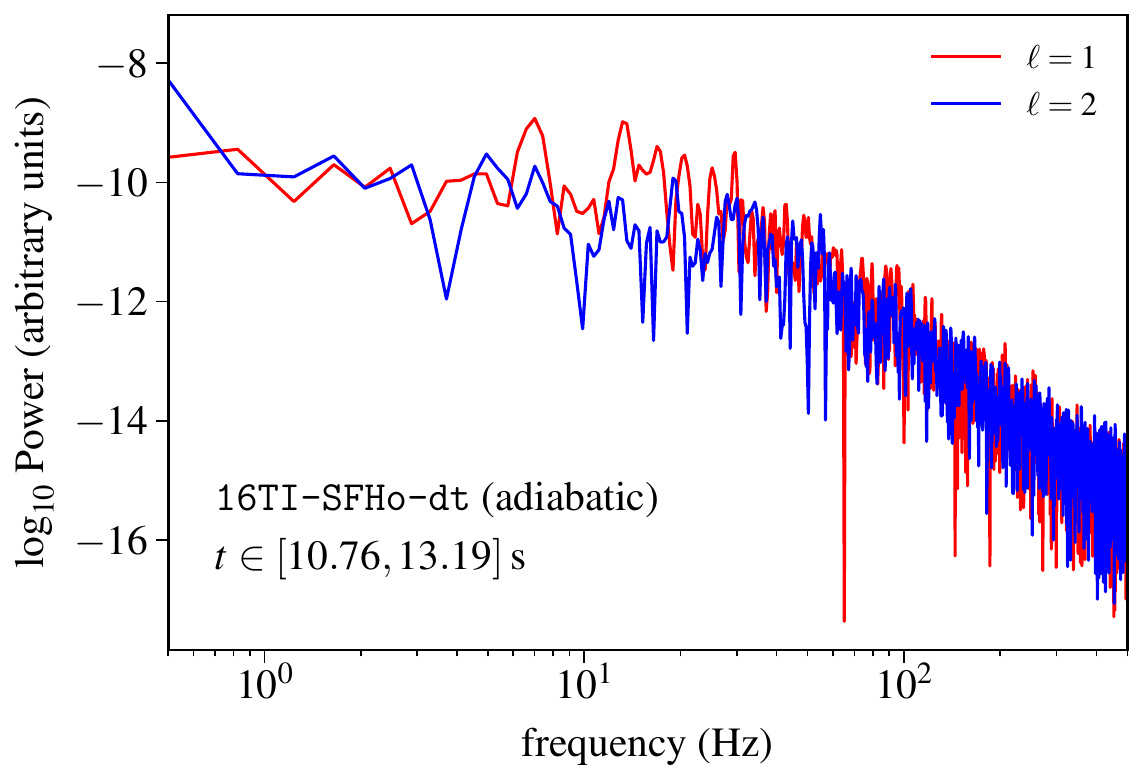}
\caption{\emph{Left:} Discrete Fourier power spectrum (periodogram estimate, Hann window) of the IceCube event rate
at $D = 10$\,kpc assuming adiabatic MSW flavor transformation and default mean energy coefficient 
for model {\tt 16TI-SFHo-dt} over the time period shown,
which covers the NDAF phase (cf. Fig.~\ref{fig:neutrino_rate}). The spectrum is normalized to the power of the IceCube noise as described
in Appendix~\ref{app:noise_power}. The dashed horizontal line shows equality between signal power
and noise power. The green curve shows a smoothing fit used to assess the detection prospects at a fiducial
frequency of $10$\,Hz (cf. Table~\ref{tab:variability}).
\emph{Right:}  Discrete Fourier power spectrum
(periodogram estimate, Hann window) of each normalized shock coefficient $a_\ell/a_0$ (multiplied by the sampling
time $\Delta t = 1$\,ms), as labeled. For both neutrino and shock coefficients, the power spectrum is nearly flat
up to $\sim 20-30$\,Hz, decreasing as a power-law for higher frequencies. }
\label{fig:neutrino_shock_spectrum}
\end{figure*}

A space-based deci-Hertz detector like DECIGO \cite{yagi_2017} would be able to better sample lower frequencies, 
particularly the explosion signal. Fig.~\ref{fig:gw_memory} shows the displacement
$Dh_+$ over the entire simulation for the original models from Paper I, which ran for $\Delta T_{\rm sim}>100$\,s.
The amplitude of the memory signal correlates with the degree of asymmetry and duration of
the explosion. Model {\tt 16TI-SFHo} explodes slowly and asymmetrically, thus the displacement
is still not converged by the end of the simulation. 
Model {\tt 16TI-lo$\alpha$} also explodes slowly and asymmetrically, but the displacement
saturates at around $t\sim 200$\,s. The high-viscosity model {\tt 16TI-hi$\alpha$} does not
undergo a significant NDAF phase and explodes quasi-spherically, consistent with a GW displacement
that saturates early and with low amplitude. The detection horizons with DECIGO (Table~\ref{tab:gw})
range from the entire Milky Way in the case of the high-viscosity model to the Andromeda galaxy
for the 35OC model.

The energy emitted in GW over the entire simulation (Eq.~\ref{eq:Egw_def}) is shown in Table~\ref{tab:gw}. 
Since the high-frequency component dominates the GW power, 
the {\tt -dt} models restricted to the NDAF phase ($\sim 2$\,s) emit comparable amounts or even more
energy than their counterparts that run for $>10$ times longer but sample
$10$ times less frequently. 
The 35OC progenitor emits $20-200$ more energy than the 16TI progenitor, depending
on the frequency range considered. {Since the high-frequency contributions are 
important for GW energy production, our emitted energies based on the quadrupole formula 
are most informative as a relative measure between models instead of absolute values.}

\subsection{Time Variability in the NDAF Phase \label{sec:time_variations}}

\begin{figure*}
\includegraphics*[width=\textwidth]{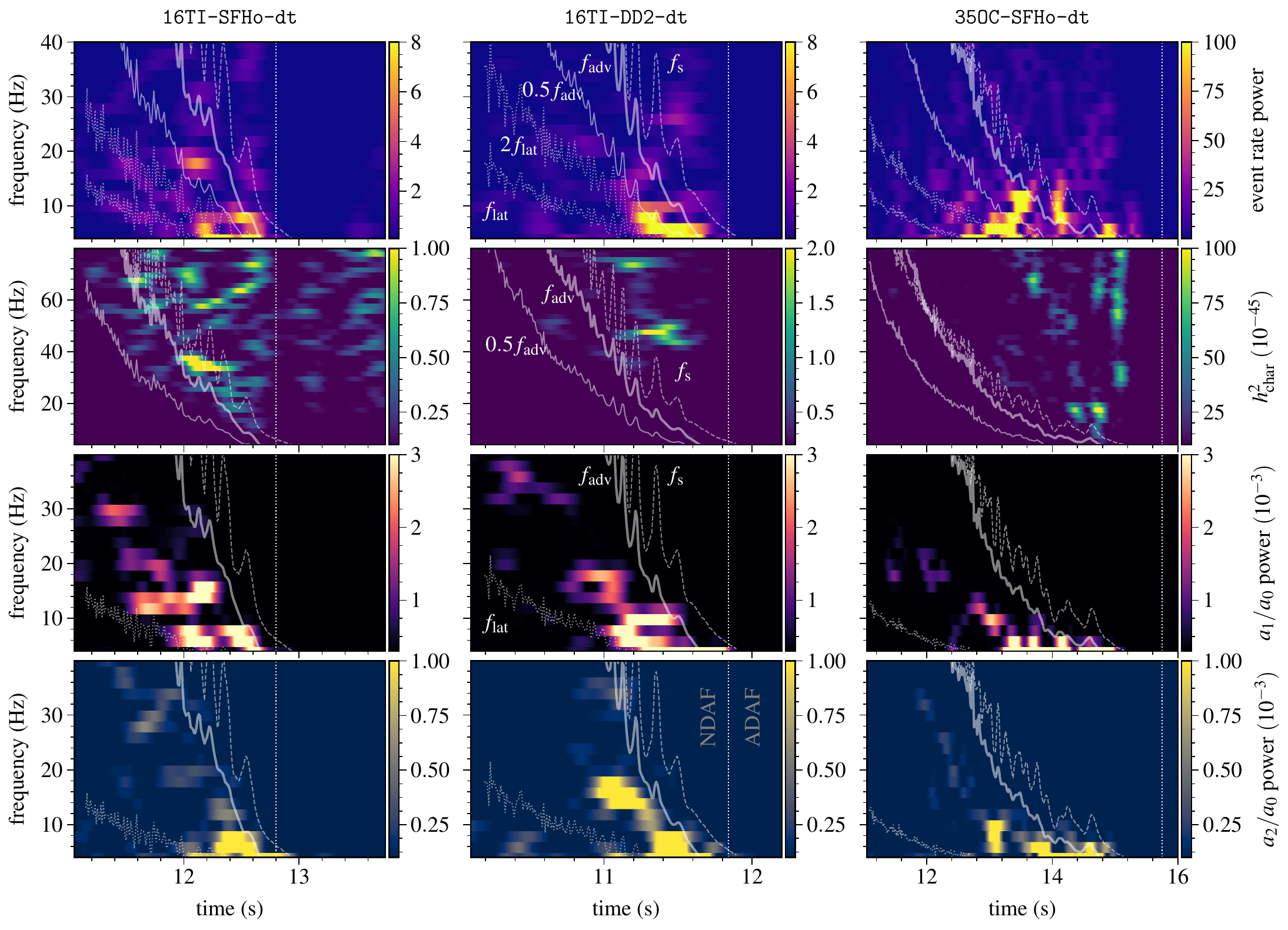}
\caption{Concurrent spectrograms of key observables and shock radius coefficients around the NDAF phase
for a collapsar disk at a distance $D=10$\,kpc, for the {\tt -dt} models as labeled above each column.
\emph{Top row:} Spectral power of the IceCube detection rate (adiabatic MSW flavor transformation, 
default mean energy), normalized to the noise power (Appendix~\ref{app:noise_power}). 
\emph{Second row:} Characteristic GW strain (Eq.~\ref{eq:hchar_definition}) squared
for an equatorial observer ($\sin\theta=1$). \emph{Third and bottom row:} Spectral power of normalized 
dipole and quadrupole shock coefficient (Eq.~\ref{eq:legendre_coeff}), 
respectively. The vertical dotted line marks the NDAF/ADAF transition.
The semi-transparent white curves show frequencies associated
with characteristic timescales in the system: advection frequency 
$f_{\rm adv} = 1/t_{\rm adv}$ (Eq.~\ref{eq:tadv}, thick solid)
and $0.5f_{\rm adv}$ (thin solid), 
radial sound crossing frequency $f_{\rm s} = 1/t_{\rm s}$ (Eq.~\ref{eq:ts}, dashed),
and lateral sound crossing at 
$r = a_0$, $f_{\rm lat}(a_0) = 1/t_{\rm lat}(a_0)$ (Eq.~\ref{eq:tlat}, dotted).
The timescales are obtained from time-averaged data over a $200$\,ms window and plotted in steps of $10$\,ms.
}
\label{fig:spectrogram_grid_DD2-35OC}
\end{figure*}

\begin{figure*}
\includegraphics*[width=\textwidth]{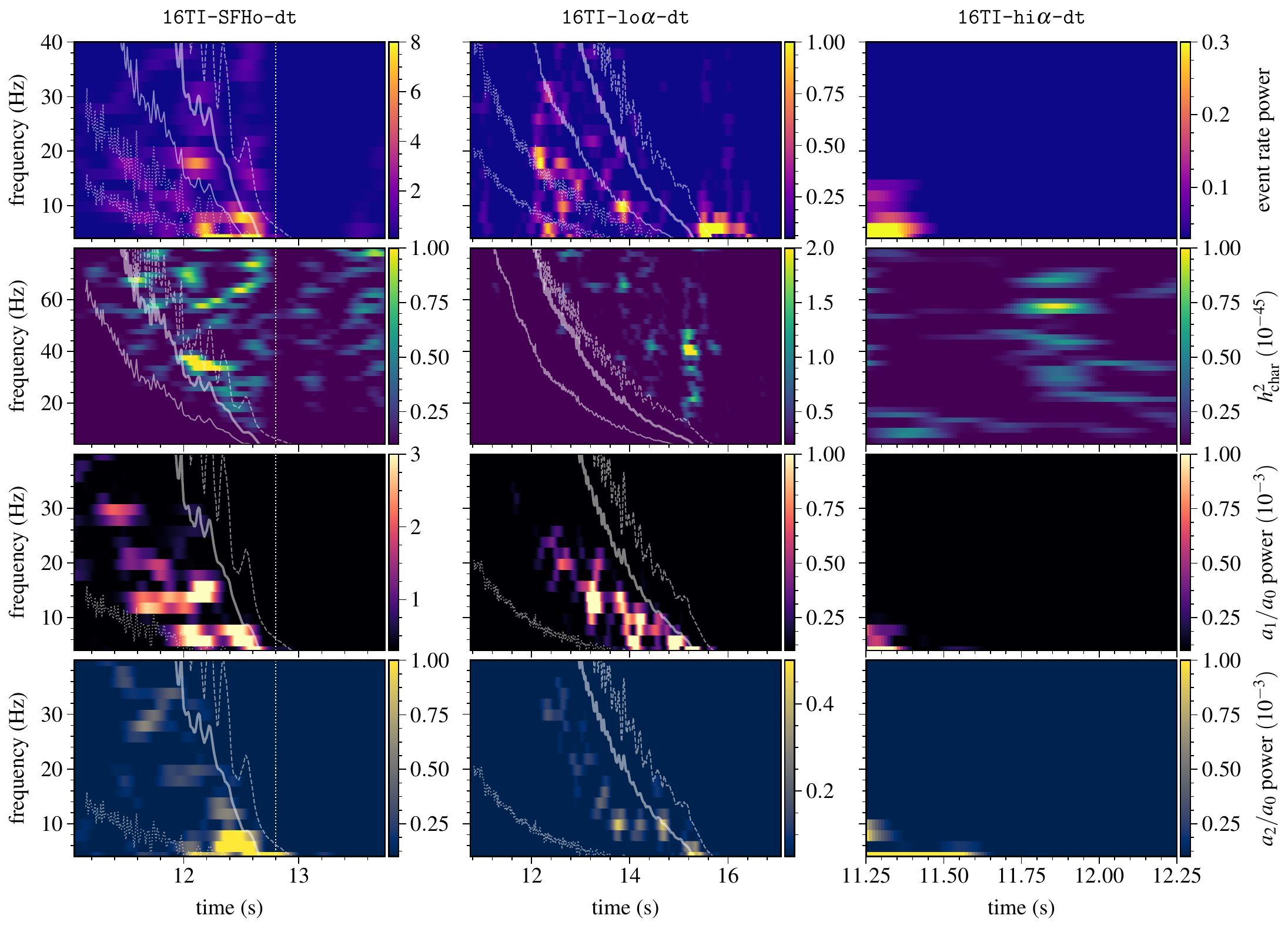}
\caption{Same as Figure~\ref{fig:spectrogram_grid_DD2-35OC}, but for models that vary the viscosity
relative to the baseline model, as labeled at the top of each column. The low-viscosity model 
(middle column) is entirely within the NDAF phase over the period shown, and the high-viscosity model (right column)
does not experience an NDAF phase. The semi-transparent curves have the same meaning as in Figure~\ref{fig:spectrogram_grid_DD2-35OC}.
}
\label{fig:spectrogram_grid_alpha}
\end{figure*}

\begin{figure}
\includegraphics*[width=\columnwidth]{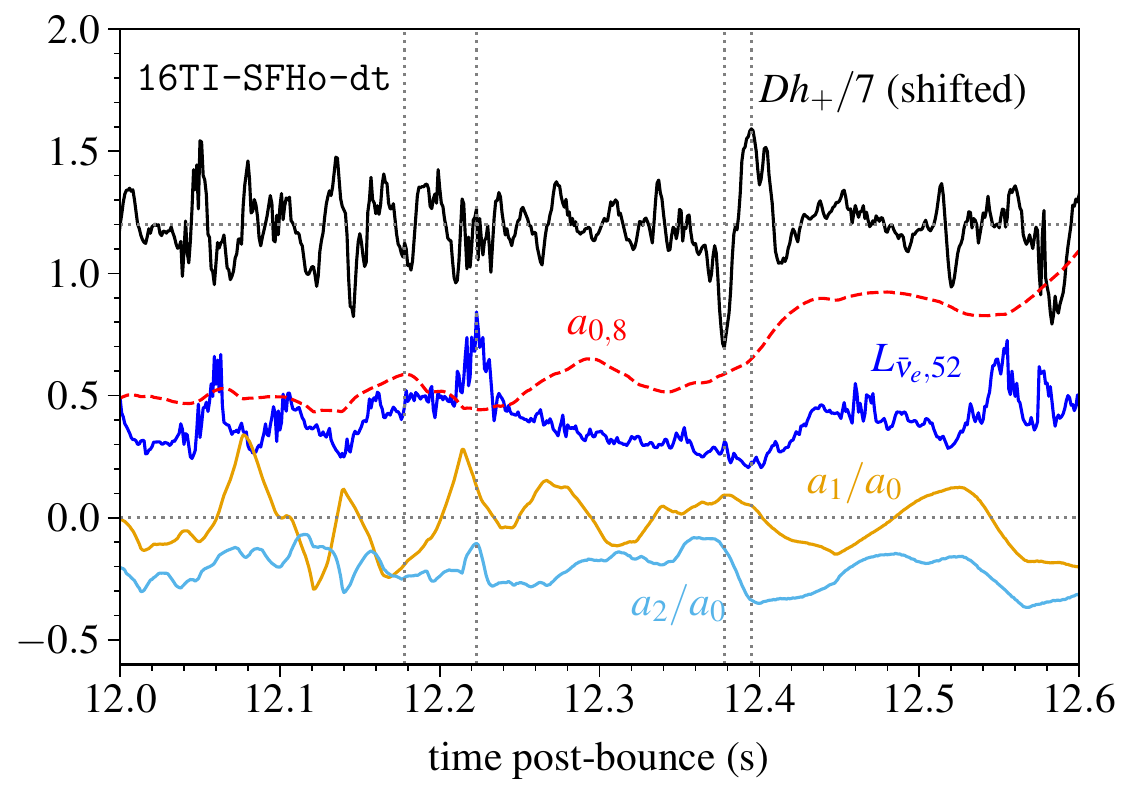}
\caption{Evolution of key quantities from model {\tt 16TI\_SFHo-dt} 
(cf. Figs.~\ref{fig:spectrogram_grid_DD2-35OC}-\ref{fig:spectrogram_grid_alpha}). Shown are the
GW displacement $Dh_+$ (black; shifted by +10 and divided by 7, for clarity), the electron antineutrino
luminosity in units of $10^{52}$\,erg\,s$^{-1}$ $L_{\bar{\nu}_e,52}$ (blue),
the average shock radius in units of $10^8$\,cm $a_{0,8}$ (red), and the normalized dipole (orange) and quadrupole
(light blue) shock coefficients $a_1/a_0$ and $a_2/a_0$, respectively. 
The vertical dotted lines mark the times of the snapshots shown in Fig.~\ref{fig:dens-source_snapshot}.
The horizontal dotted lines show reference levels.
Over the time interval shown,
the average shock radius increases by a factor of $2$, with an associated decrease in the oscillation
frequency of non-spherical shock coefficients as apparent from 
Figs.\ref{fig:spectrogram_grid_DD2-35OC}-\ref{fig:spectrogram_grid_alpha}.
}
\label{fig:time_series_all}
\end{figure}

The oscillations of the collapsar shock during the NDAF phase have striking resemblance to those experienced
by the stalled shock in non-rotating CCSNe during the PNS accretion phase prior to explosion and/or BH formation. 
These stalled-shock oscillations imprint a characteristic signature in the neutrino and GW signal, which
has been studied extensively over the last decade \cite{lund_2010,lund_2012,tamborra_2013,tamborra_2014,mueller_2014_neutrinos,walk_2018,
walk_2019,vartanyan_2019,walk_2020,lin_2020,nagakura_2021,lin_2023,beise_2024,beise_2025}. Here we borrow these 
analysis techniques and apply them to time variations in the neutrino and GW signal during the NDAF phase in collapsars.

To get an overview of the temporal behavior, Fig.~\ref{fig:neutrino_shock_spectrum} 
shows the power spectrum of the IceCube event rate for model {\tt 16TI-SFHo-dt}, assuming
adiabatic MSW 
flavor transformation, as well as the power in $\ell=1$ and $\ell = 2$ shock Legendre coefficients.
The spectrum is computed using the periodogram estimate, applying a Hann window, and 
is carried out over the time range covering the NDAF phase.
Following the convention adopted in \cite{lund_2010}, the power of the IceCube rate is normalized to the noise power 
(Appendix~\ref{app:noise_power}).
For both the shock coefficients and neutrino event rate,
most of the power is contained at low frequencies, but the spectrum flattens in the 
range $1-30$\,Hz before decreasing as a power-law at higher frequencies 
The GW spectrum $f^{-1/2} h_{\rm char}$ for the same model 
(Fig.~\ref{fig:gw_spec}) is also nearly
flat in frequency in the range $\sim 1-50$\,Hz. For all quantities, the spectrum shows distinctive peaks
in the frequency range $1-10$\,Hz. This overall spectral shape in neutrinos and GWs is common to
all models that experience an NDAF phase. The corresponding spectra of the high-viscosity model, on the other hand, 
is essentially a continuous power-law in frequency when restricted to the period
around and following shock formation.

For further insight, Figs.~\ref{fig:spectrogram_grid_DD2-35OC} and \ref{fig:spectrogram_grid_alpha} 
show concurrent spectrograms of the IceCube event rate, squared characteristic
GW strain $h_{\rm char}^2$, as well as power of $\ell=1$ and $\ell=2$ shock coefficients ($a_1/a_0$ and $a_2/a_0$, respectively), 
for all {\tt -dt} models. The spectrograms
are computed using a $500$\,ms sliding window in steps of $10$\,ms, and covering the time period around the entire NDAF phase. 
Analogous to Fig.~\ref{fig:neutrino_shock_spectrum}, the IceCube event rate power is normalized to the
noise power (Appendix~\ref{app:noise_power}). For clarity, we only show frequencies higher than $4$\,Hz
to remove the signature of slower and larger amplitude motions due to the explosion itself.

The neutrino and shock coefficient spectrograms clearly show the end of the NDAF phase,  
after which power is significantly smaller. Consistent with Fig.~\ref{fig:neutrino_shock_spectrum}, most
of the power is concentrated at low frequencies. The GW power, on the other hand, is spread out over
a wider range of frequencies, and is much smaller over low frequencies, except around the end of the NDAF phase 
(note that the quantity shown in Fig.~\ref{fig:gw_spec} is $f^{-1/2}h_{\rm char}$, whereas the characteristic
strain $h_{\rm char}$ by itself rises with frequency).

For all models that experience an NDAF phase, there is a clear pattern in the 
power of the shock coefficients: a
diagonal band that drifts to lower frequencies and with increasing power, corresponding to increasingly slower and larger
amplitude shock oscillations (as in the upper-right panel of Fig.~\ref{fig:shock_strain_rate}).
The GW signal has most of its power at frequencies above this band, while the IceCube rate
shows a broader power distribution that overlaps with that of the shock coefficients. 
The drift to lower frequencies in the GW emission as the disk evolved 
was also seen in the simulations of \cite{gottlieb_2024}.
The high-viscosity model {\tt 16TI-hi$\alpha$-dt} has a very brief episode of substantial
neutrino emission, with variations mostly at low frequencies (cf. Fig.~\ref{fig:neutrino_rate}).

To elucidate the physical meaning of other features in the spectrograms, we focus on model {\tt 16TI-SFHo-dt}
over the period $12.2-12.4$\,s, where peaks in power are present for all obsevables.
Fig.~\ref{fig:time_series_all} shows a segment of the time evolution 
of electron antineutrino luminosity, GW displacement, and shock coefficients.  
There is a spike in the neutrino luminosity around $12.2$\,s 
and a large swing in the GW displacement around $12.4$\,s. The spike in neutrino emission is
concurrent with a noticeable swing in the $\ell=1$ shock coefficient, while the GW displacement swing is concurrent
with an oscillation in both the $\ell=1$ and $\ell=2$ shock coefficients. 
While these specific features are unique to this model, and stochastic in origin, they nevertheless 
allow us to identify a clear
correlation between the shock dynamics and both the neutrino and GW signals. The spectrograms of observables
therefore contain information about the physics of the disk.

\begin{figure*}
\includegraphics*[width=0.49\textwidth]{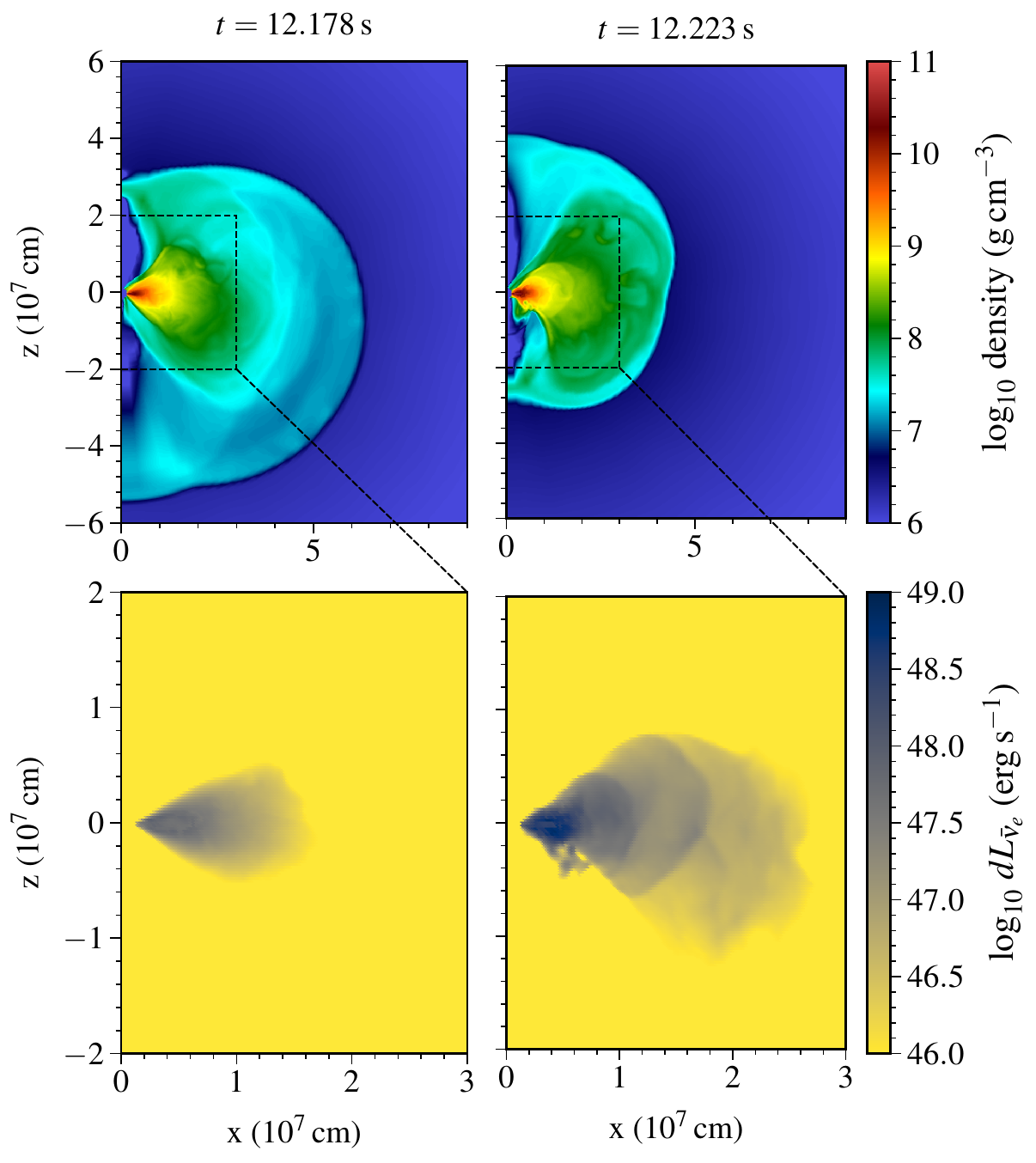}
\includegraphics*[width=0.49\textwidth]{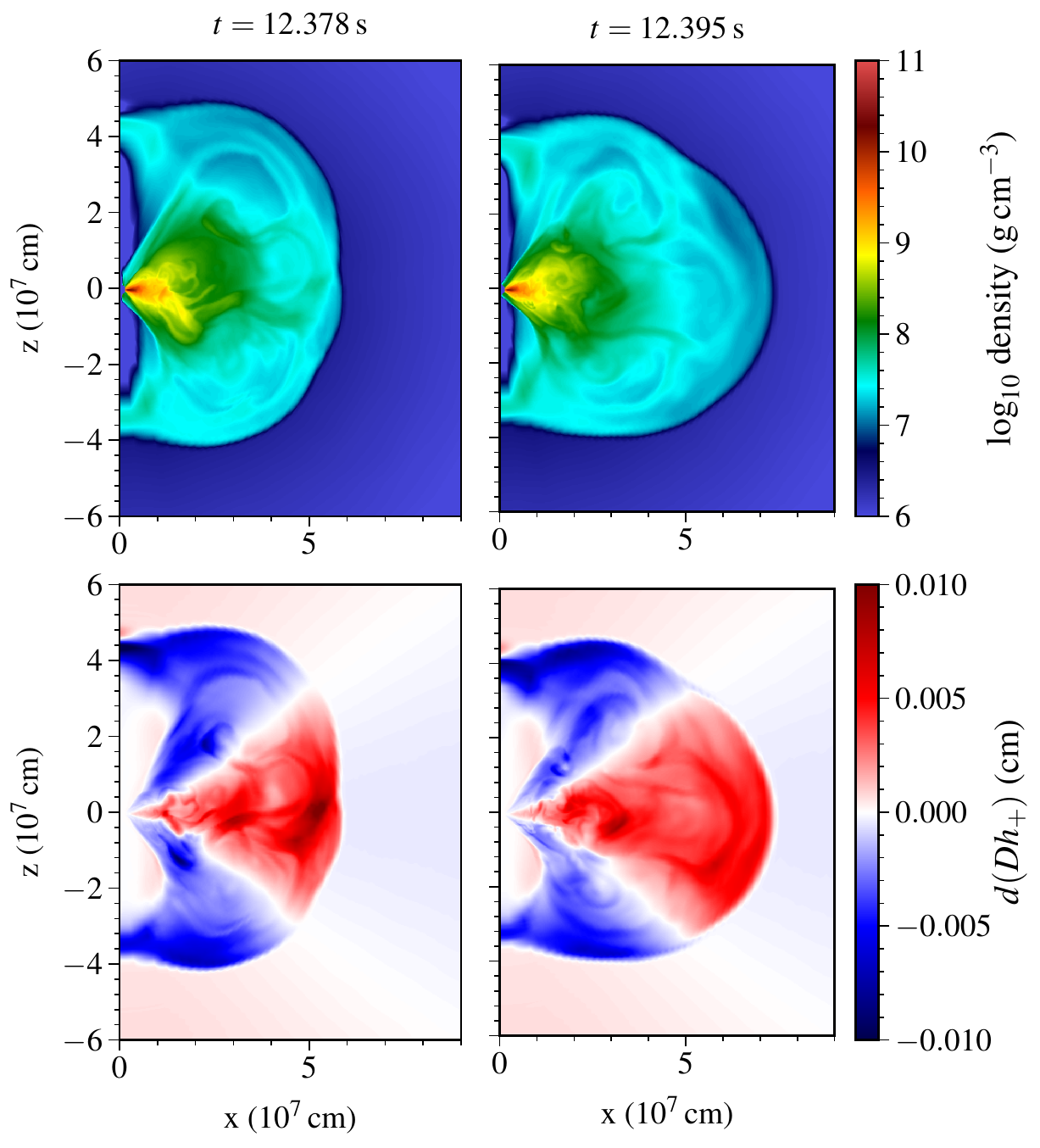}
\caption{Snapshots in the evolution of model {\tt 16TI-SFHo-dt} around the time of a
spike in neutrino emission ($t=12.2$\,s, left two panels) and a large swing in GW displacement 
($t=12.4$\,s, right two panels),
as shown in Fig.~\ref{fig:time_series_all} (vertical dotted lines).
The top row shows the density, and the lower row shows the net electron antineutrino energy
source (Eq.~\ref{eq:neutrino_source}) in the left two panels, and the GW source (Eq.~\ref{eq:gw_source}) 
on the right panels. In both cases, the integrated quantity (luminosity or displacement) is
the sum of all the cells shown. Neutrino emission is concentrated in the innermost, denser
regions of the disk, while GW emission depends on a cancellation of two terms of opposite
sign that are broadly distributed inside the shock cavity.
}
\label{fig:dens-source_snapshot}
\end{figure*}

To trace the origin of these correlations in model {\tt 16TI-SFHo-dt}, Fig.~\ref{fig:dens-source_snapshot} shows snapshots
of the shocked disk density around $t=12.2$ and $12.4$\,s. The shock cavity is executing oscillations
at these times, consistent with the evolution of $a_1/a_0$ and $a_2/a_0$ in Fig.~\ref{fig:time_series_all}.
The bottom panels show the neutrino source term around the time of the neutrino luminosity spike ($t=12.2$\,s)
and the GW source term at the time of the large swing in $Dh_+$ ($t=12.4$\,s).

The neutrino source term shown is the argument of the integral that defines the neutrino luminosity
\begin{equation}
\label{eq:neutrino_source}
dL_{\bar{\nu}_e} \equiv (\mathcal{C}-\mathcal{H}) \rho d^3x,
\end{equation}
where $\mathcal{C}$ and $\mathcal{H}$ are the total electron antineutrino energy 
emission and absorption rate per unit mass, and $d^3 x$ is the cell volume. The
net neutrino emission shown in Fig.~\ref{fig:dens-source_snapshot} is thus
proportional to the number of cells. It is apparent that neutrino emission
is dominated by the innermost regions of the disk ($\lesssim 50$\,km radius).

Figure~\ref{fig:dens-source_snapshot} shows that the spike in neutrino emission
follows an asymmetric enhacement in accretion from the south pole, associated
with an $\ell=1$ sloshing. Fig.~\ref{fig:time_series_all} shows however that
there is a time delay between the shock sloshing and the spike in emission,
which is associated with the average advection time (as is the case for shock
oscillations during the PNS phase of CCSNe). Thus, we
expect correlations in the spectrogram signatures between shock and neutrino
emission, with the signature of a characteristic timescale.

Regarding GW emission, the quantity shown on the rightmost panels of Fig.~\ref{fig:dens-source_snapshot}
is the argument of the integral that defines $Dh_+$ (Eqns.~\ref{eq:Dhplus}-\ref{eq:ddIzz_reduced}):
\begin{equation}
\label{eq:gw_source}
d(Dh_+)  \equiv  \frac{3G}{c^4}\,\sin^2\theta\, \rho d^3 x\left[v_z^2-\frac{1}{3}\mathbf{v}^2 
   -z\frac{\partial \Phi}{\partial z} + \frac{1}{3}\mathbf{x}\cdot\nabla \Phi\right]\\
\end{equation}
The GW source is broadly distributed in space, and in fact the strain produced
by the shock cavity is the difference between two terms with opposite signs.
This means that the GW signal is sensitive to global properties of the
shocked cavity, in contrast to the neutrino emission. In particular, 
the GW strain should contain direct signature of large-scale shock
oscillations as well as smaller scale turbulent motions.

The rightmost two panels of Fig.~\ref{fig:dens-source_snapshot} show that
the large swing in $Dh_+$ at $t=12.4$ is the consequence of a quadrupolar
oscillation, in which the cavity goes from prolate to oblate. This is
also reflected in Fig.~\ref{fig:time_series_all} by the evolution
of the quadrupolar shock coefficient $a_2/a_0$, which shows a concurrent
oscillation.

Also shown in Fig.~\ref{fig:time_series_all} is the evolution of the average shock radius $a_0$, 
which increases by a factor of $2$ over the time shown.
In the case of the SASI during the PNS accretion phase of CCSNe, 
the oscillation period is related to the
advection time, which changes as the average shock radius increases or
decreases, resulting in a frequency drift in the spectrogram,
which can be fitted for (e.g., \cite{mueller_2014_neutrinos}).

We now examine whether the collapsar disk spectrograms contain similar 
signatures of basic timescales in the system as the PNS accretion phase of CCSNe. 
In particular, we calculate the average radial advection time in the shocked cavity
\begin{equation}
\label{eq:tadv}
t_{\rm adv} = \int_{a_0}^{\rm r_{\rm in}} \frac{dr}{\langle v_r\rangle},
\end{equation}
where $\langle v_r\rangle < 0$ is the angle-averaged radial velocity and $r_{\rm in}$ is
the inner radial boundary, the
average radial sound-crossing time
\begin{equation}
\label{eq:ts}
t_{\rm s} = \int_{r_{\rm in}}^{a_0} \frac{dr}{\sqrt{\langle c^2_s\rangle}},
\end{equation}
where $\langle c^2_s\rangle=\langle \gamma p/\rho\rangle$ is the angle-averaged squared 
sound speed, and the lateral sound-crossing time
\begin{equation}
\label{eq:tlat}
t_{\rm lat}(r) = \frac{2\pi r}{\sqrt{\langle c^2_s\rangle}}.
\end{equation}
We then define characteristic frequencies associated with 
each these timescales, $f_{\rm adv} \equiv 1/t_{\rm adv}$, $f_{\rm s} \equiv 1/t_{\rm s}$,
and $f_{\rm lat}(r) \equiv 1/t_{\rm lat} (r)$.
Figs.~\ref{fig:spectrogram_grid_DD2-35OC} and \ref{fig:spectrogram_grid_alpha} show
these characteristic frequencies as functions of time,
for all models that experience an NDAF phase.
The timescales in Eqs.~(\ref{eq:tadv})-(\ref{eq:tlat}) are computed by time averaging data in a window of $200$\,ms before
calculating the integrals, to smooth-out high-frequency variations, then the frequencies are obtained
as the inverse of these timescales.

The diagonal band in the spectrum of $\ell=1$ and $\ell=2$ shock oscillations 
is bracketed by the radial advection frequency $f_{\rm adv}$ and the lateral
sound crossing frequency at the average shock radius $f_{\rm lat}(a_0)$. The latter
is associated with the longest lateral sound-crossing time, all smaller radii
where the disk exist have faster lateral sound-crossing times.
In the case of the (non-rotating) SASI during the PNS accretion phase of
CCSNe, the most unstable modes $\ell>0$ are such that a radius $r$ exists at which $t_{\rm adv} = t_{\rm lat}(r)/\ell$
\cite{FT09_sasi}. In our models, this radius indeed exists and
is typically $r \sim a_0/2$ (i.e. the middle of the disk). No significant
power in the shock coefficients is apparent at the radial sound crossing
frequency $f_{\rm s}$.

Regarding neutrino observables, the IceCube event rate spectrogram shows 
power at all characteristic frequencies. A significat amount of power is visible
between $0.5f_{\rm adv}$ and $f_{\rm adv}$. 
When the SASI manifests during the PNS accretion phase in CCSNe, 
the $\ell=0$ mode has an oscillation period 
close to twice the radial advection time, in which the average shock radius expands and contracts
alongside the neutrino cooling rate (e.g., \cite{foglizzo_2007}).
In our simulations, this is a possible signature of neutrino modulation associated
with the $\ell=0$ as well as higher multipole modes. Additional power bands
overlap the radial sound crossing frequency $f_{\rm s}>f_{\rm adv}$, as
well as the frequencies associated to the lateral sound crossing time at the
shock [$f_{\rm lat}(a_0)$ and $2f_{\rm lat}(a_0)$].
This shows that the neutrino event rate spectrogram contains direct
signatures of the shock oscillation and its associated timescales.

The GW spectrogram during the NDAF phase contains most of the power 
in frequencies larger than $f_{\rm adv}$, although some power extends to $0.5 f_{\rm adv}$.
In model {\tt 16TI-SFHo-dt} the peak in power at $t=12.2-12.4$\,s is bracketed between 
the advection frequency and the radial sound crossing frequency $f_{\rm s}$, although
this is not the case in other models. We surmise that power at higher frequencies is associated with turbulence,
as indicated by the presence of power bands in model {\tt 16TI-SFHo-dt} during the ADAF phase.
The limitations of the quadrupole formula for higher frequencies make the investigation of this turbulent
component unreliable with our methods.

\begin{table}
\caption{Detection horizon $D_{\rm max}$ for time variations in neutrino emission by IceCube during the NDAF phase,
for {\tt -dt} models.
The ratio of IceCube event rate power to noise power $P_{\rm signal}/P_{\rm noise}$ 
is obtained by smoothing the normalized event rate power spectrum  (as in Fig.~\ref{fig:neutrino_shock_spectrum}) and reading off
the value at $f=10$\,Hz. The number in the second column assumes a distance $10$\,kpc, adiabatic flavor transformation,
and default mean energy coefficient. The {detection} horizon is obtained by requiring $P_{\rm signal}/P_{\rm noise}=5$
and using the fact that the event rate power $\propto D^{-4}$. The high-viscosity model is not shown, as it does
not experience an NDAF phase.
\label{tab:variability}} 
\begin{ruledtabular}
\begin{tabular}{lcc}
Model & $P_{\rm signal}/P_{\rm noise}$ &  $D_{\rm max}$ (kpc)   \\
\noalign{\smallskip}
\hline
{\tt 16TI-SFHo-dt}              & 0.90 & 6.5 \\
{\tt 16TI-SFHo-lo$\alpha$-dt}   & 0.09 & 3.7 \\
{\tt 16TI-DD2-dt}               & 0.37 & 5.2 \\
{\tt 35OC-SFHo-dt}              & 10  & 12 \\
\end{tabular}
\end{ruledtabular}
\end{table}

Finally, we consider the detectability of time variations in the neutrino signal. 
We do not use the spectrograms for this purpose, because the size of the moving window is
chosen to maximize visibility of features, making the maximum power a window-dependent quantity. 
Instead, we use the entire power spectrum over the NDAF phase, as in Fig.~\ref{fig:neutrino_shock_spectrum},
normalized to the IceCube noise power, and apply a Gaussian smoothing filter. We choose a fiducial
frequency of $10$\,Hz, where most features in the spectrogram are present. The resulting value at this
frequency yields a ratio of signal power to noise power $P_{\rm signal}/P_{\rm noise}$ at a distance of $10$\,kpc. 
For simplicity, we assume adiabatic MSW flavor transformation and the default mean energy coefficient in 
Eq.~\ref{eq:mean_energy_temp}. The resulting $P_{\rm signal}/P_{\rm noise}$ is reported in Table~\ref{tab:variability}.

We estimate the detection horizon for time variability by demanding that the power ratio is $P_{\rm signal}/P_{\rm noise}=5$.
Since the event rate scales as neutrino flux ($\propto D^{-2}$), the event rate power scales as $\propto D^{-4}$.
The resulting detection horizon $D_{\rm max}$ is also shown in Table~\ref{tab:variability}, for all
{\tt -dt} models that experience an NDAF phase. For models that employ the 16TI progenitor, this maximum distance
is a few kpc, while for the 35OC progenitor is extends to $12$\,kpc. In all cases, this is confined within
the Milky Way galaxy.

\section{Summary and Discussion \label{s:summary}}

We have explored the neutrino and GW signals from collapsar accretion disks
and their outflows, in the absence of a jet,
by post-processing the simulations presented in Paper I and re-running
new models that output spatial data at a higher frequency. Our coarse time sampled 
hydrodynamic simulations are axisymmetric, span the period from the onset of 
core-collapse to shock breakout, and include angular momentum transport via shear viscosity,
neutrino emission and absorption, a 19-isotope nuclear reaction network, self-gravity, and a pseudo-Newtonian BH.
The finer time sampled simulations have the same physics and cover the period around 
which the disk is neutrino-cooled (NDAF phase). Our main results are the following:
\newline

\noindent
1. -- Neutrinos from collapsar disks have typical energies of $\mathcal{O}(10-20)$~MeV, 
and the neutrino event rate is expected to be  the strongest during the
NDAF phase (lasting for few s), which IceCube could easily detect within the galaxy and its satellites 
(Fig.~\ref{fig:neutrino_rate}, Table~\ref{tab:neutrino}).
Additional signatures include a step-like increase in neutrino emission $\sim 0.3$\,s prior to 
shocked disk formation, and
stochastic bursts of emission due to turbulence in 
the ADAF phase for a period lasting $\sim 10$\,s. 
These two additional signatures are detectable within few to several kpc, depending on stellar
progenitor. The large difference (factor $\sim 100$) 
between electron-flavor and heavy lepton neutrino/antineutrino luminosities implies an important
dependence of detectability on flavor transformation.
\newline

\noindent
2. -- The GW signal of collapsar disks during the NDAF phase 
{(at frequencies $\lesssim 100$\,Hz)}
is comparable to that of non-rotating CCSNe that experience SASI
oscillations during the PNS accretion phase (Fig.~\ref{fig:shock_strain_rate}). 
This signal alone can be detected by LIGO A+ 
over distances ranging from a few kpc out to the Milky Way satellites, 
depending on stellar progenitor (Fig.~\ref{fig:gw_spec}, Table~\ref{tab:gw}). Planned third generation 
GW observatories will provide more sensitivity and coverage at lower frequencies, 
significantly extending the detectability horizon relative to current detectors 
(with Cosmic Explorer in low frequency mode being the best at $\sim 1$\,Mpc). 
The explosion signature (GW memory signal, Fig.~\ref{fig:gw_memory}) is comparable to that of normal CCSNe 
and it could be picked up to distances reaching $\sim 1$\,Mpc by a space-based detector such
as DECIGO.
\newline

\noindent
3. -- Time variations in the neutrino and GW signal during the NDAF phase 
in the frequency range $10-50$\,Hz
encode information about 
dynamics of the disk and evolution of the shock that encloses it.
Neutrino emission is dominated by the inner disk
and is sensitive to fluctuations in the accretion rate, which in turn relate to 
large-scale shock oscillations (Figs.~\ref{fig:time_series_all}-\ref{fig:dens-source_snapshot}). 
GW emission, on the other hand, is set by the entire disk cavity
and is thus sensitive to the global dynamics of the shocked disk.
The time-frequency evolution of the neutrino and GW signal tracks
characteristic frequencies associated with key timescales in the system 
(Fig.~\ref{fig:spectrogram_grid_DD2-35OC}-\ref{fig:spectrogram_grid_alpha}),
including radial advection time, radial sound-crossing time, and lateral sound-crossing time.
These timescales are closely related to the evolution of the shocked disk. 
Time variations in the neutrino signal are detectable within $1-10$\,kpc by IceCube
depending on stellar progenitor, and flavor transformation scenario (Table~\ref{tab:variability}).
\newline

Since collapsars are a subset of CCSNe, they should occur at a lower 
rate than stellar explosions driven by the (default) neutrino mechanism. As a rough estimate, we 
can take the rate of Ic-BL supernovae in the local universe, which is $\sim 1\%$ of
all CCSNe (e.g., \cite{graur_2017}). For the Milky Way, instead of $\sim 1$ CCSNe 
per century (e.g., \cite{adams_2013}), the expected rate would be one collapsar per $10^4$\,yr.

In our neutrino detectability forecast for
the shock, NDAF, and ADAF phases, we focused on IceCube because of its large event rate.  
Another Cherenkov neutrino telescope is Super-Kamiokande. Although the  
latter  has smaller statistics than IceCube, it is  essentially  background-free  \cite{mirizzi_2016,scholberg_2018}. 
The upcoming Hyper-Kamiokande neutrino telescope \cite{HyperK} is expected to have an effective volume about ten times larger than Super-Kamiokande. Although the event rate from collapsar disks expected in Hyper-Kamiokande should be smaller than that in 
IceCube,  Hyper-Kamiokande may perform better than IceCube for collapsars occurring beyond the galaxy due to the absence of background, 
with an event rate similar to that of the soon-to-be-operative liquid scintillator detector JUNO \cite{JUNO:2015zny}. 

Super- and Hyper-Kamiokande would allow for an exquisite reconstruction of the spectral energy distribution of neutrinos reaching Earth. We do not consider the information carried by the detectable energy spectrum in this paper because of the approximations 
made in the leakage scheme employed. Nevertheless, in our models,  the average energies of neutrinos rapidly increase in the phase of shock formation, reaching $\mathcal{O}(20)$~MeV and then sharply dropping to $\mathcal{O}(15)$~MeV during the 
NDAF and ADAF phases. Such increase in the typical neutrino energies marking shock formation can be used 
as another observational diagnostic.

The axisymmetry of our simulations suppresses additional oscillation modes, such as 
the non-axisymmetric shock instability in the ADAF regime, which has been studied analytically \cite{molteni_1999,gu_foglizzo_2003,gu_lu_2006,
nagakura_yamada_2009}
and seen in global simulations \cite{nagakura_yamada_2008,taylor_2011,gottlieb_2022}. 
These additional modes leave an imprint in the GW signal \cite{gottlieb_2024},
although their impact on neutrino emission in the NDAF phase has not been systematically studied yet. 
The neutrino signal can thus display variability not captured in the present study.

The time-frequency dependence of the neutrino signal during the PNS accretion phase has 
been investigated
in the non-rotating as well as the rapidly rotating cases for CCSNe. A frequency drift in the neutrino
event spectrogram due to SASI oscillations, tracking the radial advection frequency and thus the shock radius,
was identified by \cite{mueller_2014_neutrinos} for non-rotating models. 
Exploding models showed a pattern of decreasing
SASI frequency and thus increasing shock radius. Increasing the rotation rate shows a broader
distribution of power relative to the non-rotating case, in which the SASI shows less sharp peaks
in frequency \cite{walk_2018}. At very rapid rotation, the low-T/$|$W$|$ instability sets in,
resulting in a strong frequency modulation of the neutrino signal due to deformation
of the PNS, manifesting as a drift to higher frequencies in the neutrino event histogram \cite{shibagaki_2021}.
The time-frequency signatures of collapsar disks we found are distinct from those described above, although
since a PNS accretion phase always precedes BH formation (except for pair-instability explosions in very massive stars), these standard
SASI signatures could still be present when analyzing the combined pre- and post-BH neutrino signal.

The time variations of the neutrino signal during the PNS accretion phase of 
CCSNe have a strong dependence on the viewing angle \cite{lund_2010,tamborra_2013}. 
Furthermore, evidence for a relative asymmetry in the emergent lepton number flux (LESA phenomenon)
has been observed in three-dimensional CCSNe simulations \cite{tamborra_2014b}. 
While our neutrino scheme does not provide directional information in the neutrino fluxes,
the intrinsic geometry of a BH accretion disk, which is not spherically symmetric, suggests that the 
observational signatures studied here should also have a viewing angle dependence.
In particular, if a relativistic jet is produced, the associated polar cavity 
can lead to  important differences between the emission properties of neutrinos and GWs along the polar and equatorial directions (cf.~directional differences in neutron star merger remnants \cite{foucart_2023}).

Regarding {the viewing angle dependence of the} GW emission, 
we have shown results for an equatorial observer ($\sin\theta=1$), as the
angular dependence {of the quadrupole formula in axisymmetry} is straightforward. 
{The global 3D, general-relativistic MHD simulations of \cite{gottlieb_2024} found the appearance of high-density,
non-axisymmetric Rossby vortices in collapsar disks with significant cooling
(i.e., the equivalent of the NDAF phase in our simulations). These vortices produce strong coherent
GW emission over a broad spectrum peaking at $\sim 100$\,Hz, with the largest
amplitudes toward the polar direction. The overdensities associated with
these vortices likely lead to correlated variations in the neutrino emission. While these
simulations did not include self-gravity or a microphysics-based cooling mechanism 
(which impact the evolution of overdensities) and are not directly comparable to ours, they
highlight qualitative changes to be expected in the GW signals when the assumption of axial symmetry is relaxed, which deserve further investigation.}

While many of the observable features we have studied here should be robust, detailed
quantitative predictions will depend on the physics used to model the problem. The formation of a
shocked bubble containing an accretion disk, the plateau nature of neutrino emission during the NDAF phase (if it occurs),
the correlated variability in neutrino emission, GW, and shock surface during the NDAF phase, and the episodic
spikes in emission during the ADAF phase should all be robust features, as they depend on basic
properties of the star. 
Likewise, the frequency of shock oscillations is set by the shock radius and the radial advection time.
While the detailed value of the latter is related to the rate of angular momentum transport, 
the shock radius is largely set by the accretion rate and angular momentum distribution in the star. The formation
of a neutrino-cooled accretion disk occurs only for high-enough accretion rates \cite{chen_2007,de_2021}, and a finite
range of stellar rotation rates, thus oscillation frequencies of collapsar disks in the
NDAF phase should not be too different than the ones found here.

On othe other hand, factors that can lead to quantitative modifications include
the presence of a relativistic jet, which can result in ejection of the outer layers of the star,
changing the mass available to feed the accretion disk and hence its overall lifetime.
The existence and duration of an NDAF phase depends on the 
nature of angular momentum transport: when using shear viscosity,
faster transport with higher dissipation results in direct transition to the 
ADAF phase (e.g. Fig~\ref{fig:neutrino_rate}). In MHD simulations,
the transition to the ADAF phase occurs when the disk reaches the magnetically-arrested regime \cite{issa_2025},
which is related to the initial magnetic field distribution in the stellar progenitor.
Predictions for the quantitative neutrino event rate 
are contingent on the quality of the neutrino transport used, 
including flavor conversion effects.
The nature of the shock precursor is dependent on the
microphysics used to model the problem (e.g., excluding the nuclear binding energy and neutrino
cooling skips the formation of a dwarf disk; Paper I) and on the angular momentum distribution
of the progenitor (which determines where and when does the disk form). Turbulence during the ADAF
phase also depends on the mechanism transporting angular momentum and dissipating turbulent
kinetic energy as thermal energy. Future global simulations of collapsars should be able improve predictions for all these observable signatures.

\appendix

\begin{acknowledgments}

This research was supported by the Natural Sciences and Engineering Research
Council of Canada (NSERC) through Discovery Grant RGPIN-2022-03463. 
Part of this work was performed at the Aspen Center for Physics, which is supported by 
National Science Foundation grant PHY-2210452.
SJ was supported by a fellowship of the German Academic Exchange Service (DAAD)
and by the Mitacs Globalink Research Intership program.
CD was supported in part by the Alberta Graduate Excellence Scholarship program.
IT acknowledges support from the Villum Foundation (Project No.~13164) and the European Union (ERC, ANET, Project No.~101087058). Views and opinions expressed are those of the authors only and do not necessarily reflect those of the European Union or the European Research Council. Neither the European Union nor the granting authority can be held responsible for them.
The software used in this work was in part developed by the U.S Department of Energy (DOE)
NNSA-ASC OASCR Flash Center at the University of Chicago.
Data visualization was done in part using {\tt VisIt} \cite{VisIt}, which is supported
by DOE with funding from the Advanced Simulation and Computing Program
and the Scientific Discovery through Advanced Computing Program.
Graphics were developed with Matplotlib \cite{hunter2007}. 
This research used storage resources of the
National Energy Research Scientific Computing Center
(NERSC), which is supported by the DOE Office of Science
under Contract No. DE-AC02-05CH11231 (repository m2058).
This research was enabled in part by computing and storage support
provided by Prairies DRI, BC DRI Group, Compute Ontario (computeontario.ca),
Calcul Qu\'ebec (www.calculquebec.ca) and the Digital Research Alliance of Canada (alliancecan.ca). 
Computations were performed on the Niagara supercomputer at the SciNet HPC Consortium \cite{SciNet,Niagara}. SciNet is 
funded by Innovation, Science and Economic Development Canada; the Digital Research Alliance of Canada; 
the Ontario Research Fund: Research Excellence; and the University of Toronto. 
\end{acknowledgments}

\section{Spectral Power of Poisson Noise \label{app:noise_power}}

Here we provide our calculation of the spectral power of the Poissonian noise in the IceCube detector.
Using time bins of $\Delta t = 1$\,ms, the mean expected number of events per bin is $\lambda = R_p \Delta t = 2,790$,
where dark current noise rate is $R_p = 1.5\times 10^6$\,s$^{-1}$ (\S\ref{s:neutrinos}).

Consider $N$ time bins, and let $h_k = h(t_k)$ be the number of noise events for a bin at time $t_k$, with $k=0,...,N-1$.
For a Poisson distribution, the number of events in each bin is independent of one another, and for $N\gg 1$
we can write the mean and variance in the number of events as
\begin{eqnarray}
\label{eq:Poisson_mean}
N^{-1}\sum_{k=0}^{N-1} h_k    & \simeq & \lambda\\
\label{eq:Poisson_variance}
N^{-1}\,\sum_{k=0}^{N-1} h_k^2 - \left( N^{-1}\sum_{k=0}^{N-1} h_k\right)^2 & \simeq & \lambda.
\end{eqnarray}
Combining Eq.~(\ref{eq:Poisson_mean}) with the definition of the discrete Fourier transform (DFT; e.g., \cite{NR92}) 
\begin{equation}
\label{eq:dft}
\tilde{h}_n = \sum_{k=0}^{N-1} h_k\, e^{2\pi i k n / N},
\end{equation}
we can find the power at zero frequency
\begin{equation}
\label{eq:power_zero-freq}
|\tilde{h}_0|^2 \simeq N^2\lambda^2. 
\end{equation}
The average discrete power at non-zero frequencies can be found from the discrete version of Parseval's theorem
\begin{equation}
\frac{1}{N}\sum_{n=0}^{N-1} |\tilde{h}_n|^2 = \sum_{k=0}^{N-1} |h_k|^2.
\end{equation}
Using Eqns.~(\ref{eq:Poisson_mean})-(\ref{eq:Poisson_variance}) and the fact that $h_k$ are real, we can write
\begin{equation}
|\tilde{h}_0|^2 + 2\sum_{n=1}^{N/2-1}|\tilde{h}_n|^2 + |\tilde{h}_{N/2}|^2 \simeq N^2 (\lambda + \lambda^2).
\end{equation}
Using Eq.~(\ref{eq:power_zero-freq}) to eliminate the power at zero frequency, we have
\begin{equation}
 2\sum_{n=1}^{N/2-1}|\tilde{h}_n|^2 + |\tilde{h}_{N/2}|^2 \equiv  N\langle |h_{n\ne 0}|^2\rangle \simeq N^2\lambda,
\end{equation}
where $\langle |h_{n\ne 0}|^2\rangle$ is the frequency-averaged power at non-zero frequencies. The periodogram
estimate (e.g., \cite{NR92}) of the Poisson noise power spectrum at non-zero frequencies is then
\begin{equation}
\label{eq:noise_power}
P_{\rm noise}(f\ne 0) \simeq \frac{2}{N^2} \langle |h_{n\ne 0}|^2\rangle \simeq \frac{2\lambda}{N},
\end{equation}
where the factor two comes from the fact that the power at a given frequency (except Nyquist) is $2|\tilde{h}_{n\neq 0}|^2$.

Ref.~\cite{lund_2010} showed that when using a Hann window, the noise power in non-zero frequencies  
changes\footnote{Numerical experiments show that this is a good approximation for frequencies $f\ge 2/(N\Delta t$).} 
by a factor equal to the average of the window squared $\langle w^2\rangle$. For our definition of
the Hann window
\begin{equation}
w_k = \frac{1}{2}\left[1 - \cos\left(\frac{2\pi k}{N}\right) \right]
\end{equation}
we have $\langle w^2\rangle \simeq 3/8$. Since we use the periodogram estimate for the signal power spectrum,
which normalizes the square amplitude of the DFT coefficients by a factor $N^2\langle w^2\rangle$ for windowed data, 
the window factor cancels out from Eq.~(\ref{eq:noise_power}). Aside from this window factor, our
noise power estimate is identical to that in Ref.~\cite{lund_2010} (in their notation, $N_{\rm bkgd}=N\lambda$). 
Since we work with events per bin, so that
we can use Eqns.~(\ref{eq:Poisson_mean})-(\ref{eq:Poisson_variance}), there is a 
factor $\Delta t$ already contained in Eq.~(\ref{eq:dft}) through $\lambda$.

\bibliographystyle{apsrev4-2}
\bibliography{collapsar,gw,neutrinos}

\begin{thebibliography}{138}%
\makeatletter
\providecommand \@ifxundefined [1]{%
 \@ifx{#1\undefined}
}%
\providecommand \@ifnum [1]{%
 \ifnum #1\expandafter \@firstoftwo
 \else \expandafter \@secondoftwo
 \fi
}%
\providecommand \@ifx [1]{%
 \ifx #1\expandafter \@firstoftwo
 \else \expandafter \@secondoftwo
 \fi
}%
\providecommand \natexlab [1]{#1}%
\providecommand \enquote  [1]{``#1''}%
\providecommand \bibnamefont  [1]{#1}%
\providecommand \bibfnamefont [1]{#1}%
\providecommand \citenamefont [1]{#1}%
\providecommand \href@noop [0]{\@secondoftwo}%
\providecommand \href [0]{\begingroup \@sanitize@url \@href}%
\providecommand \@href[1]{\@@startlink{#1}\@@href}%
\providecommand \@@href[1]{\endgroup#1\@@endlink}%
\providecommand \@sanitize@url [0]{\catcode `\\12\catcode `\$12\catcode
  `\&12\catcode `\#12\catcode `\^12\catcode `\_12\catcode `\%12\relax}%
\providecommand \@@startlink[1]{}%
\providecommand \@@endlink[0]{}%
\providecommand \url  [0]{\begingroup\@sanitize@url \@url }%
\providecommand \@url [1]{\endgroup\@href {#1}{\urlprefix }}%
\providecommand \urlprefix  [0]{URL }%
\providecommand \Eprint [0]{\href }%
\providecommand \doibase [0]{https://doi.org/}%
\providecommand \selectlanguage [0]{\@gobble}%
\providecommand \bibinfo  [0]{\@secondoftwo}%
\providecommand \bibfield  [0]{\@secondoftwo}%
\providecommand \translation [1]{[#1]}%
\providecommand \BibitemOpen [0]{}%
\providecommand \bibitemStop [0]{}%
\providecommand \bibitemNoStop [0]{.\EOS\space}%
\providecommand \EOS [0]{\spacefactor3000\relax}%
\providecommand \BibitemShut  [1]{\csname bibitem#1\endcsname}%
\let\auto@bib@innerbib\@empty
\bibitem [{\citenamefont {{Adams}}\ \emph {et~al.}(2013)\citenamefont
  {{Adams}}, \citenamefont {{Kochanek}}, \citenamefont {{Beacom}},
  \citenamefont {{Vagins}},\ and\ \citenamefont {{Stanek}}}]{adams_2013}%
  \BibitemOpen
  \bibfield  {author} {\bibinfo {author} {\bibfnamefont {S.~M.}\ \bibnamefont
  {{Adams}}}, \bibinfo {author} {\bibfnamefont {C.~S.}\ \bibnamefont
  {{Kochanek}}}, \bibinfo {author} {\bibfnamefont {J.~F.}\ \bibnamefont
  {{Beacom}}}, \bibinfo {author} {\bibfnamefont {M.~R.}\ \bibnamefont
  {{Vagins}}},\ and\ \bibinfo {author} {\bibfnamefont {K.~Z.}\ \bibnamefont
  {{Stanek}}},\ }\href {https://doi.org/10.1088/0004-637X/778/2/164} {\bibfield
   {journal} {\bibinfo  {journal} {\apj}\ }\textbf {\bibinfo {volume} {778}},\
  \bibinfo {eid} {164} (\bibinfo {year} {2013})},\ \Eprint
  {https://arxiv.org/abs/1306.0559} {arXiv:1306.0559 [astro-ph.HE]}
  \BibitemShut {NoStop}%
\bibitem [{\citenamefont {{Tamborra}}(2025)}]{tamborra_2025}%
  \BibitemOpen
  \bibfield  {author} {\bibinfo {author} {\bibfnamefont {I.}~\bibnamefont
  {{Tamborra}}},\ }\href {https://doi.org/10.1038/s42254-025-00828-2}
  {\bibfield  {journal} {\bibinfo  {journal} {Nature Reviews Physics}\ }\textbf
  {\bibinfo {volume} {7}},\ \bibinfo {pages} {285} (\bibinfo {year} {2025})},\
  \Eprint {https://arxiv.org/abs/2412.09699} {arXiv:2412.09699 [astro-ph.HE]}
  \BibitemShut {NoStop}%
\bibitem [{\citenamefont {{Hirata}}\ \emph {et~al.}(1987)\citenamefont
  {{Hirata}} \emph {et~al.}}]{hirata_1987}%
  \BibitemOpen
  \bibfield  {author} {\bibinfo {author} {\bibfnamefont {K.}~\bibnamefont
  {{Hirata}}} \emph {et~al.},\ }\href
  {https://doi.org/10.1103/PhysRevLett.58.1490} {\bibfield  {journal} {\bibinfo
   {journal} {\prl}\ }\textbf {\bibinfo {volume} {58}},\ \bibinfo {pages}
  {1490} (\bibinfo {year} {1987})}\BibitemShut {NoStop}%
\bibitem [{\citenamefont {{Bionta}}\ \emph {et~al.}(1987)\citenamefont
  {{Bionta}} \emph {et~al.}}]{bionta_1987}%
  \BibitemOpen
  \bibfield  {author} {\bibinfo {author} {\bibfnamefont {R.~M.}\ \bibnamefont
  {{Bionta}}} \emph {et~al.},\ }\href
  {https://doi.org/10.1103/PhysRevLett.58.1494} {\bibfield  {journal} {\bibinfo
   {journal} {\prl}\ }\textbf {\bibinfo {volume} {58}},\ \bibinfo {pages}
  {1494} (\bibinfo {year} {1987})}\BibitemShut {NoStop}%
\bibitem [{\citenamefont {{Alekseev}}\ \emph {et~al.}(1987)\citenamefont
  {{Alekseev}}, \citenamefont {{Alekseeva}}, \citenamefont {{Volchenko}},\ and\
  \citenamefont {{Krivosheina}}}]{alekseev_1987}%
  \BibitemOpen
  \bibfield  {author} {\bibinfo {author} {\bibfnamefont {E.~N.}\ \bibnamefont
  {{Alekseev}}}, \bibinfo {author} {\bibfnamefont {L.~N.}\ \bibnamefont
  {{Alekseeva}}}, \bibinfo {author} {\bibfnamefont {V.~I.}\ \bibnamefont
  {{Volchenko}}},\ and\ \bibinfo {author} {\bibfnamefont {I.~V.}\ \bibnamefont
  {{Krivosheina}}},\ }\href@noop {} {\bibfield  {journal} {\bibinfo  {journal}
  {Soviet Journal of Experimental and Theoretical Physics Letters}\ }\textbf
  {\bibinfo {volume} {45}},\ \bibinfo {pages} {589} (\bibinfo {year}
  {1987})}\BibitemShut {NoStop}%
\bibitem [{\citenamefont {{Arnett}}\ \emph {et~al.}(1989)\citenamefont
  {{Arnett}}, \citenamefont {{Bahcall}}, \citenamefont {{Kirshner}},\ and\
  \citenamefont {{Woosley}}}]{arnett_1989}%
  \BibitemOpen
  \bibfield  {author} {\bibinfo {author} {\bibfnamefont {W.~D.}\ \bibnamefont
  {{Arnett}}}, \bibinfo {author} {\bibfnamefont {J.~N.}\ \bibnamefont
  {{Bahcall}}}, \bibinfo {author} {\bibfnamefont {R.~P.}\ \bibnamefont
  {{Kirshner}}},\ and\ \bibinfo {author} {\bibfnamefont {S.~E.}\ \bibnamefont
  {{Woosley}}},\ }\href {https://doi.org/10.1146/annurev.aa.27.090189.003213}
  {\bibfield  {journal} {\bibinfo  {journal} {Annual Review of Astronomy and
  Astrophysics}\ }\textbf {\bibinfo {volume} {27}},\ \bibinfo {pages} {629}
  (\bibinfo {year} {1989})}\BibitemShut {NoStop}%
\bibitem [{\citenamefont {{Mirizzi}}\ \emph {et~al.}(2016)\citenamefont
  {{Mirizzi}}, \citenamefont {{Tamborra}}, \citenamefont {{Janka}},
  \citenamefont {{Saviano}}, \citenamefont {{Scholberg}}, \citenamefont
  {{Bollig}}, \citenamefont {{H{\"u}depohl}},\ and\ \citenamefont
  {{Chakraborty}}}]{mirizzi_2016}%
  \BibitemOpen
  \bibfield  {author} {\bibinfo {author} {\bibfnamefont {A.}~\bibnamefont
  {{Mirizzi}}}, \bibinfo {author} {\bibfnamefont {I.}~\bibnamefont
  {{Tamborra}}}, \bibinfo {author} {\bibfnamefont {H.~T.}\ \bibnamefont
  {{Janka}}}, \bibinfo {author} {\bibfnamefont {N.}~\bibnamefont {{Saviano}}},
  \bibinfo {author} {\bibfnamefont {K.}~\bibnamefont {{Scholberg}}}, \bibinfo
  {author} {\bibfnamefont {R.}~\bibnamefont {{Bollig}}}, \bibinfo {author}
  {\bibfnamefont {L.}~\bibnamefont {{H{\"u}depohl}}},\ and\ \bibinfo {author}
  {\bibfnamefont {S.}~\bibnamefont {{Chakraborty}}},\ }\href
  {https://doi.org/10.1393/ncr/i2016-10120-8} {\bibfield  {journal} {\bibinfo
  {journal} {Nuovo Cimento Rivista Serie}\ }\textbf {\bibinfo {volume} {39}},\
  \bibinfo {pages} {1} (\bibinfo {year} {2016})},\ \Eprint
  {https://arxiv.org/abs/1508.00785} {arXiv:1508.00785 [astro-ph.HE]}
  \BibitemShut {NoStop}%
\bibitem [{\citenamefont {{Janka}}(2017)}]{janka_2017}%
  \BibitemOpen
  \bibfield  {author} {\bibinfo {author} {\bibfnamefont {H.-T.}\ \bibnamefont
  {{Janka}}},\ }in\ \href {https://doi.org/10.1007/978-3-319-21846-5_4} {\emph
  {\bibinfo {booktitle} {Handbook of Supernovae}}},\ \bibinfo {editor} {edited
  by\ \bibinfo {editor} {\bibfnamefont {A.~W.}\ \bibnamefont {{Alsabti}}}\ and\
  \bibinfo {editor} {\bibfnamefont {P.}~\bibnamefont {{Murdin}}}}\ (\bibinfo
  {year} {2017})\ p.\ \bibinfo {pages} {1575}\BibitemShut {NoStop}%
\bibitem [{\citenamefont {{Horiuchi}}\ and\ \citenamefont
  {{Kneller}}(2018)}]{horiuchi_2018}%
  \BibitemOpen
  \bibfield  {author} {\bibinfo {author} {\bibfnamefont {S.}~\bibnamefont
  {{Horiuchi}}}\ and\ \bibinfo {author} {\bibfnamefont {J.~P.}\ \bibnamefont
  {{Kneller}}},\ }\href {https://doi.org/10.1088/1361-6471/aaa90a} {\bibfield
  {journal} {\bibinfo  {journal} {Journal of Physics G Nuclear Physics}\
  }\textbf {\bibinfo {volume} {45}},\ \bibinfo {pages} {043002} (\bibinfo
  {year} {2018})},\ \Eprint {https://arxiv.org/abs/1709.01515}
  {arXiv:1709.01515 [astro-ph.HE]} \BibitemShut {NoStop}%
\bibitem [{\citenamefont {{Scholberg}}(2018)}]{scholberg_2018}%
  \BibitemOpen
  \bibfield  {author} {\bibinfo {author} {\bibfnamefont {K.}~\bibnamefont
  {{Scholberg}}},\ }\href {https://doi.org/10.1088/1361-6471/aa97be} {\bibfield
   {journal} {\bibinfo  {journal} {Journal of Physics G Nuclear Physics}\
  }\textbf {\bibinfo {volume} {45}},\ \bibinfo {pages} {014002} (\bibinfo
  {year} {2018})},\ \Eprint {https://arxiv.org/abs/1707.06384}
  {arXiv:1707.06384 [hep-ex]} \BibitemShut {NoStop}%
\bibitem [{\citenamefont {{M{\"u}ller}}(2019)}]{mueller_2019}%
  \BibitemOpen
  \bibfield  {author} {\bibinfo {author} {\bibfnamefont {B.}~\bibnamefont
  {{M{\"u}ller}}},\ }\href {https://doi.org/10.1146/annurev-nucl-101918-023434}
  {\bibfield  {journal} {\bibinfo  {journal} {Annual Review of Nuclear and
  Particle Science}\ }\textbf {\bibinfo {volume} {69}},\ \bibinfo {pages} {253}
  (\bibinfo {year} {2019})},\ \Eprint {https://arxiv.org/abs/1904.11067}
  {arXiv:1904.11067 [astro-ph.HE]} \BibitemShut {NoStop}%
\bibitem [{\citenamefont {{Abdikamalov}}\ \emph {et~al.}(2022)\citenamefont
  {{Abdikamalov}}, \citenamefont {{Pagliaroli}},\ and\ \citenamefont
  {{Radice}}}]{abdikamalov_2022}%
  \BibitemOpen
  \bibfield  {author} {\bibinfo {author} {\bibfnamefont {E.}~\bibnamefont
  {{Abdikamalov}}}, \bibinfo {author} {\bibfnamefont {G.}~\bibnamefont
  {{Pagliaroli}}},\ and\ \bibinfo {author} {\bibfnamefont {D.}~\bibnamefont
  {{Radice}}},\ }in\ \href {https://doi.org/10.1007/978-981-15-4702-7_21-1}
  {\emph {\bibinfo {booktitle} {Handbook of Gravitational Wave Astronomy}}},\
  \bibinfo {editor} {edited by\ \bibinfo {editor} {\bibfnamefont
  {C.}~\bibnamefont {{Bambi}}}, \bibinfo {editor} {\bibfnamefont
  {S.}~\bibnamefont {{Katsanevas}}},\ and\ \bibinfo {editor} {\bibfnamefont
  {K.~D.}\ \bibnamefont {{Kokkotas}}}}\ (\bibinfo {year} {2022})\ p.~\bibinfo
  {pages} {21}\BibitemShut {NoStop}%
\bibitem [{\citenamefont {{Christensen}}\ and\ \citenamefont
  {{Meyer}}(2022)}]{christensen_2022}%
  \BibitemOpen
  \bibfield  {author} {\bibinfo {author} {\bibfnamefont {N.}~\bibnamefont
  {{Christensen}}}\ and\ \bibinfo {author} {\bibfnamefont {R.}~\bibnamefont
  {{Meyer}}},\ }\href {https://doi.org/10.1103/RevModPhys.94.025001} {\bibfield
   {journal} {\bibinfo  {journal} {Reviews of Modern Physics}\ }\textbf
  {\bibinfo {volume} {94}},\ \bibinfo {eid} {025001} (\bibinfo {year}
  {2022})},\ \Eprint {https://arxiv.org/abs/2204.04449} {arXiv:2204.04449
  [gr-qc]} \BibitemShut {NoStop}%
\bibitem [{\citenamefont {{Powell}}\ and\ \citenamefont
  {{M{\"u}ller}}(2022)}]{powell_2022}%
  \BibitemOpen
  \bibfield  {author} {\bibinfo {author} {\bibfnamefont {J.}~\bibnamefont
  {{Powell}}}\ and\ \bibinfo {author} {\bibfnamefont {B.}~\bibnamefont
  {{M{\"u}ller}}},\ }\href {https://doi.org/10.1103/PhysRevD.105.063018}
  {\bibfield  {journal} {\bibinfo  {journal} {\prd}\ }\textbf {\bibinfo
  {volume} {105}},\ \bibinfo {eid} {063018} (\bibinfo {year} {2022})},\ \Eprint
  {https://arxiv.org/abs/2201.01397} {arXiv:2201.01397 [astro-ph.HE]}
  \BibitemShut {NoStop}%
\bibitem [{\citenamefont {{Woosley}}(1993)}]{woosley_1993}%
  \BibitemOpen
  \bibfield  {author} {\bibinfo {author} {\bibfnamefont {S.~E.}\ \bibnamefont
  {{Woosley}}},\ }\href {https://doi.org/10.1086/172359} {\bibfield  {journal}
  {\bibinfo  {journal} {ApJ}\ }\textbf {\bibinfo {volume} {405}},\ \bibinfo
  {pages} {273} (\bibinfo {year} {1993})}\BibitemShut {NoStop}%
\bibitem [{\citenamefont {{Popham}}\ \emph {et~al.}(1999)\citenamefont
  {{Popham}}, \citenamefont {{Woosley}},\ and\ \citenamefont
  {{Fryer}}}]{popham1999}%
  \BibitemOpen
  \bibfield  {author} {\bibinfo {author} {\bibfnamefont {R.}~\bibnamefont
  {{Popham}}}, \bibinfo {author} {\bibfnamefont {S.~E.}\ \bibnamefont
  {{Woosley}}},\ and\ \bibinfo {author} {\bibfnamefont {C.}~\bibnamefont
  {{Fryer}}},\ }\href@noop {} {\bibfield  {journal} {\bibinfo  {journal} {ApJ}\
  }\textbf {\bibinfo {volume} {518}},\ \bibinfo {pages} {356} (\bibinfo {year}
  {1999})}\BibitemShut {NoStop}%
\bibitem [{\citenamefont {{Chen}}\ and\ \citenamefont
  {{Beloborodov}}(2007)}]{chen_2007}%
  \BibitemOpen
  \bibfield  {author} {\bibinfo {author} {\bibfnamefont {W.-X.}\ \bibnamefont
  {{Chen}}}\ and\ \bibinfo {author} {\bibfnamefont {A.~M.}\ \bibnamefont
  {{Beloborodov}}},\ }\href {https://doi.org/10.1086/508923} {\bibfield
  {journal} {\bibinfo  {journal} {\apj}\ }\textbf {\bibinfo {volume} {657}},\
  \bibinfo {pages} {383} (\bibinfo {year} {2007})},\ \Eprint
  {https://arxiv.org/abs/astro-ph/0607145} {arXiv:astro-ph/0607145 [astro-ph]}
  \BibitemShut {NoStop}%
\bibitem [{\citenamefont {{MacFadyen}}\ and\ \citenamefont
  {{Woosley}}(1999)}]{macfadyen_1999}%
  \BibitemOpen
  \bibfield  {author} {\bibinfo {author} {\bibfnamefont {A.~I.}\ \bibnamefont
  {{MacFadyen}}}\ and\ \bibinfo {author} {\bibfnamefont {S.~E.}\ \bibnamefont
  {{Woosley}}},\ }\href {https://doi.org/10.1086/307790} {\bibfield  {journal}
  {\bibinfo  {journal} {ApJ}\ }\textbf {\bibinfo {volume} {524}},\ \bibinfo
  {pages} {262} (\bibinfo {year} {1999})},\ \Eprint
  {https://arxiv.org/abs/astro-ph/9810274} {astro-ph/9810274} \BibitemShut
  {NoStop}%
\bibitem [{\citenamefont {{MacFadyen}}(2003)}]{macfadyen_2003}%
  \BibitemOpen
  \bibfield  {author} {\bibinfo {author} {\bibfnamefont {A.~I.}\ \bibnamefont
  {{MacFadyen}}},\ }in\ \href {https://doi.org/10.1007/10828549_14} {\emph
  {\bibinfo {booktitle} {From Twilight to Highlight: The Physics of
  Supernovae}}},\ \bibinfo {editor} {edited by\ \bibinfo {editor}
  {\bibfnamefont {W.}~\bibnamefont {{Hillebrandt}}}\ and\ \bibinfo {editor}
  {\bibfnamefont {B.}~\bibnamefont {{Leibundgut}}}}\ (\bibinfo {year} {2003})\
  p.~\bibinfo {pages} {97},\ \Eprint {https://arxiv.org/abs/astro-ph/0301425}
  {arXiv:astro-ph/0301425 [astro-ph]} \BibitemShut {NoStop}%
\bibitem [{\citenamefont {{Woosley}}\ and\ \citenamefont
  {{Bloom}}(2006)}]{woosley_2006}%
  \BibitemOpen
  \bibfield  {author} {\bibinfo {author} {\bibfnamefont {S.~E.}\ \bibnamefont
  {{Woosley}}}\ and\ \bibinfo {author} {\bibfnamefont {J.~S.}\ \bibnamefont
  {{Bloom}}},\ }\href {https://doi.org/10.1146/annurev.astro.43.072103.150558}
  {\bibfield  {journal} {\bibinfo  {journal} {ARA\&A}\ }\textbf {\bibinfo
  {volume} {44}},\ \bibinfo {pages} {507} (\bibinfo {year} {2006})},\ \Eprint
  {https://arxiv.org/abs/astro-ph/0609142} {arXiv:astro-ph/0609142 [astro-ph]}
  \BibitemShut {NoStop}%
\bibitem [{\citenamefont {{Pruet}}\ \emph {et~al.}(2003)\citenamefont
  {{Pruet}}, \citenamefont {{Woosley}},\ and\ \citenamefont
  {{Hoffman}}}]{pruet_2003}%
  \BibitemOpen
  \bibfield  {author} {\bibinfo {author} {\bibfnamefont {J.}~\bibnamefont
  {{Pruet}}}, \bibinfo {author} {\bibfnamefont {S.~E.}\ \bibnamefont
  {{Woosley}}},\ and\ \bibinfo {author} {\bibfnamefont {R.~D.}\ \bibnamefont
  {{Hoffman}}},\ }\href {https://doi.org/10.1086/367957} {\bibfield  {journal}
  {\bibinfo  {journal} {ApJ}\ }\textbf {\bibinfo {volume} {586}},\ \bibinfo
  {pages} {1254} (\bibinfo {year} {2003})},\ \Eprint
  {https://arxiv.org/abs/astro-ph/0209412} {arXiv:astro-ph/0209412 [astro-ph]}
  \BibitemShut {NoStop}%
\bibitem [{\citenamefont {{Kohri}}\ \emph {et~al.}(2005)\citenamefont
  {{Kohri}}, \citenamefont {{Narayan}},\ and\ \citenamefont
  {{Piran}}}]{kohri_2005}%
  \BibitemOpen
  \bibfield  {author} {\bibinfo {author} {\bibfnamefont {K.}~\bibnamefont
  {{Kohri}}}, \bibinfo {author} {\bibfnamefont {R.}~\bibnamefont {{Narayan}}},\
  and\ \bibinfo {author} {\bibfnamefont {T.}~\bibnamefont {{Piran}}},\ }\href
  {https://doi.org/10.1086/431354} {\bibfield  {journal} {\bibinfo  {journal}
  {ApJ}\ }\textbf {\bibinfo {volume} {629}},\ \bibinfo {pages} {341} (\bibinfo
  {year} {2005})},\ \Eprint {https://arxiv.org/abs/astro-ph/0502470}
  {arXiv:astro-ph/0502470 [astro-ph]} \BibitemShut {NoStop}%
\bibitem [{\citenamefont {{Siegel}}\ \emph {et~al.}(2019)\citenamefont
  {{Siegel}}, \citenamefont {{Barnes}},\ and\ \citenamefont
  {{Metzger}}}]{siegel_2019}%
  \BibitemOpen
  \bibfield  {author} {\bibinfo {author} {\bibfnamefont {D.~M.}\ \bibnamefont
  {{Siegel}}}, \bibinfo {author} {\bibfnamefont {J.}~\bibnamefont {{Barnes}}},\
  and\ \bibinfo {author} {\bibfnamefont {B.~D.}\ \bibnamefont {{Metzger}}},\
  }\href {https://doi.org/10.1038/s41586-019-1136-0} {\bibfield  {journal}
  {\bibinfo  {journal} {Nature}\ }\textbf {\bibinfo {volume} {569}},\ \bibinfo
  {pages} {241} (\bibinfo {year} {2019})},\ \Eprint
  {https://arxiv.org/abs/1810.00098} {arXiv:1810.00098 [astro-ph.HE]}
  \BibitemShut {NoStop}%
\bibitem [{\citenamefont {{Dean}}\ and\ \citenamefont
  {{Fern{\'a}ndez}}(2024{\natexlab{a}})}]{DF24b}%
  \BibitemOpen
  \bibfield  {author} {\bibinfo {author} {\bibfnamefont {C.}~\bibnamefont
  {{Dean}}}\ and\ \bibinfo {author} {\bibfnamefont {R.}~\bibnamefont
  {{Fern{\'a}ndez}}},\ }\href {https://doi.org/10.1103/PhysRevD.110.083024}
  {\bibfield  {journal} {\bibinfo  {journal} {\prd}\ }\textbf {\bibinfo
  {volume} {110}},\ \bibinfo {eid} {083024} (\bibinfo {year}
  {2024}{\natexlab{a}})},\ \Eprint {https://arxiv.org/abs/2408.15338}
  {arXiv:2408.15338 [astro-ph.HE]} \BibitemShut {NoStop}%
\bibitem [{\citenamefont {{Lopez-Camara}}\ \emph {et~al.}(2009)\citenamefont
  {{Lopez-Camara}}, \citenamefont {{Lee}},\ and\ \citenamefont
  {{Ramirez-Ruiz}}}]{lopezcamara_2009}%
  \BibitemOpen
  \bibfield  {author} {\bibinfo {author} {\bibfnamefont {D.}~\bibnamefont
  {{Lopez-Camara}}}, \bibinfo {author} {\bibfnamefont {W.~H.}\ \bibnamefont
  {{Lee}}},\ and\ \bibinfo {author} {\bibfnamefont {E.}~\bibnamefont
  {{Ramirez-Ruiz}}},\ }\href {https://doi.org/10.1088/0004-637X/692/1/804}
  {\bibfield  {journal} {\bibinfo  {journal} {ApJ}\ }\textbf {\bibinfo {volume}
  {692}},\ \bibinfo {pages} {804} (\bibinfo {year} {2009})},\ \Eprint
  {https://arxiv.org/abs/0808.0462} {arXiv:0808.0462 [astro-ph]} \BibitemShut
  {NoStop}%
\bibitem [{\citenamefont {{Sekiguchi}}\ and\ \citenamefont
  {{Shibata}}(2011)}]{sekiguchi_2011}%
  \BibitemOpen
  \bibfield  {author} {\bibinfo {author} {\bibfnamefont {Y.}~\bibnamefont
  {{Sekiguchi}}}\ and\ \bibinfo {author} {\bibfnamefont {M.}~\bibnamefont
  {{Shibata}}},\ }\href {https://doi.org/10.1088/0004-637X/737/1/6} {\bibfield
  {journal} {\bibinfo  {journal} {ApJ}\ }\textbf {\bibinfo {volume} {737}},\
  \bibinfo {eid} {6} (\bibinfo {year} {2011})},\ \Eprint
  {https://arxiv.org/abs/1009.5303} {arXiv:1009.5303 [astro-ph.HE]}
  \BibitemShut {NoStop}%
\bibitem [{\citenamefont {{Just}}\ \emph {et~al.}(2022)\citenamefont {{Just}},
  \citenamefont {{Aloy}}, \citenamefont {{Obergaulinger}},\ and\ \citenamefont
  {{Nagataki}}}]{just_2022}%
  \BibitemOpen
  \bibfield  {author} {\bibinfo {author} {\bibfnamefont {O.}~\bibnamefont
  {{Just}}}, \bibinfo {author} {\bibfnamefont {M.~A.}\ \bibnamefont {{Aloy}}},
  \bibinfo {author} {\bibfnamefont {M.}~\bibnamefont {{Obergaulinger}}},\ and\
  \bibinfo {author} {\bibfnamefont {S.}~\bibnamefont {{Nagataki}}},\ }\href
  {https://doi.org/10.3847/2041-8213/ac83a1} {\bibfield  {journal} {\bibinfo
  {journal} {\apj}\ }\textbf {\bibinfo {volume} {934}},\ \bibinfo {eid} {L30}
  (\bibinfo {year} {2022})},\ \Eprint {https://arxiv.org/abs/2205.14158}
  {arXiv:2205.14158 [astro-ph.HE]} \BibitemShut {NoStop}%
\bibitem [{\citenamefont {{Taylor}}\ \emph {et~al.}(2011)\citenamefont
  {{Taylor}}, \citenamefont {{Miller}},\ and\ \citenamefont
  {{Podsiadlowski}}}]{taylor_2011}%
  \BibitemOpen
  \bibfield  {author} {\bibinfo {author} {\bibfnamefont {P.~A.}\ \bibnamefont
  {{Taylor}}}, \bibinfo {author} {\bibfnamefont {J.~C.}\ \bibnamefont
  {{Miller}}},\ and\ \bibinfo {author} {\bibfnamefont {P.}~\bibnamefont
  {{Podsiadlowski}}},\ }\href
  {https://doi.org/10.1111/j.1365-2966.2010.17618.x} {\bibfield  {journal}
  {\bibinfo  {journal} {MNRAS}\ }\textbf {\bibinfo {volume} {410}},\ \bibinfo
  {pages} {2385} (\bibinfo {year} {2011})},\ \Eprint
  {https://arxiv.org/abs/1006.4624} {arXiv:1006.4624 [astro-ph.HE]}
  \BibitemShut {NoStop}%
\bibitem [{\citenamefont {{Gottlieb}}\ \emph {et~al.}(2022)\citenamefont
  {{Gottlieb}}, \citenamefont {{Lalakos}}, \citenamefont {{Bromberg}},
  \citenamefont {{Liska}},\ and\ \citenamefont
  {{Tchekhovskoy}}}]{gottlieb_2022}%
  \BibitemOpen
  \bibfield  {author} {\bibinfo {author} {\bibfnamefont {O.}~\bibnamefont
  {{Gottlieb}}}, \bibinfo {author} {\bibfnamefont {A.}~\bibnamefont
  {{Lalakos}}}, \bibinfo {author} {\bibfnamefont {O.}~\bibnamefont
  {{Bromberg}}}, \bibinfo {author} {\bibfnamefont {M.}~\bibnamefont
  {{Liska}}},\ and\ \bibinfo {author} {\bibfnamefont {A.}~\bibnamefont
  {{Tchekhovskoy}}},\ }\href {https://doi.org/10.1093/mnras/stab3784}
  {\bibfield  {journal} {\bibinfo  {journal} {MNRAS}\ }\textbf {\bibinfo
  {volume} {510}},\ \bibinfo {pages} {4962} (\bibinfo {year} {2022})},\ \Eprint
  {https://arxiv.org/abs/2109.14619} {arXiv:2109.14619 [astro-ph.HE]}
  \BibitemShut {NoStop}%
\bibitem [{\citenamefont {{Blondin}}\ \emph {et~al.}(2003)\citenamefont
  {{Blondin}}, \citenamefont {{Mezzacappa}},\ and\ \citenamefont
  {{DeMarino}}}]{blondin_2003}%
  \BibitemOpen
  \bibfield  {author} {\bibinfo {author} {\bibfnamefont {J.~M.}\ \bibnamefont
  {{Blondin}}}, \bibinfo {author} {\bibfnamefont {A.}~\bibnamefont
  {{Mezzacappa}}},\ and\ \bibinfo {author} {\bibfnamefont {C.}~\bibnamefont
  {{DeMarino}}},\ }\href {https://doi.org/10.1086/345812} {\bibfield  {journal}
  {\bibinfo  {journal} {\apj}\ }\textbf {\bibinfo {volume} {584}},\ \bibinfo
  {pages} {971} (\bibinfo {year} {2003})},\ \Eprint
  {https://arxiv.org/abs/astro-ph/0210634} {arXiv:astro-ph/0210634 [astro-ph]}
  \BibitemShut {NoStop}%
\bibitem [{\citenamefont {{Foglizzo}}\ \emph {et~al.}(2007)\citenamefont
  {{Foglizzo}}, \citenamefont {{Galletti}}, \citenamefont {{Scheck}},\ and\
  \citenamefont {{Janka}}}]{foglizzo_2007}%
  \BibitemOpen
  \bibfield  {author} {\bibinfo {author} {\bibfnamefont {T.}~\bibnamefont
  {{Foglizzo}}}, \bibinfo {author} {\bibfnamefont {P.}~\bibnamefont
  {{Galletti}}}, \bibinfo {author} {\bibfnamefont {L.}~\bibnamefont
  {{Scheck}}},\ and\ \bibinfo {author} {\bibfnamefont {H.~T.}\ \bibnamefont
  {{Janka}}},\ }\href {https://doi.org/10.1086/509612} {\bibfield  {journal}
  {\bibinfo  {journal} {\apj}\ }\textbf {\bibinfo {volume} {654}},\ \bibinfo
  {pages} {1006} (\bibinfo {year} {2007})},\ \Eprint
  {https://arxiv.org/abs/astro-ph/0606640} {arXiv:astro-ph/0606640 [astro-ph]}
  \BibitemShut {NoStop}%
\bibitem [{\citenamefont {{Molteni}}\ \emph {et~al.}(1999)\citenamefont
  {{Molteni}}, \citenamefont {{T{\'o}th}},\ and\ \citenamefont
  {{Kuznetsov}}}]{molteni_1999}%
  \BibitemOpen
  \bibfield  {author} {\bibinfo {author} {\bibfnamefont {D.}~\bibnamefont
  {{Molteni}}}, \bibinfo {author} {\bibfnamefont {G.}~\bibnamefont
  {{T{\'o}th}}},\ and\ \bibinfo {author} {\bibfnamefont {O.~A.}\ \bibnamefont
  {{Kuznetsov}}},\ }\href {https://doi.org/10.1086/307079} {\bibfield
  {journal} {\bibinfo  {journal} {\apj}\ }\textbf {\bibinfo {volume} {516}},\
  \bibinfo {pages} {411} (\bibinfo {year} {1999})},\ \Eprint
  {https://arxiv.org/abs/astro-ph/9812453} {arXiv:astro-ph/9812453 [astro-ph]}
  \BibitemShut {NoStop}%
\bibitem [{\citenamefont {{Gu}}\ and\ \citenamefont
  {{Foglizzo}}(2003)}]{gu_foglizzo_2003}%
  \BibitemOpen
  \bibfield  {author} {\bibinfo {author} {\bibfnamefont {W.-M.}\ \bibnamefont
  {{Gu}}}\ and\ \bibinfo {author} {\bibfnamefont {T.}~\bibnamefont
  {{Foglizzo}}},\ }\href {https://doi.org/10.1051/0004-6361:20031080}
  {\bibfield  {journal} {\bibinfo  {journal} {Astronomy \& Astrophysics}\
  }\textbf {\bibinfo {volume} {409}},\ \bibinfo {pages} {1} (\bibinfo {year}
  {2003})},\ \Eprint {https://arxiv.org/abs/astro-ph/0307047}
  {arXiv:astro-ph/0307047 [astro-ph]} \BibitemShut {NoStop}%
\bibitem [{\citenamefont {{Gu}}\ and\ \citenamefont {{Lu}}(2006)}]{gu_lu_2006}%
  \BibitemOpen
  \bibfield  {author} {\bibinfo {author} {\bibfnamefont {W.-M.}\ \bibnamefont
  {{Gu}}}\ and\ \bibinfo {author} {\bibfnamefont {J.-F.}\ \bibnamefont
  {{Lu}}},\ }\href {https://doi.org/10.1111/j.1365-2966.2005.09750.x}
  {\bibfield  {journal} {\bibinfo  {journal} {MNRAS}\ }\textbf {\bibinfo
  {volume} {365}},\ \bibinfo {pages} {647} (\bibinfo {year} {2006})},\ \Eprint
  {https://arxiv.org/abs/astro-ph/0511211} {arXiv:astro-ph/0511211 [astro-ph]}
  \BibitemShut {NoStop}%
\bibitem [{\citenamefont {{Nagakura}}\ and\ \citenamefont
  {{Yamada}}(2008)}]{nagakura_yamada_2008}%
  \BibitemOpen
  \bibfield  {author} {\bibinfo {author} {\bibfnamefont {H.}~\bibnamefont
  {{Nagakura}}}\ and\ \bibinfo {author} {\bibfnamefont {S.}~\bibnamefont
  {{Yamada}}},\ }\href {https://doi.org/10.1086/590325} {\bibfield  {journal}
  {\bibinfo  {journal} {\apj}\ }\textbf {\bibinfo {volume} {689}},\ \bibinfo
  {pages} {391} (\bibinfo {year} {2008})},\ \Eprint
  {https://arxiv.org/abs/0808.4141} {arXiv:0808.4141 [astro-ph]} \BibitemShut
  {NoStop}%
\bibitem [{\citenamefont {{Nagakura}}\ and\ \citenamefont
  {{Yamada}}(2009)}]{nagakura_yamada_2009}%
  \BibitemOpen
  \bibfield  {author} {\bibinfo {author} {\bibfnamefont {H.}~\bibnamefont
  {{Nagakura}}}\ and\ \bibinfo {author} {\bibfnamefont {S.}~\bibnamefont
  {{Yamada}}},\ }\href {https://doi.org/10.1088/0004-637X/696/2/2026}
  {\bibfield  {journal} {\bibinfo  {journal} {\apj}\ }\textbf {\bibinfo
  {volume} {696}},\ \bibinfo {pages} {2026} (\bibinfo {year} {2009})},\ \Eprint
  {https://arxiv.org/abs/0901.4053} {arXiv:0901.4053 [astro-ph.HE]}
  \BibitemShut {NoStop}%
\bibitem [{\citenamefont {{Lund}}\ \emph {et~al.}(2010)\citenamefont {{Lund}},
  \citenamefont {{Marek}}, \citenamefont {{Lunardini}}, \citenamefont
  {{Janka}},\ and\ \citenamefont {{Raffelt}}}]{lund_2010}%
  \BibitemOpen
  \bibfield  {author} {\bibinfo {author} {\bibfnamefont {T.}~\bibnamefont
  {{Lund}}}, \bibinfo {author} {\bibfnamefont {A.}~\bibnamefont {{Marek}}},
  \bibinfo {author} {\bibfnamefont {C.}~\bibnamefont {{Lunardini}}}, \bibinfo
  {author} {\bibfnamefont {H.-T.}\ \bibnamefont {{Janka}}},\ and\ \bibinfo
  {author} {\bibfnamefont {G.}~\bibnamefont {{Raffelt}}},\ }\href
  {https://doi.org/10.1103/PhysRevD.82.063007} {\bibfield  {journal} {\bibinfo
  {journal} {\prd}\ }\textbf {\bibinfo {volume} {82}},\ \bibinfo {eid} {063007}
  (\bibinfo {year} {2010})},\ \Eprint {https://arxiv.org/abs/1006.1889}
  {arXiv:1006.1889 [astro-ph.HE]} \BibitemShut {NoStop}%
\bibitem [{\citenamefont {{Lund}}\ \emph {et~al.}(2012)\citenamefont {{Lund}},
  \citenamefont {{Wongwathanarat}}, \citenamefont {{Janka}}, \citenamefont
  {{M{\"u}ller}},\ and\ \citenamefont {{Raffelt}}}]{lund_2012}%
  \BibitemOpen
  \bibfield  {author} {\bibinfo {author} {\bibfnamefont {T.}~\bibnamefont
  {{Lund}}}, \bibinfo {author} {\bibfnamefont {A.}~\bibnamefont
  {{Wongwathanarat}}}, \bibinfo {author} {\bibfnamefont {H.-T.}\ \bibnamefont
  {{Janka}}}, \bibinfo {author} {\bibfnamefont {E.}~\bibnamefont
  {{M{\"u}ller}}},\ and\ \bibinfo {author} {\bibfnamefont {G.}~\bibnamefont
  {{Raffelt}}},\ }\href {https://doi.org/10.1103/PhysRevD.86.105031} {\bibfield
   {journal} {\bibinfo  {journal} {\prd}\ }\textbf {\bibinfo {volume} {86}},\
  \bibinfo {eid} {105031} (\bibinfo {year} {2012})},\ \Eprint
  {https://arxiv.org/abs/1208.0043} {arXiv:1208.0043 [astro-ph.HE]}
  \BibitemShut {NoStop}%
\bibitem [{\citenamefont {{Tamborra}}\ \emph {et~al.}(2013)\citenamefont
  {{Tamborra}}, \citenamefont {{Hanke}}, \citenamefont {{M{\"u}ller}},
  \citenamefont {{Janka}},\ and\ \citenamefont {{Raffelt}}}]{tamborra_2013}%
  \BibitemOpen
  \bibfield  {author} {\bibinfo {author} {\bibfnamefont {I.}~\bibnamefont
  {{Tamborra}}}, \bibinfo {author} {\bibfnamefont {F.}~\bibnamefont {{Hanke}}},
  \bibinfo {author} {\bibfnamefont {B.}~\bibnamefont {{M{\"u}ller}}}, \bibinfo
  {author} {\bibfnamefont {H.-T.}\ \bibnamefont {{Janka}}},\ and\ \bibinfo
  {author} {\bibfnamefont {G.}~\bibnamefont {{Raffelt}}},\ }\href
  {https://doi.org/10.1103/PhysRevLett.111.121104} {\bibfield  {journal}
  {\bibinfo  {journal} {\prl}\ }\textbf {\bibinfo {volume} {111}},\ \bibinfo
  {eid} {121104} (\bibinfo {year} {2013})},\ \Eprint
  {https://arxiv.org/abs/1307.7936} {arXiv:1307.7936 [astro-ph.SR]}
  \BibitemShut {NoStop}%
\bibitem [{\citenamefont {{Tamborra}}\ \emph
  {et~al.}(2014{\natexlab{a}})\citenamefont {{Tamborra}}, \citenamefont
  {{Raffelt}}, \citenamefont {{Hanke}}, \citenamefont {{Janka}},\ and\
  \citenamefont {{M{\"u}ller}}}]{tamborra_2014}%
  \BibitemOpen
  \bibfield  {author} {\bibinfo {author} {\bibfnamefont {I.}~\bibnamefont
  {{Tamborra}}}, \bibinfo {author} {\bibfnamefont {G.}~\bibnamefont
  {{Raffelt}}}, \bibinfo {author} {\bibfnamefont {F.}~\bibnamefont {{Hanke}}},
  \bibinfo {author} {\bibfnamefont {H.-T.}\ \bibnamefont {{Janka}}},\ and\
  \bibinfo {author} {\bibfnamefont {B.}~\bibnamefont {{M{\"u}ller}}},\ }\href
  {https://doi.org/10.1103/PhysRevD.90.045032} {\bibfield  {journal} {\bibinfo
  {journal} {\prd}\ }\textbf {\bibinfo {volume} {90}},\ \bibinfo {eid} {045032}
  (\bibinfo {year} {2014}{\natexlab{a}})},\ \Eprint
  {https://arxiv.org/abs/1406.0006} {arXiv:1406.0006 [astro-ph.SR]}
  \BibitemShut {NoStop}%
\bibitem [{\citenamefont {{M{\"u}ller}}\ and\ \citenamefont
  {{Janka}}(2014)}]{mueller_2014_neutrinos}%
  \BibitemOpen
  \bibfield  {author} {\bibinfo {author} {\bibfnamefont {B.}~\bibnamefont
  {{M{\"u}ller}}}\ and\ \bibinfo {author} {\bibfnamefont {H.-T.}\ \bibnamefont
  {{Janka}}},\ }\href {https://doi.org/10.1088/0004-637X/788/1/82} {\bibfield
  {journal} {\bibinfo  {journal} {\apj}\ }\textbf {\bibinfo {volume} {788}},\
  \bibinfo {eid} {82} (\bibinfo {year} {2014})},\ \Eprint
  {https://arxiv.org/abs/1402.3415} {arXiv:1402.3415 [astro-ph.SR]}
  \BibitemShut {NoStop}%
\bibitem [{\citenamefont {{Walk}}\ \emph {et~al.}(2018)\citenamefont {{Walk}},
  \citenamefont {{Tamborra}}, \citenamefont {{Janka}},\ and\ \citenamefont
  {{Summa}}}]{walk_2018}%
  \BibitemOpen
  \bibfield  {author} {\bibinfo {author} {\bibfnamefont {L.}~\bibnamefont
  {{Walk}}}, \bibinfo {author} {\bibfnamefont {I.}~\bibnamefont {{Tamborra}}},
  \bibinfo {author} {\bibfnamefont {H.-T.}\ \bibnamefont {{Janka}}},\ and\
  \bibinfo {author} {\bibfnamefont {A.}~\bibnamefont {{Summa}}},\ }\href
  {https://doi.org/10.1103/PhysRevD.98.123001} {\bibfield  {journal} {\bibinfo
  {journal} {\prd}\ }\textbf {\bibinfo {volume} {98}},\ \bibinfo {eid} {123001}
  (\bibinfo {year} {2018})},\ \Eprint {https://arxiv.org/abs/1807.02366}
  {arXiv:1807.02366 [astro-ph.HE]} \BibitemShut {NoStop}%
\bibitem [{\citenamefont {{Walk}}\ \emph {et~al.}(2019)\citenamefont {{Walk}},
  \citenamefont {{Tamborra}}, \citenamefont {{Janka}},\ and\ \citenamefont
  {{Summa}}}]{walk_2019}%
  \BibitemOpen
  \bibfield  {author} {\bibinfo {author} {\bibfnamefont {L.}~\bibnamefont
  {{Walk}}}, \bibinfo {author} {\bibfnamefont {I.}~\bibnamefont {{Tamborra}}},
  \bibinfo {author} {\bibfnamefont {H.-T.}\ \bibnamefont {{Janka}}},\ and\
  \bibinfo {author} {\bibfnamefont {A.}~\bibnamefont {{Summa}}},\ }\href
  {https://doi.org/10.1103/PhysRevD.100.063018} {\bibfield  {journal} {\bibinfo
   {journal} {\prd}\ }\textbf {\bibinfo {volume} {100}},\ \bibinfo {eid}
  {063018} (\bibinfo {year} {2019})},\ \Eprint
  {https://arxiv.org/abs/1901.06235} {arXiv:1901.06235 [astro-ph.HE]}
  \BibitemShut {NoStop}%
\bibitem [{\citenamefont {{Vartanyan}}\ \emph {et~al.}(2019)\citenamefont
  {{Vartanyan}}, \citenamefont {{Burrows}},\ and\ \citenamefont
  {{Radice}}}]{vartanyan_2019}%
  \BibitemOpen
  \bibfield  {author} {\bibinfo {author} {\bibfnamefont {D.}~\bibnamefont
  {{Vartanyan}}}, \bibinfo {author} {\bibfnamefont {A.}~\bibnamefont
  {{Burrows}}},\ and\ \bibinfo {author} {\bibfnamefont {D.}~\bibnamefont
  {{Radice}}},\ }\href {https://doi.org/10.1093/mnras/stz2307} {\bibfield
  {journal} {\bibinfo  {journal} {MNRAS}\ }\textbf {\bibinfo {volume} {489}},\
  \bibinfo {pages} {2227} (\bibinfo {year} {2019})},\ \Eprint
  {https://arxiv.org/abs/1906.08787} {arXiv:1906.08787 [astro-ph.HE]}
  \BibitemShut {NoStop}%
\bibitem [{\citenamefont {{Walk}}\ \emph {et~al.}(2020)\citenamefont {{Walk}},
  \citenamefont {{Tamborra}}, \citenamefont {{Janka}}, \citenamefont
  {{Summa}},\ and\ \citenamefont {{Kresse}}}]{walk_2020}%
  \BibitemOpen
  \bibfield  {author} {\bibinfo {author} {\bibfnamefont {L.}~\bibnamefont
  {{Walk}}}, \bibinfo {author} {\bibfnamefont {I.}~\bibnamefont {{Tamborra}}},
  \bibinfo {author} {\bibfnamefont {H.-T.}\ \bibnamefont {{Janka}}}, \bibinfo
  {author} {\bibfnamefont {A.}~\bibnamefont {{Summa}}},\ and\ \bibinfo {author}
  {\bibfnamefont {D.}~\bibnamefont {{Kresse}}},\ }\href
  {https://doi.org/10.1103/PhysRevD.101.123013} {\bibfield  {journal} {\bibinfo
   {journal} {\prd}\ }\textbf {\bibinfo {volume} {101}},\ \bibinfo {eid}
  {123013} (\bibinfo {year} {2020})},\ \Eprint
  {https://arxiv.org/abs/1910.12971} {arXiv:1910.12971 [astro-ph.HE]}
  \BibitemShut {NoStop}%
\bibitem [{\citenamefont {{Lin}}\ \emph {et~al.}(2020)\citenamefont {{Lin}},
  \citenamefont {{Lunardini}}, \citenamefont {{Zanolin}}, \citenamefont
  {{Kotake}},\ and\ \citenamefont {{Richardson}}}]{lin_2020}%
  \BibitemOpen
  \bibfield  {author} {\bibinfo {author} {\bibfnamefont {Z.}~\bibnamefont
  {{Lin}}}, \bibinfo {author} {\bibfnamefont {C.}~\bibnamefont {{Lunardini}}},
  \bibinfo {author} {\bibfnamefont {M.}~\bibnamefont {{Zanolin}}}, \bibinfo
  {author} {\bibfnamefont {K.}~\bibnamefont {{Kotake}}},\ and\ \bibinfo
  {author} {\bibfnamefont {C.}~\bibnamefont {{Richardson}}},\ }\href
  {https://doi.org/10.1103/PhysRevD.101.123028} {\bibfield  {journal} {\bibinfo
   {journal} {\prd}\ }\textbf {\bibinfo {volume} {101}},\ \bibinfo {eid}
  {123028} (\bibinfo {year} {2020})}\BibitemShut {NoStop}%
\bibitem [{\citenamefont {{Nagakura}}\ \emph {et~al.}(2021)\citenamefont
  {{Nagakura}}, \citenamefont {{Burrows}}, \citenamefont {{Vartanyan}},\ and\
  \citenamefont {{Radice}}}]{nagakura_2021}%
  \BibitemOpen
  \bibfield  {author} {\bibinfo {author} {\bibfnamefont {H.}~\bibnamefont
  {{Nagakura}}}, \bibinfo {author} {\bibfnamefont {A.}~\bibnamefont
  {{Burrows}}}, \bibinfo {author} {\bibfnamefont {D.}~\bibnamefont
  {{Vartanyan}}},\ and\ \bibinfo {author} {\bibfnamefont {D.}~\bibnamefont
  {{Radice}}},\ }\href {https://doi.org/10.1093/mnras/staa2691} {\bibfield
  {journal} {\bibinfo  {journal} {MNRAS}\ }\textbf {\bibinfo {volume} {500}},\
  \bibinfo {pages} {696} (\bibinfo {year} {2021})},\ \Eprint
  {https://arxiv.org/abs/2007.05000} {arXiv:2007.05000 [astro-ph.HE]}
  \BibitemShut {NoStop}%
\bibitem [{\citenamefont {{Lin}}\ \emph {et~al.}(2023)\citenamefont {{Lin}},
  \citenamefont {{Rijal}}, \citenamefont {{Lunardini}}, \citenamefont
  {{Morales}},\ and\ \citenamefont {{Zanolin}}}]{lin_2023}%
  \BibitemOpen
  \bibfield  {author} {\bibinfo {author} {\bibfnamefont {Z.}~\bibnamefont
  {{Lin}}}, \bibinfo {author} {\bibfnamefont {A.}~\bibnamefont {{Rijal}}},
  \bibinfo {author} {\bibfnamefont {C.}~\bibnamefont {{Lunardini}}}, \bibinfo
  {author} {\bibfnamefont {M.~D.}\ \bibnamefont {{Morales}}},\ and\ \bibinfo
  {author} {\bibfnamefont {M.}~\bibnamefont {{Zanolin}}},\ }\href
  {https://doi.org/10.1103/PhysRevD.107.083017} {\bibfield  {journal} {\bibinfo
   {journal} {\prd}\ }\textbf {\bibinfo {volume} {107}},\ \bibinfo {eid}
  {083017} (\bibinfo {year} {2023})},\ \Eprint
  {https://arxiv.org/abs/2211.07878} {arXiv:2211.07878 [astro-ph.HE]}
  \BibitemShut {NoStop}%
\bibitem [{\citenamefont {Beise}(2024)}]{beise_2024}%
  \BibitemOpen
  \bibfield  {author} {\bibinfo {author} {\bibfnamefont {J.}~\bibnamefont
  {Beise}},\ }in\ \href {https://doi.org/10.22323/1.441.0159} {\emph {\bibinfo
  {booktitle} {Proceedings of XVIII International Conference on Topics in
  Astroparticle and Underground Physics {\textemdash} PoS(TAUP2023)}}},\ Vol.\
  \bibinfo {volume} {441}\ (\bibinfo {year} {2024})\ p.\ \bibinfo {pages}
  {159}\BibitemShut {NoStop}%
\bibitem [{\citenamefont {{Beise}}\ \emph {et~al.}(2025)\citenamefont
  {{Beise}}, \citenamefont {{BenZvi}}, \citenamefont {{Griswold}},
  \citenamefont {{Valtonen-Mattila}},\ and\ \citenamefont
  {{O'Sullivan}}}]{beise_2025}%
  \BibitemOpen
  \bibfield  {author} {\bibinfo {author} {\bibfnamefont {J.}~\bibnamefont
  {{Beise}}}, \bibinfo {author} {\bibfnamefont {S.}~\bibnamefont {{BenZvi}}},
  \bibinfo {author} {\bibfnamefont {S.}~\bibnamefont {{Griswold}}}, \bibinfo
  {author} {\bibfnamefont {N.}~\bibnamefont {{Valtonen-Mattila}}},\ and\
  \bibinfo {author} {\bibfnamefont {E.}~\bibnamefont {{O'Sullivan}}},\ }\href
  {https://doi.org/10.48550/arXiv.2502.05024} {\bibfield  {journal} {\bibinfo
  {journal} {arXiv:2502.05024}\ ,\ \bibinfo {eid} {arXiv:2502.05024}} (\bibinfo
  {year} {2025})},\ \Eprint {https://arxiv.org/abs/2502.05024}
  {arXiv:2502.05024 [astro-ph.HE]} \BibitemShut {NoStop}%
\bibitem [{\citenamefont {{Kotake}}\ \emph {et~al.}(2009)\citenamefont
  {{Kotake}}, \citenamefont {{Iwakami}}, \citenamefont {{Ohnishi}},\ and\
  \citenamefont {{Yamada}}}]{kotake_2009}%
  \BibitemOpen
  \bibfield  {author} {\bibinfo {author} {\bibfnamefont {K.}~\bibnamefont
  {{Kotake}}}, \bibinfo {author} {\bibfnamefont {W.}~\bibnamefont {{Iwakami}}},
  \bibinfo {author} {\bibfnamefont {N.}~\bibnamefont {{Ohnishi}}},\ and\
  \bibinfo {author} {\bibfnamefont {S.}~\bibnamefont {{Yamada}}},\ }\href
  {https://doi.org/10.1088/0004-637X/697/2/L133} {\bibfield  {journal}
  {\bibinfo  {journal} {\apj}\ }\textbf {\bibinfo {volume} {697}},\ \bibinfo
  {pages} {L133} (\bibinfo {year} {2009})},\ \Eprint
  {https://arxiv.org/abs/0904.4300} {arXiv:0904.4300 [astro-ph.HE]}
  \BibitemShut {NoStop}%
\bibitem [{\citenamefont {{Murphy}}\ \emph {et~al.}(2009)\citenamefont
  {{Murphy}}, \citenamefont {{Ott}},\ and\ \citenamefont
  {{Burrows}}}]{murphy_2009}%
  \BibitemOpen
  \bibfield  {author} {\bibinfo {author} {\bibfnamefont {J.~W.}\ \bibnamefont
  {{Murphy}}}, \bibinfo {author} {\bibfnamefont {C.~D.}\ \bibnamefont
  {{Ott}}},\ and\ \bibinfo {author} {\bibfnamefont {A.}~\bibnamefont
  {{Burrows}}},\ }\href {https://doi.org/10.1088/0004-637X/707/2/1173}
  {\bibfield  {journal} {\bibinfo  {journal} {\apj}\ }\textbf {\bibinfo
  {volume} {707}},\ \bibinfo {pages} {1173} (\bibinfo {year} {2009})},\ \Eprint
  {https://arxiv.org/abs/0907.4762} {arXiv:0907.4762 [astro-ph.SR]}
  \BibitemShut {NoStop}%
\bibitem [{\citenamefont {{Cerd{\'a}-Dur{\'a}n}}\ \emph
  {et~al.}(2013)\citenamefont {{Cerd{\'a}-Dur{\'a}n}}, \citenamefont
  {{DeBrye}}, \citenamefont {{Aloy}}, \citenamefont {{Font}},\ and\
  \citenamefont {{Obergaulinger}}}]{cerda-duran_2013}%
  \BibitemOpen
  \bibfield  {author} {\bibinfo {author} {\bibfnamefont {P.}~\bibnamefont
  {{Cerd{\'a}-Dur{\'a}n}}}, \bibinfo {author} {\bibfnamefont {N.}~\bibnamefont
  {{DeBrye}}}, \bibinfo {author} {\bibfnamefont {M.~A.}\ \bibnamefont
  {{Aloy}}}, \bibinfo {author} {\bibfnamefont {J.~A.}\ \bibnamefont {{Font}}},\
  and\ \bibinfo {author} {\bibfnamefont {M.}~\bibnamefont {{Obergaulinger}}},\
  }\href {https://doi.org/10.1088/2041-8205/779/2/L18} {\bibfield  {journal}
  {\bibinfo  {journal} {\apj}\ }\textbf {\bibinfo {volume} {779}},\ \bibinfo
  {eid} {L18} (\bibinfo {year} {2013})},\ \Eprint
  {https://arxiv.org/abs/1310.8290} {arXiv:1310.8290 [astro-ph.SR]}
  \BibitemShut {NoStop}%
\bibitem [{\citenamefont {{Yakunin}}\ \emph {et~al.}(2015)\citenamefont
  {{Yakunin}}, \citenamefont {{Mezzacappa}}, \citenamefont {{Marronetti}},
  \citenamefont {{Yoshida}}, \citenamefont {{Bruenn}}, \citenamefont {{Hix}},
  \citenamefont {{Lentz}}, \citenamefont {{Bronson Messer}}, \citenamefont
  {{Harris}}, \citenamefont {{Endeve}}, \citenamefont {{Blondin}},\ and\
  \citenamefont {{Lingerfelt}}}]{yakunin_2015}%
  \BibitemOpen
  \bibfield  {author} {\bibinfo {author} {\bibfnamefont {K.~N.}\ \bibnamefont
  {{Yakunin}}}, \bibinfo {author} {\bibfnamefont {A.}~\bibnamefont
  {{Mezzacappa}}}, \bibinfo {author} {\bibfnamefont {P.}~\bibnamefont
  {{Marronetti}}}, \bibinfo {author} {\bibfnamefont {S.}~\bibnamefont
  {{Yoshida}}}, \bibinfo {author} {\bibfnamefont {S.~W.}\ \bibnamefont
  {{Bruenn}}}, \bibinfo {author} {\bibfnamefont {W.~R.}\ \bibnamefont {{Hix}}},
  \bibinfo {author} {\bibfnamefont {E.~J.}\ \bibnamefont {{Lentz}}}, \bibinfo
  {author} {\bibfnamefont {O.~E.}\ \bibnamefont {{Bronson Messer}}}, \bibinfo
  {author} {\bibfnamefont {J.~A.}\ \bibnamefont {{Harris}}}, \bibinfo {author}
  {\bibfnamefont {E.}~\bibnamefont {{Endeve}}}, \bibinfo {author}
  {\bibfnamefont {J.~M.}\ \bibnamefont {{Blondin}}},\ and\ \bibinfo {author}
  {\bibfnamefont {E.~J.}\ \bibnamefont {{Lingerfelt}}},\ }\href
  {https://doi.org/10.1103/PhysRevD.92.084040} {\bibfield  {journal} {\bibinfo
  {journal} {\prd}\ }\textbf {\bibinfo {volume} {92}},\ \bibinfo {eid} {084040}
  (\bibinfo {year} {2015})},\ \Eprint {https://arxiv.org/abs/1505.05824}
  {arXiv:1505.05824 [astro-ph.HE]} \BibitemShut {NoStop}%
\bibitem [{\citenamefont {{Andresen}}\ \emph {et~al.}(2017)\citenamefont
  {{Andresen}}, \citenamefont {{M{\"u}ller}}, \citenamefont {{M{\"u}ller}},\
  and\ \citenamefont {{Janka}}}]{andresen_2017}%
  \BibitemOpen
  \bibfield  {author} {\bibinfo {author} {\bibfnamefont {H.}~\bibnamefont
  {{Andresen}}}, \bibinfo {author} {\bibfnamefont {B.}~\bibnamefont
  {{M{\"u}ller}}}, \bibinfo {author} {\bibfnamefont {E.}~\bibnamefont
  {{M{\"u}ller}}},\ and\ \bibinfo {author} {\bibfnamefont {H.~T.}\ \bibnamefont
  {{Janka}}},\ }\href {https://doi.org/10.1093/mnras/stx618} {\bibfield
  {journal} {\bibinfo  {journal} {MNRAS}\ }\textbf {\bibinfo {volume} {468}},\
  \bibinfo {pages} {2032} (\bibinfo {year} {2017})},\ \Eprint
  {https://arxiv.org/abs/1607.05199} {arXiv:1607.05199 [astro-ph.HE]}
  \BibitemShut {NoStop}%
\bibitem [{\citenamefont {{Pajkos}}\ \emph {et~al.}(2019)\citenamefont
  {{Pajkos}}, \citenamefont {{Couch}}, \citenamefont {{Pan}},\ and\
  \citenamefont {{O'Connor}}}]{pajkos_2019}%
  \BibitemOpen
  \bibfield  {author} {\bibinfo {author} {\bibfnamefont {M.~A.}\ \bibnamefont
  {{Pajkos}}}, \bibinfo {author} {\bibfnamefont {S.~M.}\ \bibnamefont
  {{Couch}}}, \bibinfo {author} {\bibfnamefont {K.-C.}\ \bibnamefont {{Pan}}},\
  and\ \bibinfo {author} {\bibfnamefont {E.~P.}\ \bibnamefont {{O'Connor}}},\
  }\href {https://doi.org/10.3847/1538-4357/ab1de2} {\bibfield  {journal}
  {\bibinfo  {journal} {\apj}\ }\textbf {\bibinfo {volume} {878}},\ \bibinfo
  {eid} {13} (\bibinfo {year} {2019})},\ \Eprint
  {https://arxiv.org/abs/1901.09055} {arXiv:1901.09055 [astro-ph.HE]}
  \BibitemShut {NoStop}%
\bibitem [{\citenamefont {{Kuroda}}\ \emph {et~al.}(2017)\citenamefont
  {{Kuroda}}, \citenamefont {{Kotake}}, \citenamefont {{Hayama}},\ and\
  \citenamefont {{Takiwaki}}}]{kuroda_2017}%
  \BibitemOpen
  \bibfield  {author} {\bibinfo {author} {\bibfnamefont {T.}~\bibnamefont
  {{Kuroda}}}, \bibinfo {author} {\bibfnamefont {K.}~\bibnamefont {{Kotake}}},
  \bibinfo {author} {\bibfnamefont {K.}~\bibnamefont {{Hayama}}},\ and\
  \bibinfo {author} {\bibfnamefont {T.}~\bibnamefont {{Takiwaki}}},\ }\href
  {https://doi.org/10.3847/1538-4357/aa988d} {\bibfield  {journal} {\bibinfo
  {journal} {\apj}\ }\textbf {\bibinfo {volume} {851}},\ \bibinfo {eid} {62}
  (\bibinfo {year} {2017})},\ \Eprint {https://arxiv.org/abs/1708.05252}
  {arXiv:1708.05252 [astro-ph.HE]} \BibitemShut {NoStop}%
\bibitem [{\citenamefont {{Takiwaki}}\ and\ \citenamefont
  {{Kotake}}(2018)}]{takiwaki_2018}%
  \BibitemOpen
  \bibfield  {author} {\bibinfo {author} {\bibfnamefont {T.}~\bibnamefont
  {{Takiwaki}}}\ and\ \bibinfo {author} {\bibfnamefont {K.}~\bibnamefont
  {{Kotake}}},\ }\href {https://doi.org/10.1093/mnrasl/sly008} {\bibfield
  {journal} {\bibinfo  {journal} {MNRAS}\ }\textbf {\bibinfo {volume} {475}},\
  \bibinfo {pages} {L91} (\bibinfo {year} {2018})},\ \Eprint
  {https://arxiv.org/abs/1711.01905} {arXiv:1711.01905 [astro-ph.HE]}
  \BibitemShut {NoStop}%
\bibitem [{\citenamefont {{Shibagaki}}\ \emph {et~al.}(2021)\citenamefont
  {{Shibagaki}}, \citenamefont {{Kuroda}}, \citenamefont {{Kotake}},\ and\
  \citenamefont {{Takiwaki}}}]{shibagaki_2021}%
  \BibitemOpen
  \bibfield  {author} {\bibinfo {author} {\bibfnamefont {S.}~\bibnamefont
  {{Shibagaki}}}, \bibinfo {author} {\bibfnamefont {T.}~\bibnamefont
  {{Kuroda}}}, \bibinfo {author} {\bibfnamefont {K.}~\bibnamefont {{Kotake}}},\
  and\ \bibinfo {author} {\bibfnamefont {T.}~\bibnamefont {{Takiwaki}}},\
  }\href {https://doi.org/10.1093/mnras/stab228} {\bibfield  {journal}
  {\bibinfo  {journal} {MNRAS}\ }\textbf {\bibinfo {volume} {502}},\ \bibinfo
  {pages} {3066} (\bibinfo {year} {2021})},\ \Eprint
  {https://arxiv.org/abs/2010.03882} {arXiv:2010.03882 [astro-ph.HE]}
  \BibitemShut {NoStop}%
\bibitem [{\citenamefont {{Drago}}\ \emph {et~al.}(2023)\citenamefont
  {{Drago}}, \citenamefont {{Andresen}}, \citenamefont {{Di Palma}},
  \citenamefont {{Tamborra}},\ and\ \citenamefont
  {{Torres-Forn{\'e}}}}]{drago_2023}%
  \BibitemOpen
  \bibfield  {author} {\bibinfo {author} {\bibfnamefont {M.}~\bibnamefont
  {{Drago}}}, \bibinfo {author} {\bibfnamefont {H.}~\bibnamefont {{Andresen}}},
  \bibinfo {author} {\bibfnamefont {I.}~\bibnamefont {{Di Palma}}}, \bibinfo
  {author} {\bibfnamefont {I.}~\bibnamefont {{Tamborra}}},\ and\ \bibinfo
  {author} {\bibfnamefont {A.}~\bibnamefont {{Torres-Forn{\'e}}}},\ }\href
  {https://doi.org/10.1103/PhysRevD.108.103036} {\bibfield  {journal} {\bibinfo
   {journal} {\prd}\ }\textbf {\bibinfo {volume} {108}},\ \bibinfo {eid}
  {103036} (\bibinfo {year} {2023})},\ \Eprint
  {https://arxiv.org/abs/2305.07688} {arXiv:2305.07688 [astro-ph.HE]}
  \BibitemShut {NoStop}%
\bibitem [{\citenamefont {{Lee}}\ and\ \citenamefont
  {{Ramirez-Ruiz}}(2006)}]{lee_2006}%
  \BibitemOpen
  \bibfield  {author} {\bibinfo {author} {\bibfnamefont {W.~H.}\ \bibnamefont
  {{Lee}}}\ and\ \bibinfo {author} {\bibfnamefont {E.}~\bibnamefont
  {{Ramirez-Ruiz}}},\ }\href {https://doi.org/10.1086/500533} {\bibfield
  {journal} {\bibinfo  {journal} {\apj}\ }\textbf {\bibinfo {volume} {641}},\
  \bibinfo {pages} {961} (\bibinfo {year} {2006})},\ \Eprint
  {https://arxiv.org/abs/astro-ph/0509307} {arXiv:astro-ph/0509307 [astro-ph]}
  \BibitemShut {NoStop}%
\bibitem [{\citenamefont {{Nagataki}}\ \emph {et~al.}(2007)\citenamefont
  {{Nagataki}}, \citenamefont {{Takahashi}}, \citenamefont {{Mizuta}},\ and\
  \citenamefont {{Takiwaki}}}]{nagataki_2007}%
  \BibitemOpen
  \bibfield  {author} {\bibinfo {author} {\bibfnamefont {S.}~\bibnamefont
  {{Nagataki}}}, \bibinfo {author} {\bibfnamefont {R.}~\bibnamefont
  {{Takahashi}}}, \bibinfo {author} {\bibfnamefont {A.}~\bibnamefont
  {{Mizuta}}},\ and\ \bibinfo {author} {\bibfnamefont {T.}~\bibnamefont
  {{Takiwaki}}},\ }\href {https://doi.org/10.1086/512057} {\bibfield  {journal}
  {\bibinfo  {journal} {ApJ}\ }\textbf {\bibinfo {volume} {659}},\ \bibinfo
  {pages} {512} (\bibinfo {year} {2007})},\ \Eprint
  {https://arxiv.org/abs/astro-ph/0608233} {arXiv:astro-ph/0608233 [astro-ph]}
  \BibitemShut {NoStop}%
\bibitem [{\citenamefont {{Harikae}}\ \emph {et~al.}(2009)\citenamefont
  {{Harikae}}, \citenamefont {{Takiwaki}},\ and\ \citenamefont
  {{Kotake}}}]{harikae_2009a}%
  \BibitemOpen
  \bibfield  {author} {\bibinfo {author} {\bibfnamefont {S.}~\bibnamefont
  {{Harikae}}}, \bibinfo {author} {\bibfnamefont {T.}~\bibnamefont
  {{Takiwaki}}},\ and\ \bibinfo {author} {\bibfnamefont {K.}~\bibnamefont
  {{Kotake}}},\ }\href {https://doi.org/10.1088/0004-637X/704/1/354} {\bibfield
   {journal} {\bibinfo  {journal} {ApJ}\ }\textbf {\bibinfo {volume} {704}},\
  \bibinfo {pages} {354} (\bibinfo {year} {2009})},\ \Eprint
  {https://arxiv.org/abs/0905.2006} {arXiv:0905.2006 [astro-ph.HE]}
  \BibitemShut {NoStop}%
\bibitem [{\citenamefont {{Obergaulinger}}\ and\ \citenamefont
  {{Aloy}}(2017)}]{obergaulinger_2017}%
  \BibitemOpen
  \bibfield  {author} {\bibinfo {author} {\bibfnamefont {M.}~\bibnamefont
  {{Obergaulinger}}}\ and\ \bibinfo {author} {\bibfnamefont {M.~{\'A}.}\
  \bibnamefont {{Aloy}}},\ }\href {https://doi.org/10.1093/mnrasl/slx046}
  {\bibfield  {journal} {\bibinfo  {journal} {MNRAS}\ }\textbf {\bibinfo
  {volume} {469}},\ \bibinfo {pages} {L43} (\bibinfo {year} {2017})},\ \Eprint
  {https://arxiv.org/abs/1703.09893} {arXiv:1703.09893 [astro-ph.SR]}
  \BibitemShut {NoStop}%
\bibitem [{\citenamefont {{Fujibayashi}}\ \emph {et~al.}(2023)\citenamefont
  {{Fujibayashi}}, \citenamefont {{Sekiguchi}}, \citenamefont {{Shibata}},\
  and\ \citenamefont {{Wanajo}}}]{fujibayashi_2022}%
  \BibitemOpen
  \bibfield  {author} {\bibinfo {author} {\bibfnamefont {S.}~\bibnamefont
  {{Fujibayashi}}}, \bibinfo {author} {\bibfnamefont {Y.}~\bibnamefont
  {{Sekiguchi}}}, \bibinfo {author} {\bibfnamefont {M.}~\bibnamefont
  {{Shibata}}},\ and\ \bibinfo {author} {\bibfnamefont {S.}~\bibnamefont
  {{Wanajo}}},\ }\href {https://doi.org/10.3847/1538-4357/acf5e5} {\bibfield
  {journal} {\bibinfo  {journal} {\apj}\ }\textbf {\bibinfo {volume} {956}},\
  \bibinfo {eid} {100} (\bibinfo {year} {2023})},\ \Eprint
  {https://arxiv.org/abs/2212.03958} {arXiv:2212.03958 [astro-ph.HE]}
  \BibitemShut {NoStop}%
\bibitem [{\citenamefont {{Issa}}\ \emph {et~al.}(2025)\citenamefont {{Issa}},
  \citenamefont {{Gottlieb}}, \citenamefont {{Metzger}}, \citenamefont
  {{Jacquemin-Ide}}, \citenamefont {{Liska}}, \citenamefont {{Foucart}},
  \citenamefont {{Halevi}},\ and\ \citenamefont {{Tchekhovskoy}}}]{issa_2025}%
  \BibitemOpen
  \bibfield  {author} {\bibinfo {author} {\bibfnamefont {D.}~\bibnamefont
  {{Issa}}}, \bibinfo {author} {\bibfnamefont {O.}~\bibnamefont {{Gottlieb}}},
  \bibinfo {author} {\bibfnamefont {B.~D.}\ \bibnamefont {{Metzger}}}, \bibinfo
  {author} {\bibfnamefont {J.}~\bibnamefont {{Jacquemin-Ide}}}, \bibinfo
  {author} {\bibfnamefont {M.}~\bibnamefont {{Liska}}}, \bibinfo {author}
  {\bibfnamefont {F.}~\bibnamefont {{Foucart}}}, \bibinfo {author}
  {\bibfnamefont {G.}~\bibnamefont {{Halevi}}},\ and\ \bibinfo {author}
  {\bibfnamefont {A.}~\bibnamefont {{Tchekhovskoy}}},\ }\href
  {https://doi.org/10.3847/2041-8213/adc694} {\bibfield  {journal} {\bibinfo
  {journal} {\apj}\ }\textbf {\bibinfo {volume} {985}},\ \bibinfo {eid} {L26}
  (\bibinfo {year} {2025})}\BibitemShut {NoStop}%
\bibitem [{\citenamefont {{Shibata}}\ \emph {et~al.}(2025)\citenamefont
  {{Shibata}}, \citenamefont {{Fujibayashi}}, \citenamefont {{Wanajo}},
  \citenamefont {{Ioka}}, \citenamefont {{Lam}},\ and\ \citenamefont
  {{Sekiguchi}}}]{shibata_2025}%
  \BibitemOpen
  \bibfield  {author} {\bibinfo {author} {\bibfnamefont {M.}~\bibnamefont
  {{Shibata}}}, \bibinfo {author} {\bibfnamefont {S.}~\bibnamefont
  {{Fujibayashi}}}, \bibinfo {author} {\bibfnamefont {S.}~\bibnamefont
  {{Wanajo}}}, \bibinfo {author} {\bibfnamefont {K.}~\bibnamefont {{Ioka}}},
  \bibinfo {author} {\bibfnamefont {A.~T.-L.}\ \bibnamefont {{Lam}}},\ and\
  \bibinfo {author} {\bibfnamefont {Y.}~\bibnamefont {{Sekiguchi}}},\ }\href
  {https://doi.org/10.1103/msy2-fwhx} {\bibfield  {journal} {\bibinfo
  {journal} {\prd}\ }\textbf {\bibinfo {volume} {111}},\ \bibinfo {eid}
  {123017} (\bibinfo {year} {2025})},\ \Eprint
  {https://arxiv.org/abs/2502.02077} {arXiv:2502.02077 [astro-ph.HE]}
  \BibitemShut {NoStop}%
\bibitem [{\citenamefont {{Liu}}\ \emph {et~al.}(2016)\citenamefont {{Liu}},
  \citenamefont {{Zhang}}, \citenamefont {{Li}}, \citenamefont {{Ma}},\ and\
  \citenamefont {{Xue}}}]{liu_2016}%
  \BibitemOpen
  \bibfield  {author} {\bibinfo {author} {\bibfnamefont {T.}~\bibnamefont
  {{Liu}}}, \bibinfo {author} {\bibfnamefont {B.}~\bibnamefont {{Zhang}}},
  \bibinfo {author} {\bibfnamefont {Y.}~\bibnamefont {{Li}}}, \bibinfo {author}
  {\bibfnamefont {R.-Y.}\ \bibnamefont {{Ma}}},\ and\ \bibinfo {author}
  {\bibfnamefont {L.}~\bibnamefont {{Xue}}},\ }\href
  {https://doi.org/10.1103/PhysRevD.93.123004} {\bibfield  {journal} {\bibinfo
  {journal} {\prd}\ }\textbf {\bibinfo {volume} {93}},\ \bibinfo {eid} {123004}
  (\bibinfo {year} {2016})},\ \Eprint {https://arxiv.org/abs/1512.07203}
  {arXiv:1512.07203 [astro-ph.HE]} \BibitemShut {NoStop}%
\bibitem [{\citenamefont {{Schilbach}}\ \emph {et~al.}(2019)\citenamefont
  {{Schilbach}}, \citenamefont {{Caballero}},\ and\ \citenamefont
  {{McLaughlin}}}]{schilbach_2019}%
  \BibitemOpen
  \bibfield  {author} {\bibinfo {author} {\bibfnamefont {T.~S.~H.}\
  \bibnamefont {{Schilbach}}}, \bibinfo {author} {\bibfnamefont {O.~L.}\
  \bibnamefont {{Caballero}}},\ and\ \bibinfo {author} {\bibfnamefont {G.~C.}\
  \bibnamefont {{McLaughlin}}},\ }\href
  {https://doi.org/10.1103/PhysRevD.100.043008} {\bibfield  {journal} {\bibinfo
   {journal} {\prd}\ }\textbf {\bibinfo {volume} {100}},\ \bibinfo {eid}
  {043008} (\bibinfo {year} {2019})},\ \Eprint
  {https://arxiv.org/abs/1808.03627} {arXiv:1808.03627 [astro-ph.HE]}
  \BibitemShut {NoStop}%
\bibitem [{\citenamefont {{Wei}}\ \emph {et~al.}(2024)\citenamefont {{Wei}},
  \citenamefont {{Liu}},\ and\ \citenamefont {{Song}}}]{wei_2024}%
  \BibitemOpen
  \bibfield  {author} {\bibinfo {author} {\bibfnamefont {Y.-F.}\ \bibnamefont
  {{Wei}}}, \bibinfo {author} {\bibfnamefont {T.}~\bibnamefont {{Liu}}},\ and\
  \bibinfo {author} {\bibfnamefont {C.-Y.}\ \bibnamefont {{Song}}},\ }\href
  {https://doi.org/10.3847/1538-4357/ad3824} {\bibfield  {journal} {\bibinfo
  {journal} {\apj}\ }\textbf {\bibinfo {volume} {966}},\ \bibinfo {eid} {101}
  (\bibinfo {year} {2024})},\ \Eprint {https://arxiv.org/abs/2403.16856}
  {arXiv:2403.16856 [astro-ph.HE]} \BibitemShut {NoStop}%
\bibitem [{\citenamefont {{Mart{\'\i}nez-Mirav{\'e}}}\ \emph
  {et~al.}(2024)\citenamefont {{Mart{\'\i}nez-Mirav{\'e}}}, \citenamefont
  {{Tamborra}}, \citenamefont {{Aloy}},\ and\ \citenamefont
  {{Obergaulinger}}}]{martinez-mirave_2024}%
  \BibitemOpen
  \bibfield  {author} {\bibinfo {author} {\bibfnamefont {P.}~\bibnamefont
  {{Mart{\'\i}nez-Mirav{\'e}}}}, \bibinfo {author} {\bibfnamefont
  {I.}~\bibnamefont {{Tamborra}}}, \bibinfo {author} {\bibfnamefont
  {M.~{\'A}.}\ \bibnamefont {{Aloy}}},\ and\ \bibinfo {author} {\bibfnamefont
  {M.}~\bibnamefont {{Obergaulinger}}},\ }\href
  {https://doi.org/10.1103/PhysRevD.110.103029} {\bibfield  {journal} {\bibinfo
   {journal} {\prd}\ }\textbf {\bibinfo {volume} {110}},\ \bibinfo {eid}
  {103029} (\bibinfo {year} {2024})},\ \Eprint
  {https://arxiv.org/abs/2409.09126} {arXiv:2409.09126 [astro-ph.HE]}
  \BibitemShut {NoStop}%
\bibitem [{\citenamefont {{van Putten}}(2001)}]{vanputten_2001}%
  \BibitemOpen
  \bibfield  {author} {\bibinfo {author} {\bibfnamefont {M.~H.~P.~M.}\
  \bibnamefont {{van Putten}}},\ }\href {https://doi.org/10.1086/338120}
  {\bibfield  {journal} {\bibinfo  {journal} {\apj}\ }\textbf {\bibinfo
  {volume} {562}},\ \bibinfo {pages} {L51} (\bibinfo {year} {2001})},\ \Eprint
  {https://arxiv.org/abs/astro-ph/0110422} {arXiv:astro-ph/0110422 [astro-ph]}
  \BibitemShut {NoStop}%
\bibitem [{\citenamefont {{Fryer}}\ \emph {et~al.}(2002)\citenamefont
  {{Fryer}}, \citenamefont {{Holz}},\ and\ \citenamefont
  {{Hughes}}}]{fryer_2002}%
  \BibitemOpen
  \bibfield  {author} {\bibinfo {author} {\bibfnamefont {C.~L.}\ \bibnamefont
  {{Fryer}}}, \bibinfo {author} {\bibfnamefont {D.~E.}\ \bibnamefont
  {{Holz}}},\ and\ \bibinfo {author} {\bibfnamefont {S.~A.}\ \bibnamefont
  {{Hughes}}},\ }\href {https://doi.org/10.1086/324034} {\bibfield  {journal}
  {\bibinfo  {journal} {\apj}\ }\textbf {\bibinfo {volume} {565}},\ \bibinfo
  {pages} {430} (\bibinfo {year} {2002})},\ \Eprint
  {https://arxiv.org/abs/astro-ph/0106113} {arXiv:astro-ph/0106113 [astro-ph]}
  \BibitemShut {NoStop}%
\bibitem [{\citenamefont {{Kobayashi}}\ and\ \citenamefont
  {{M{\'e}sz{\'a}ros}}(2003)}]{kobayashi_2003}%
  \BibitemOpen
  \bibfield  {author} {\bibinfo {author} {\bibfnamefont {S.}~\bibnamefont
  {{Kobayashi}}}\ and\ \bibinfo {author} {\bibfnamefont {P.}~\bibnamefont
  {{M{\'e}sz{\'a}ros}}},\ }\href {https://doi.org/10.1086/374733} {\bibfield
  {journal} {\bibinfo  {journal} {\apj}\ }\textbf {\bibinfo {volume} {589}},\
  \bibinfo {pages} {861} (\bibinfo {year} {2003})},\ \Eprint
  {https://arxiv.org/abs/astro-ph/0210211} {arXiv:astro-ph/0210211 [astro-ph]}
  \BibitemShut {NoStop}%
\bibitem [{\citenamefont {{Piro}}\ and\ \citenamefont
  {{Pfahl}}(2007)}]{piro_2007}%
  \BibitemOpen
  \bibfield  {author} {\bibinfo {author} {\bibfnamefont {A.~L.}\ \bibnamefont
  {{Piro}}}\ and\ \bibinfo {author} {\bibfnamefont {E.}~\bibnamefont
  {{Pfahl}}},\ }\href {https://doi.org/10.1086/511672} {\bibfield  {journal}
  {\bibinfo  {journal} {\apj}\ }\textbf {\bibinfo {volume} {658}},\ \bibinfo
  {pages} {1173} (\bibinfo {year} {2007})},\ \Eprint
  {https://arxiv.org/abs/astro-ph/0610696} {arXiv:astro-ph/0610696 [astro-ph]}
  \BibitemShut {NoStop}%
\bibitem [{\citenamefont {{Romero}}\ \emph {et~al.}(2010)\citenamefont
  {{Romero}}, \citenamefont {{Reynoso}},\ and\ \citenamefont
  {{Christiansen}}}]{romero_2010}%
  \BibitemOpen
  \bibfield  {author} {\bibinfo {author} {\bibfnamefont {G.~E.}\ \bibnamefont
  {{Romero}}}, \bibinfo {author} {\bibfnamefont {M.~M.}\ \bibnamefont
  {{Reynoso}}},\ and\ \bibinfo {author} {\bibfnamefont {H.~R.}\ \bibnamefont
  {{Christiansen}}},\ }\href {https://doi.org/10.1051/0004-6361/201014882}
  {\bibfield  {journal} {\bibinfo  {journal} {A\&A}\ }\textbf {\bibinfo
  {volume} {524}},\ \bibinfo {eid} {A4} (\bibinfo {year} {2010})},\ \Eprint
  {https://arxiv.org/abs/1009.3679} {arXiv:1009.3679 [astro-ph.HE]}
  \BibitemShut {NoStop}%
\bibitem [{\citenamefont {{Sun}}\ \emph {et~al.}(2012)\citenamefont {{Sun}},
  \citenamefont {{Liu}}, \citenamefont {{Gu}},\ and\ \citenamefont
  {{Lu}}}]{sun_2012}%
  \BibitemOpen
  \bibfield  {author} {\bibinfo {author} {\bibfnamefont {M.-Y.}\ \bibnamefont
  {{Sun}}}, \bibinfo {author} {\bibfnamefont {T.}~\bibnamefont {{Liu}}},
  \bibinfo {author} {\bibfnamefont {W.-M.}\ \bibnamefont {{Gu}}},\ and\
  \bibinfo {author} {\bibfnamefont {J.-F.}\ \bibnamefont {{Lu}}},\ }\href
  {https://doi.org/10.1088/0004-637X/752/1/31} {\bibfield  {journal} {\bibinfo
  {journal} {\apj}\ }\textbf {\bibinfo {volume} {752}},\ \bibinfo {eid} {31}
  (\bibinfo {year} {2012})},\ \Eprint {https://arxiv.org/abs/1204.2028}
  {arXiv:1204.2028 [astro-ph.HE]} \BibitemShut {NoStop}%
\bibitem [{\citenamefont {{Hiramatsu}}\ \emph {et~al.}(2005)\citenamefont
  {{Hiramatsu}}, \citenamefont {{Kotake}}, \citenamefont {{Kudoh}},\ and\
  \citenamefont {{Taruya}}}]{hiramatsu_2005}%
  \BibitemOpen
  \bibfield  {author} {\bibinfo {author} {\bibfnamefont {T.}~\bibnamefont
  {{Hiramatsu}}}, \bibinfo {author} {\bibfnamefont {K.}~\bibnamefont
  {{Kotake}}}, \bibinfo {author} {\bibfnamefont {H.}~\bibnamefont {{Kudoh}}},\
  and\ \bibinfo {author} {\bibfnamefont {A.}~\bibnamefont {{Taruya}}},\ }\href
  {https://doi.org/10.1111/j.1365-2966.2005.09643.x} {\bibfield  {journal}
  {\bibinfo  {journal} {MNRAS}\ }\textbf {\bibinfo {volume} {364}},\ \bibinfo
  {pages} {1063} (\bibinfo {year} {2005})},\ \Eprint
  {https://arxiv.org/abs/astro-ph/0509787} {arXiv:astro-ph/0509787 [astro-ph]}
  \BibitemShut {NoStop}%
\bibitem [{\citenamefont {{Suwa}}\ and\ \citenamefont
  {{Murase}}(2009)}]{suwa_2009}%
  \BibitemOpen
  \bibfield  {author} {\bibinfo {author} {\bibfnamefont {Y.}~\bibnamefont
  {{Suwa}}}\ and\ \bibinfo {author} {\bibfnamefont {K.}~\bibnamefont
  {{Murase}}},\ }\href {https://doi.org/10.1103/PhysRevD.80.123008} {\bibfield
  {journal} {\bibinfo  {journal} {\prd}\ }\textbf {\bibinfo {volume} {80}},\
  \bibinfo {eid} {123008} (\bibinfo {year} {2009})},\ \Eprint
  {https://arxiv.org/abs/0906.3833} {arXiv:0906.3833 [astro-ph.HE]}
  \BibitemShut {NoStop}%
\bibitem [{\citenamefont {{Ott}}\ \emph {et~al.}(2011)\citenamefont {{Ott}},
  \citenamefont {{Reisswig}}, \citenamefont {{Schnetter}}, \citenamefont
  {{O'Connor}}, \citenamefont {{Sperhake}}, \citenamefont {{L{\"o}ffler}},
  \citenamefont {{Diener}}, \citenamefont {{Abdikamalov}}, \citenamefont
  {{Hawke}},\ and\ \citenamefont {{Burrows}}}]{ott_2011}%
  \BibitemOpen
  \bibfield  {author} {\bibinfo {author} {\bibfnamefont {C.~D.}\ \bibnamefont
  {{Ott}}}, \bibinfo {author} {\bibfnamefont {C.}~\bibnamefont {{Reisswig}}},
  \bibinfo {author} {\bibfnamefont {E.}~\bibnamefont {{Schnetter}}}, \bibinfo
  {author} {\bibfnamefont {E.}~\bibnamefont {{O'Connor}}}, \bibinfo {author}
  {\bibfnamefont {U.}~\bibnamefont {{Sperhake}}}, \bibinfo {author}
  {\bibfnamefont {F.}~\bibnamefont {{L{\"o}ffler}}}, \bibinfo {author}
  {\bibfnamefont {P.}~\bibnamefont {{Diener}}}, \bibinfo {author}
  {\bibfnamefont {E.}~\bibnamefont {{Abdikamalov}}}, \bibinfo {author}
  {\bibfnamefont {I.}~\bibnamefont {{Hawke}}},\ and\ \bibinfo {author}
  {\bibfnamefont {A.}~\bibnamefont {{Burrows}}},\ }\href
  {https://doi.org/10.1103/PhysRevLett.106.161103} {\bibfield  {journal}
  {\bibinfo  {journal} {\prl}\ }\textbf {\bibinfo {volume} {106}},\ \bibinfo
  {eid} {161103} (\bibinfo {year} {2011})},\ \Eprint
  {https://arxiv.org/abs/1012.1853} {arXiv:1012.1853 [astro-ph.HE]}
  \BibitemShut {NoStop}%
\bibitem [{\citenamefont {{Kotake}}\ \emph {et~al.}(2012)\citenamefont
  {{Kotake}}, \citenamefont {{Takiwaki}},\ and\ \citenamefont
  {{Harikae}}}]{kotake_2012}%
  \BibitemOpen
  \bibfield  {author} {\bibinfo {author} {\bibfnamefont {K.}~\bibnamefont
  {{Kotake}}}, \bibinfo {author} {\bibfnamefont {T.}~\bibnamefont
  {{Takiwaki}}},\ and\ \bibinfo {author} {\bibfnamefont {S.}~\bibnamefont
  {{Harikae}}},\ }\href {https://doi.org/10.1088/0004-637X/755/2/84} {\bibfield
   {journal} {\bibinfo  {journal} {\apj}\ }\textbf {\bibinfo {volume} {755}},\
  \bibinfo {eid} {84} (\bibinfo {year} {2012})},\ \Eprint
  {https://arxiv.org/abs/1205.6061} {arXiv:1205.6061 [astro-ph.HE]}
  \BibitemShut {NoStop}%
\bibitem [{\citenamefont {{Gottlieb}}\ \emph {et~al.}(2024)\citenamefont
  {{Gottlieb}}, \citenamefont {{Levinson}},\ and\ \citenamefont
  {{Levin}}}]{gottlieb_2024}%
  \BibitemOpen
  \bibfield  {author} {\bibinfo {author} {\bibfnamefont {O.}~\bibnamefont
  {{Gottlieb}}}, \bibinfo {author} {\bibfnamefont {A.}~\bibnamefont
  {{Levinson}}},\ and\ \bibinfo {author} {\bibfnamefont {Y.}~\bibnamefont
  {{Levin}}},\ }\href {https://doi.org/10.3847/2041-8213/ad697c} {\bibfield
  {journal} {\bibinfo  {journal} {\apj}\ }\textbf {\bibinfo {volume} {972}},\
  \bibinfo {eid} {L4} (\bibinfo {year} {2024})},\ \Eprint
  {https://arxiv.org/abs/2406.19452} {arXiv:2406.19452 [astro-ph.HE]}
  \BibitemShut {NoStop}%
\bibitem [{\citenamefont {{Dean}}\ and\ \citenamefont
  {{Fern{\'a}ndez}}(2024{\natexlab{b}})}]{DF24a}%
  \BibitemOpen
  \bibfield  {author} {\bibinfo {author} {\bibfnamefont {C.}~\bibnamefont
  {{Dean}}}\ and\ \bibinfo {author} {\bibfnamefont {R.}~\bibnamefont
  {{Fern{\'a}ndez}}},\ }\href {https://doi.org/10.1103/PhysRevD.109.083010}
  {\bibfield  {journal} {\bibinfo  {journal} {\prd}\ }\textbf {\bibinfo
  {volume} {109}},\ \bibinfo {eid} {083010} (\bibinfo {year}
  {2024}{\natexlab{b}})},\ \Eprint {https://arxiv.org/abs/2403.08877}
  {arXiv:2403.08877 [astro-ph.HE]} \BibitemShut {NoStop}%
\bibitem [{\citenamefont {{O'Connor}}\ and\ \citenamefont
  {{Ott}}(2010)}]{oconnor_2010}%
  \BibitemOpen
  \bibfield  {author} {\bibinfo {author} {\bibfnamefont {E.}~\bibnamefont
  {{O'Connor}}}\ and\ \bibinfo {author} {\bibfnamefont {C.~D.}\ \bibnamefont
  {{Ott}}},\ }\href {https://doi.org/10.1088/0264-9381/27/11/114103} {\bibfield
   {journal} {\bibinfo  {journal} {Classical and Quantum Gravity}\ }\textbf
  {\bibinfo {volume} {27}},\ \bibinfo {eid} {114103} (\bibinfo {year}
  {2010})},\ \Eprint {https://arxiv.org/abs/0912.2393} {arXiv:0912.2393
  [astro-ph.HE]} \BibitemShut {NoStop}%
\bibitem [{\citenamefont {{Fryxell}}\ \emph {et~al.}(2000)\citenamefont
  {{Fryxell}}, \citenamefont {{Olson}}, \citenamefont {{Ricker}}, \citenamefont
  {{Timmes}}, \citenamefont {{Zingale}}, \citenamefont {{Lamb}}, \citenamefont
  {{MacNeice}}, \citenamefont {{Rosner}}, \citenamefont {{Truran}},\ and\
  \citenamefont {{Tufo}}}]{fryxell00}%
  \BibitemOpen
  \bibfield  {author} {\bibinfo {author} {\bibfnamefont {B.}~\bibnamefont
  {{Fryxell}}}, \bibinfo {author} {\bibfnamefont {K.}~\bibnamefont {{Olson}}},
  \bibinfo {author} {\bibfnamefont {P.}~\bibnamefont {{Ricker}}}, \bibinfo
  {author} {\bibfnamefont {F.~X.}\ \bibnamefont {{Timmes}}}, \bibinfo {author}
  {\bibfnamefont {M.}~\bibnamefont {{Zingale}}}, \bibinfo {author}
  {\bibfnamefont {D.~Q.}\ \bibnamefont {{Lamb}}}, \bibinfo {author}
  {\bibfnamefont {P.}~\bibnamefont {{MacNeice}}}, \bibinfo {author}
  {\bibfnamefont {R.}~\bibnamefont {{Rosner}}}, \bibinfo {author}
  {\bibfnamefont {J.~W.}\ \bibnamefont {{Truran}}},\ and\ \bibinfo {author}
  {\bibfnamefont {H.}~\bibnamefont {{Tufo}}},\ }\href
  {https://doi.org/10.1086/317361} {\bibfield  {journal} {\bibinfo  {journal}
  {ApJS}\ }\textbf {\bibinfo {volume} {131}},\ \bibinfo {pages} {273} (\bibinfo
  {year} {2000})}\BibitemShut {NoStop}%
\bibitem [{\citenamefont {Dubey}\ \emph {et~al.}(2009)\citenamefont {Dubey},
  \citenamefont {Antypas}, \citenamefont {Ganapathy}, \citenamefont {Reid},
  \citenamefont {Riley}, \citenamefont {Sheeler}, \citenamefont {Siegel},\ and\
  \citenamefont {Weide}}]{dubey2009}%
  \BibitemOpen
  \bibfield  {author} {\bibinfo {author} {\bibfnamefont {A.}~\bibnamefont
  {Dubey}}, \bibinfo {author} {\bibfnamefont {K.}~\bibnamefont {Antypas}},
  \bibinfo {author} {\bibfnamefont {M.~K.}\ \bibnamefont {Ganapathy}}, \bibinfo
  {author} {\bibfnamefont {L.~B.}\ \bibnamefont {Reid}}, \bibinfo {author}
  {\bibfnamefont {K.}~\bibnamefont {Riley}}, \bibinfo {author} {\bibfnamefont
  {D.}~\bibnamefont {Sheeler}}, \bibinfo {author} {\bibfnamefont
  {A.}~\bibnamefont {Siegel}},\ and\ \bibinfo {author} {\bibfnamefont
  {K.}~\bibnamefont {Weide}},\ }\href {https://doi.org/DOI:
  10.1016/j.parco.2009.08.001} {\bibfield  {journal} {\bibinfo  {journal} {J.
  Par. Comp.}\ }\textbf {\bibinfo {volume} {35}},\ \bibinfo {pages} {512 }
  (\bibinfo {year} {2009})}\BibitemShut {NoStop}%
\bibitem [{\citenamefont {{Colella}}\ and\ \citenamefont
  {{Woodward}}(1984)}]{colella84}%
  \BibitemOpen
  \bibfield  {author} {\bibinfo {author} {\bibfnamefont {P.}~\bibnamefont
  {{Colella}}}\ and\ \bibinfo {author} {\bibfnamefont {P.~R.}\ \bibnamefont
  {{Woodward}}},\ }\href@noop {} {\bibfield  {journal} {\bibinfo  {journal}
  {JCP}\ }\textbf {\bibinfo {volume} {54}},\ \bibinfo {pages} {174} (\bibinfo
  {year} {1984})}\BibitemShut {NoStop}%
\bibitem [{\citenamefont {{Fern{\'a}ndez}}\ and\ \citenamefont
  {{Metzger}}(2013{\natexlab{a}})}]{FM12}%
  \BibitemOpen
  \bibfield  {author} {\bibinfo {author} {\bibfnamefont {R.}~\bibnamefont
  {{Fern{\'a}ndez}}}\ and\ \bibinfo {author} {\bibfnamefont {B.~D.}\
  \bibnamefont {{Metzger}}},\ }\href
  {https://doi.org/10.1088/0004-637X/763/2/108} {\bibfield  {journal} {\bibinfo
   {journal} {ApJ}\ }\textbf {\bibinfo {volume} {763}},\ \bibinfo {eid} {108}
  (\bibinfo {year} {2013}{\natexlab{a}})},\ \Eprint
  {https://arxiv.org/abs/1209.2712} {arXiv:1209.2712 [astro-ph.HE]}
  \BibitemShut {NoStop}%
\bibitem [{\citenamefont {{M{\"u}ller}}\ and\ \citenamefont
  {{Steinmetz}}(1995)}]{MuellerSteinmetz1995}%
  \BibitemOpen
  \bibfield  {author} {\bibinfo {author} {\bibfnamefont {E.}~\bibnamefont
  {{M{\"u}ller}}}\ and\ \bibinfo {author} {\bibfnamefont {M.}~\bibnamefont
  {{Steinmetz}}},\ }\href {https://doi.org/10.1016/0010-4655(94)00185-5}
  {\bibfield  {journal} {\bibinfo  {journal} {Computer Physics Communications}\
  }\textbf {\bibinfo {volume} {89}},\ \bibinfo {pages} {45} (\bibinfo {year}
  {1995})},\ \Eprint {https://arxiv.org/abs/astro-ph/9402070}
  {astro-ph/9402070} \BibitemShut {NoStop}%
\bibitem [{\citenamefont {{Fern{\'a}ndez}}\ \emph {et~al.}(2019)\citenamefont
  {{Fern{\'a}ndez}}, \citenamefont {{Margalit}},\ and\ \citenamefont
  {{Metzger}}}]{FMM19}%
  \BibitemOpen
  \bibfield  {author} {\bibinfo {author} {\bibfnamefont {R.}~\bibnamefont
  {{Fern{\'a}ndez}}}, \bibinfo {author} {\bibfnamefont {B.}~\bibnamefont
  {{Margalit}}},\ and\ \bibinfo {author} {\bibfnamefont {B.~D.}\ \bibnamefont
  {{Metzger}}},\ }\href {https://doi.org/10.1093/mnras/stz1701} {\bibfield
  {journal} {\bibinfo  {journal} {MNRAS}\ }\textbf {\bibinfo {volume} {488}},\
  \bibinfo {pages} {259} (\bibinfo {year} {2019})},\ \Eprint
  {https://arxiv.org/abs/1905.06343} {arXiv:1905.06343 [astro-ph.HE]}
  \BibitemShut {NoStop}%
\bibitem [{\citenamefont {{Artemova}}\ \emph {et~al.}(1996)\citenamefont
  {{Artemova}}, \citenamefont {{Bjoernsson}},\ and\ \citenamefont
  {{Novikov}}}]{artemova1996}%
  \BibitemOpen
  \bibfield  {author} {\bibinfo {author} {\bibfnamefont {I.~V.}\ \bibnamefont
  {{Artemova}}}, \bibinfo {author} {\bibfnamefont {G.}~\bibnamefont
  {{Bjoernsson}}},\ and\ \bibinfo {author} {\bibfnamefont {I.~D.}\ \bibnamefont
  {{Novikov}}},\ }\href {https://doi.org/10.1086/177084} {\bibfield  {journal}
  {\bibinfo  {journal} {ApJ}\ }\textbf {\bibinfo {volume} {461}},\ \bibinfo
  {pages} {565} (\bibinfo {year} {1996})}\BibitemShut {NoStop}%
\bibitem [{\citenamefont {{Fern{\'a}ndez}}\ \emph {et~al.}(2015)\citenamefont
  {{Fern{\'a}ndez}}, \citenamefont {{Kasen}}, \citenamefont {{Metzger}},\ and\
  \citenamefont {{Quataert}}}]{FKMQ14}%
  \BibitemOpen
  \bibfield  {author} {\bibinfo {author} {\bibfnamefont {R.}~\bibnamefont
  {{Fern{\'a}ndez}}}, \bibinfo {author} {\bibfnamefont {D.}~\bibnamefont
  {{Kasen}}}, \bibinfo {author} {\bibfnamefont {B.~D.}\ \bibnamefont
  {{Metzger}}},\ and\ \bibinfo {author} {\bibfnamefont {E.}~\bibnamefont
  {{Quataert}}},\ }\href {https://doi.org/10.1093/mnras/stu2112} {\bibfield
  {journal} {\bibinfo  {journal} {MNRAS}\ }\textbf {\bibinfo {volume} {446}},\
  \bibinfo {pages} {750} (\bibinfo {year} {2015})}\BibitemShut {NoStop}%
\bibitem [{\citenamefont {{Fern{\'a}ndez}}\ and\ \citenamefont
  {{Metzger}}(2013{\natexlab{b}})}]{FM13}%
  \BibitemOpen
  \bibfield  {author} {\bibinfo {author} {\bibfnamefont {R.}~\bibnamefont
  {{Fern{\'a}ndez}}}\ and\ \bibinfo {author} {\bibfnamefont {B.~D.}\
  \bibnamefont {{Metzger}}},\ }\href {https://doi.org/10.1093/mnras/stt1312}
  {\bibfield  {journal} {\bibinfo  {journal} {MNRAS}\ }\textbf {\bibinfo
  {volume} {435}},\ \bibinfo {pages} {502} (\bibinfo {year}
  {2013}{\natexlab{b}})},\ \Eprint {https://arxiv.org/abs/1304.6720}
  {arXiv:1304.6720 [astro-ph.HE]} \BibitemShut {NoStop}%
\bibitem [{\citenamefont {{Fern{\'a}ndez}}\ \emph {et~al.}(2022)\citenamefont
  {{Fern{\'a}ndez}}, \citenamefont {{Richers}}, \citenamefont {{Mulyk}},\ and\
  \citenamefont {{Fahlman}}}]{fernandez_2022}%
  \BibitemOpen
  \bibfield  {author} {\bibinfo {author} {\bibfnamefont {R.}~\bibnamefont
  {{Fern{\'a}ndez}}}, \bibinfo {author} {\bibfnamefont {S.}~\bibnamefont
  {{Richers}}}, \bibinfo {author} {\bibfnamefont {N.}~\bibnamefont {{Mulyk}}},\
  and\ \bibinfo {author} {\bibfnamefont {S.}~\bibnamefont {{Fahlman}}},\ }\href
  {https://doi.org/10.1103/PhysRevD.106.103003} {\bibfield  {journal} {\bibinfo
   {journal} {\prd}\ }\textbf {\bibinfo {volume} {106}},\ \bibinfo {eid}
  {103003} (\bibinfo {year} {2022})},\ \Eprint
  {https://arxiv.org/abs/2207.10680} {arXiv:2207.10680 [astro-ph.HE]}
  \BibitemShut {NoStop}%
\bibitem [{\citenamefont {{Weaver}}\ \emph {et~al.}(1978)\citenamefont
  {{Weaver}}, \citenamefont {{Zimmerman}},\ and\ \citenamefont
  {{Woosley}}}]{weaver1978}%
  \BibitemOpen
  \bibfield  {author} {\bibinfo {author} {\bibfnamefont {T.~A.}\ \bibnamefont
  {{Weaver}}}, \bibinfo {author} {\bibfnamefont {G.~B.}\ \bibnamefont
  {{Zimmerman}}},\ and\ \bibinfo {author} {\bibfnamefont {S.~E.}\ \bibnamefont
  {{Woosley}}},\ }\href {https://doi.org/10.1086/156569} {\bibfield  {journal}
  {\bibinfo  {journal} {ApJ}\ }\textbf {\bibinfo {volume} {225}},\ \bibinfo
  {pages} {1021} (\bibinfo {year} {1978})}\BibitemShut {NoStop}%
\bibitem [{\citenamefont {{Timmes}}(1999)}]{Timmes1999}%
  \BibitemOpen
  \bibfield  {author} {\bibinfo {author} {\bibfnamefont {F.~X.}\ \bibnamefont
  {{Timmes}}},\ }\href {https://doi.org/10.1086/313257} {\bibfield  {journal}
  {\bibinfo  {journal} {ApJS}\ }\textbf {\bibinfo {volume} {124}},\ \bibinfo
  {pages} {241} (\bibinfo {year} {1999})}\BibitemShut {NoStop}%
\bibitem [{\citenamefont {{Seitenzahl}}\ \emph {et~al.}(2008)\citenamefont
  {{Seitenzahl}}, \citenamefont {{Timmes}}, \citenamefont
  {{Marin-Lafl{\`e}che}}, \citenamefont {{Brown}}, \citenamefont
  {{Magkotsios}},\ and\ \citenamefont {{Truran}}}]{seitenzahl_2008}%
  \BibitemOpen
  \bibfield  {author} {\bibinfo {author} {\bibfnamefont {I.~R.}\ \bibnamefont
  {{Seitenzahl}}}, \bibinfo {author} {\bibfnamefont {F.~X.}\ \bibnamefont
  {{Timmes}}}, \bibinfo {author} {\bibfnamefont {A.}~\bibnamefont
  {{Marin-Lafl{\`e}che}}}, \bibinfo {author} {\bibfnamefont {E.}~\bibnamefont
  {{Brown}}}, \bibinfo {author} {\bibfnamefont {G.}~\bibnamefont
  {{Magkotsios}}},\ and\ \bibinfo {author} {\bibfnamefont {J.}~\bibnamefont
  {{Truran}}},\ }\href {https://doi.org/10.1086/592501} {\bibfield  {journal}
  {\bibinfo  {journal} {ApJ}\ }\textbf {\bibinfo {volume} {685}},\ \bibinfo
  {pages} {L129} (\bibinfo {year} {2008})},\ \Eprint
  {https://arxiv.org/abs/0808.2033} {arXiv:0808.2033 [astro-ph]} \BibitemShut
  {NoStop}%
\bibitem [{\citenamefont {{Timmes}}\ and\ \citenamefont
  {{Swesty}}(2000)}]{timmes2000}%
  \BibitemOpen
  \bibfield  {author} {\bibinfo {author} {\bibfnamefont {F.~X.}\ \bibnamefont
  {{Timmes}}}\ and\ \bibinfo {author} {\bibfnamefont {F.~D.}\ \bibnamefont
  {{Swesty}}},\ }\href {https://doi.org/10.1086/313304} {\bibfield  {journal}
  {\bibinfo  {journal} {ApJS}\ }\textbf {\bibinfo {volume} {126}},\ \bibinfo
  {pages} {501} (\bibinfo {year} {2000})}\BibitemShut {NoStop}%
\bibitem [{\citenamefont {{Woosley}}\ and\ \citenamefont
  {{Heger}}(2006)}]{woosley_2006b}%
  \BibitemOpen
  \bibfield  {author} {\bibinfo {author} {\bibfnamefont {S.~E.}\ \bibnamefont
  {{Woosley}}}\ and\ \bibinfo {author} {\bibfnamefont {A.}~\bibnamefont
  {{Heger}}},\ }\href {https://doi.org/10.1086/498500} {\bibfield  {journal}
  {\bibinfo  {journal} {\apj}\ }\textbf {\bibinfo {volume} {637}},\ \bibinfo
  {pages} {914} (\bibinfo {year} {2006})},\ \Eprint
  {https://arxiv.org/abs/astro-ph/0508175} {arXiv:astro-ph/0508175 [astro-ph]}
  \BibitemShut {NoStop}%
\bibitem [{\citenamefont {{Steiner}}\ \emph {et~al.}(2013)\citenamefont
  {{Steiner}}, \citenamefont {{Hempel}},\ and\ \citenamefont
  {{Fischer}}}]{steiner_2013}%
  \BibitemOpen
  \bibfield  {author} {\bibinfo {author} {\bibfnamefont {A.~W.}\ \bibnamefont
  {{Steiner}}}, \bibinfo {author} {\bibfnamefont {M.}~\bibnamefont
  {{Hempel}}},\ and\ \bibinfo {author} {\bibfnamefont {T.}~\bibnamefont
  {{Fischer}}},\ }\href {https://doi.org/10.1088/0004-637X/774/1/17} {\bibfield
   {journal} {\bibinfo  {journal} {ApJ}\ }\textbf {\bibinfo {volume} {774}},\
  \bibinfo {eid} {17} (\bibinfo {year} {2013})},\ \Eprint
  {https://arxiv.org/abs/1207.2184} {arXiv:1207.2184 [astro-ph.SR]}
  \BibitemShut {NoStop}%
\bibitem [{\citenamefont {{Hempel}}\ \emph {et~al.}(2012)\citenamefont
  {{Hempel}}, \citenamefont {{Fischer}}, \citenamefont {{Schaffner-Bielich}},\
  and\ \citenamefont {{Liebend{\"o}rfer}}}]{hempel_2012}%
  \BibitemOpen
  \bibfield  {author} {\bibinfo {author} {\bibfnamefont {M.}~\bibnamefont
  {{Hempel}}}, \bibinfo {author} {\bibfnamefont {T.}~\bibnamefont {{Fischer}}},
  \bibinfo {author} {\bibfnamefont {J.}~\bibnamefont {{Schaffner-Bielich}}},\
  and\ \bibinfo {author} {\bibfnamefont {M.}~\bibnamefont
  {{Liebend{\"o}rfer}}},\ }\href {https://doi.org/10.1088/0004-637X/748/1/70}
  {\bibfield  {journal} {\bibinfo  {journal} {ApJ}\ }\textbf {\bibinfo {volume}
  {748}},\ \bibinfo {eid} {70} (\bibinfo {year} {2012})},\ \Eprint
  {https://arxiv.org/abs/1108.0848} {arXiv:1108.0848 [astro-ph.HE]}
  \BibitemShut {NoStop}%
\bibitem [{\citenamefont {{Fern\'andez'}}\ \emph {et~al.}(2025)\citenamefont
  {{Fern\'andez'}}, \citenamefont {{Janke}}, \citenamefont {{Dean}},\ and\
  \citenamefont {{Tamborra}}}]{zenodo_repo}%
  \BibitemOpen
  \bibfield  {author} {\bibinfo {author} {\bibfnamefont {R.}~\bibnamefont
  {{Fern\'andez'}}}, \bibinfo {author} {\bibfnamefont {S.}~\bibnamefont
  {{Janke}}}, \bibinfo {author} {\bibfnamefont {C.}~\bibnamefont {{Dean}}},\
  and\ \bibinfo {author} {\bibfnamefont {I.}~\bibnamefont {{Tamborra}}},\
  }\href {https://doi.org/10.5281/zenodo.17299306.svg} {\bibinfo {title} {{Data
  for Collapsar Disk Outflows III. Detectable Neutrino and Gravitational Wave
  Signatures}}},\ \bibinfo {howpublished}
  {\href{https://doi.org/10.5281/zenodo.17299306}{https://doi.org/10.5281/zenodo.17299306}}
  (\bibinfo {year} {2025})\BibitemShut {NoStop}%
\bibitem [{\citenamefont {{Ruffert}}\ \emph {et~al.}(1996)\citenamefont
  {{Ruffert}}, \citenamefont {{Janka}},\ and\ \citenamefont
  {{Schaefer}}}]{ruffert_1996}%
  \BibitemOpen
  \bibfield  {author} {\bibinfo {author} {\bibfnamefont {M.}~\bibnamefont
  {{Ruffert}}}, \bibinfo {author} {\bibfnamefont {H.~T.}\ \bibnamefont
  {{Janka}}},\ and\ \bibinfo {author} {\bibfnamefont {G.}~\bibnamefont
  {{Schaefer}}},\ }\href {https://doi.org/10.48550/arXiv.astro-ph/9509006}
  {\bibfield  {journal} {\bibinfo  {journal} {Astronomy \& Astrophysics}\
  }\textbf {\bibinfo {volume} {311}},\ \bibinfo {pages} {532} (\bibinfo {year}
  {1996})},\ \Eprint {https://arxiv.org/abs/astro-ph/9509006}
  {arXiv:astro-ph/9509006 [astro-ph]} \BibitemShut {NoStop}%
\bibitem [{\citenamefont {{Fern{\'a}ndez}}\ and\ \citenamefont
  {{Metzger}}(2013{\natexlab{c}})}]{fernandez_2013}%
  \BibitemOpen
  \bibfield  {author} {\bibinfo {author} {\bibfnamefont {R.}~\bibnamefont
  {{Fern{\'a}ndez}}}\ and\ \bibinfo {author} {\bibfnamefont {B.~D.}\
  \bibnamefont {{Metzger}}},\ }\href {https://doi.org/10.1093/mnras/stt1312}
  {\bibfield  {journal} {\bibinfo  {journal} {MNRAS}\ }\textbf {\bibinfo
  {volume} {435}},\ \bibinfo {pages} {502} (\bibinfo {year}
  {2013}{\natexlab{c}})},\ \Eprint {https://arxiv.org/abs/1304.6720}
  {arXiv:1304.6720 [astro-ph.HE]} \BibitemShut {NoStop}%
\bibitem [{\citenamefont {{Bludman}}\ and\ \citenamefont {{van
  Riper}}(1978)}]{bludman_1978}%
  \BibitemOpen
  \bibfield  {author} {\bibinfo {author} {\bibfnamefont {S.~A.}\ \bibnamefont
  {{Bludman}}}\ and\ \bibinfo {author} {\bibfnamefont {K.~A.}\ \bibnamefont
  {{van Riper}}},\ }\href {https://doi.org/10.1086/156412} {\bibfield
  {journal} {\bibinfo  {journal} {\apj}\ }\textbf {\bibinfo {volume} {224}},\
  \bibinfo {pages} {631} (\bibinfo {year} {1978})}\BibitemShut {NoStop}%
\bibitem [{\citenamefont {{Tamborra}}\ and\ \citenamefont
  {{Shalgar}}(2021)}]{tamborra_2021}%
  \BibitemOpen
  \bibfield  {author} {\bibinfo {author} {\bibfnamefont {I.}~\bibnamefont
  {{Tamborra}}}\ and\ \bibinfo {author} {\bibfnamefont {S.}~\bibnamefont
  {{Shalgar}}},\ }\href {https://doi.org/10.1146/annurev-nucl-102920-050505}
  {\bibfield  {journal} {\bibinfo  {journal} {Annual Review of Nuclear and
  Particle Science}\ }\textbf {\bibinfo {volume} {71}},\ \bibinfo {pages} {165}
  (\bibinfo {year} {2021})},\ \Eprint {https://arxiv.org/abs/2011.01948}
  {arXiv:2011.01948 [astro-ph.HE]} \BibitemShut {NoStop}%
\bibitem [{\citenamefont {{Volpe}}(2024)}]{volpe_2024}%
  \BibitemOpen
  \bibfield  {author} {\bibinfo {author} {\bibfnamefont {M.~C.}\ \bibnamefont
  {{Volpe}}},\ }\href {https://doi.org/10.1103/RevModPhys.96.025004} {\bibfield
   {journal} {\bibinfo  {journal} {Reviews of Modern Physics}\ }\textbf
  {\bibinfo {volume} {96}},\ \bibinfo {eid} {025004} (\bibinfo {year}
  {2024})},\ \Eprint {https://arxiv.org/abs/2301.11814} {arXiv:2301.11814
  [hep-ph]} \BibitemShut {NoStop}%
\bibitem [{\citenamefont {{Dighe}}\ and\ \citenamefont
  {{Smirnov}}(2000)}]{dighe_2000}%
  \BibitemOpen
  \bibfield  {author} {\bibinfo {author} {\bibfnamefont {A.~S.}\ \bibnamefont
  {{Dighe}}}\ and\ \bibinfo {author} {\bibfnamefont {A.~Y.}\ \bibnamefont
  {{Smirnov}}},\ }\href {https://doi.org/10.1103/PhysRevD.62.033007} {\bibfield
   {journal} {\bibinfo  {journal} {\prd}\ }\textbf {\bibinfo {volume} {62}},\
  \bibinfo {eid} {033007} (\bibinfo {year} {2000})},\ \Eprint
  {https://arxiv.org/abs/hep-ph/9907423} {arXiv:hep-ph/9907423 [hep-ph]}
  \BibitemShut {NoStop}%
\bibitem [{\citenamefont {{Esteban}}\ \emph {et~al.}(2020)\citenamefont
  {{Esteban}}, \citenamefont {{Gonzalez-Garcia}}, \citenamefont {{Maltoni}},
  \citenamefont {{Schwetz}},\ and\ \citenamefont {{Zhou}}}]{esteban_2020}%
  \BibitemOpen
  \bibfield  {author} {\bibinfo {author} {\bibfnamefont {I.}~\bibnamefont
  {{Esteban}}}, \bibinfo {author} {\bibfnamefont {M.~C.}\ \bibnamefont
  {{Gonzalez-Garcia}}}, \bibinfo {author} {\bibfnamefont {M.}~\bibnamefont
  {{Maltoni}}}, \bibinfo {author} {\bibfnamefont {T.}~\bibnamefont
  {{Schwetz}}},\ and\ \bibinfo {author} {\bibfnamefont {A.}~\bibnamefont
  {{Zhou}}},\ }\href {https://doi.org/10.1007/JHEP09(2020)178} {\bibfield
  {journal} {\bibinfo  {journal} {Journal of High Energy Physics}\ }\textbf
  {\bibinfo {volume} {2020}},\ \bibinfo {eid} {178} (\bibinfo {year} {2020})},\
  \Eprint {https://arxiv.org/abs/2007.14792} {arXiv:2007.14792 [hep-ph]}
  \BibitemShut {NoStop}%
\bibitem [{\citenamefont {{Abbasi}}\ \emph {et~al.}(2011)\citenamefont
  {{Abbasi}} \emph {et~al.}}]{Abbasi_2011}%
  \BibitemOpen
  \bibfield  {author} {\bibinfo {author} {\bibfnamefont {R.}~\bibnamefont
  {{Abbasi}}} \emph {et~al.},\ }\href
  {https://doi.org/10.1051/0004-6361/201117810} {\bibfield  {journal} {\bibinfo
   {journal} {A\&A}\ }\textbf {\bibinfo {volume} {535}},\ \bibinfo {eid} {A109}
  (\bibinfo {year} {2011})},\ \Eprint {https://arxiv.org/abs/1108.0171}
  {arXiv:1108.0171 [astro-ph.HE]} \BibitemShut {NoStop}%
\bibitem [{\citenamefont {{Malmenbeck}}\ and\ \citenamefont
  {{O'Sullivan}}(2019)}]{Malmenbeck_2019}%
  \BibitemOpen
  \bibfield  {author} {\bibinfo {author} {\bibfnamefont {F.}~\bibnamefont
  {{Malmenbeck}}}\ and\ \bibinfo {author} {\bibfnamefont {E.}~\bibnamefont
  {{O'Sullivan}}},\ }\href {https://doi.org/10.48550/arXiv.1909.00886}
  {\bibfield  {journal} {\bibinfo  {journal} {arXiv e-prints}\ ,\ \bibinfo
  {eid} {arXiv:1909.00886}} (\bibinfo {year} {2019})},\ \Eprint
  {https://arxiv.org/abs/1909.00886} {arXiv:1909.00886 [astro-ph.HE]}
  \BibitemShut {NoStop}%
\bibitem [{\citenamefont {{Baxter}}\ \emph {et~al.}(2022)\citenamefont
  {{Baxter}}, \citenamefont {{Benzvi}}, \citenamefont {{Jaimes}}, \citenamefont
  {{Coleiro}}, \citenamefont {{Molla}}, \citenamefont {{Dornic}}, \citenamefont
  {{Goldhagen}}, \citenamefont {{Graf}}, \citenamefont {{Griswold}},
  \citenamefont {{Habig}}, \citenamefont {{Hill}}, \citenamefont {{Horiuchi}},
  \citenamefont {{Kneller}}, \citenamefont {{Lang}}, \citenamefont
  {{Lincetto}}, \citenamefont {{Migenda}}, \citenamefont {{Nakamura}},
  \citenamefont {{O'Connor}}, \citenamefont {{Renshaw}}, \citenamefont
  {{Scholberg}}, \citenamefont {{Tunnell}}, \citenamefont {{Uberoi}},
  \citenamefont {{Worlikar}},\ and\ \citenamefont {{SNEWS
  Collaboration}}}]{baxter_2022}%
  \BibitemOpen
  \bibfield  {author} {\bibinfo {author} {\bibfnamefont {A.~L.}\ \bibnamefont
  {{Baxter}}}, \bibinfo {author} {\bibfnamefont {S.}~\bibnamefont {{Benzvi}}},
  \bibinfo {author} {\bibfnamefont {J.~C.}\ \bibnamefont {{Jaimes}}}, \bibinfo
  {author} {\bibfnamefont {A.}~\bibnamefont {{Coleiro}}}, \bibinfo {author}
  {\bibfnamefont {M.~C.}\ \bibnamefont {{Molla}}}, \bibinfo {author}
  {\bibfnamefont {D.}~\bibnamefont {{Dornic}}}, \bibinfo {author}
  {\bibfnamefont {T.}~\bibnamefont {{Goldhagen}}}, \bibinfo {author}
  {\bibfnamefont {A.}~\bibnamefont {{Graf}}}, \bibinfo {author} {\bibfnamefont
  {S.}~\bibnamefont {{Griswold}}}, \bibinfo {author} {\bibfnamefont
  {A.}~\bibnamefont {{Habig}}}, \bibinfo {author} {\bibfnamefont
  {R.}~\bibnamefont {{Hill}}}, \bibinfo {author} {\bibfnamefont
  {S.}~\bibnamefont {{Horiuchi}}}, \bibinfo {author} {\bibfnamefont {J.~P.}\
  \bibnamefont {{Kneller}}}, \bibinfo {author} {\bibfnamefont {R.~F.}\
  \bibnamefont {{Lang}}}, \bibinfo {author} {\bibfnamefont {M.}~\bibnamefont
  {{Lincetto}}}, \bibinfo {author} {\bibfnamefont {J.}~\bibnamefont
  {{Migenda}}}, \bibinfo {author} {\bibfnamefont {K.}~\bibnamefont
  {{Nakamura}}}, \bibinfo {author} {\bibfnamefont {E.}~\bibnamefont
  {{O'Connor}}}, \bibinfo {author} {\bibfnamefont {A.}~\bibnamefont
  {{Renshaw}}}, \bibinfo {author} {\bibfnamefont {K.}~\bibnamefont
  {{Scholberg}}}, \bibinfo {author} {\bibfnamefont {C.}~\bibnamefont
  {{Tunnell}}}, \bibinfo {author} {\bibfnamefont {N.}~\bibnamefont {{Uberoi}}},
  \bibinfo {author} {\bibfnamefont {A.}~\bibnamefont {{Worlikar}}},\ and\
  \bibinfo {author} {\bibnamefont {{SNEWS Collaboration}}},\ }\href
  {https://doi.org/10.3847/1538-4357/ac350f} {\bibfield  {journal} {\bibinfo
  {journal} {\apj}\ }\textbf {\bibinfo {volume} {925}},\ \bibinfo {eid} {107}
  (\bibinfo {year} {2022})}\BibitemShut {NoStop}%
\bibitem [{\citenamefont {{Halzen}}\ and\ \citenamefont
  {{Raffelt}}(2009)}]{halzen_2009}%
  \BibitemOpen
  \bibfield  {author} {\bibinfo {author} {\bibfnamefont {F.}~\bibnamefont
  {{Halzen}}}\ and\ \bibinfo {author} {\bibfnamefont {G.~G.}\ \bibnamefont
  {{Raffelt}}},\ }\href {https://doi.org/10.1103/PhysRevD.80.087301} {\bibfield
   {journal} {\bibinfo  {journal} {\prd}\ }\textbf {\bibinfo {volume} {80}},\
  \bibinfo {eid} {087301} (\bibinfo {year} {2009})},\ \Eprint
  {https://arxiv.org/abs/0908.2317} {arXiv:0908.2317 [astro-ph.HE]}
  \BibitemShut {NoStop}%
\bibitem [{\citenamefont {{Oohara}}\ and\ \citenamefont
  {{Nakamura}}(1989)}]{oohara_1989}%
  \BibitemOpen
  \bibfield  {author} {\bibinfo {author} {\bibfnamefont {K.}~\bibnamefont
  {{Oohara}}}\ and\ \bibinfo {author} {\bibfnamefont {T.}~\bibnamefont
  {{Nakamura}}},\ }\href {https://doi.org/10.1143/PTP.82.535} {\bibfield
  {journal} {\bibinfo  {journal} {Progress of Theoretical Physics}\ }\textbf
  {\bibinfo {volume} {82}},\ \bibinfo {pages} {535} (\bibinfo {year}
  {1989})}\BibitemShut {NoStop}%
\bibitem [{\citenamefont {{Rasio}}\ and\ \citenamefont
  {{Shapiro}}(1992)}]{rasio_1992}%
  \BibitemOpen
  \bibfield  {author} {\bibinfo {author} {\bibfnamefont {F.~A.}\ \bibnamefont
  {{Rasio}}}\ and\ \bibinfo {author} {\bibfnamefont {S.~L.}\ \bibnamefont
  {{Shapiro}}},\ }\href {https://doi.org/10.1086/172055} {\bibfield  {journal}
  {\bibinfo  {journal} {\apj}\ }\textbf {\bibinfo {volume} {401}},\ \bibinfo
  {pages} {226} (\bibinfo {year} {1992})}\BibitemShut {NoStop}%
\bibitem [{\citenamefont {{Finn}}\ and\ \citenamefont
  {{Evans}}(1990)}]{finn_1990}%
  \BibitemOpen
  \bibfield  {author} {\bibinfo {author} {\bibfnamefont {L.~S.}\ \bibnamefont
  {{Finn}}}\ and\ \bibinfo {author} {\bibfnamefont {C.~R.}\ \bibnamefont
  {{Evans}}},\ }\href {https://doi.org/10.1086/168497} {\bibfield  {journal}
  {\bibinfo  {journal} {\apj}\ }\textbf {\bibinfo {volume} {351}},\ \bibinfo
  {pages} {588} (\bibinfo {year} {1990})}\BibitemShut {NoStop}%
\bibitem [{\citenamefont {{Blanchet}}\ \emph {et~al.}(1990)\citenamefont
  {{Blanchet}}, \citenamefont {{Damour}},\ and\ \citenamefont
  {{Schaefer}}}]{blanchet_1990}%
  \BibitemOpen
  \bibfield  {author} {\bibinfo {author} {\bibfnamefont {L.}~\bibnamefont
  {{Blanchet}}}, \bibinfo {author} {\bibfnamefont {T.}~\bibnamefont
  {{Damour}}},\ and\ \bibinfo {author} {\bibfnamefont {G.}~\bibnamefont
  {{Schaefer}}},\ }\href {https://doi.org/10.1093/mnras/242.3.289} {\bibfield
  {journal} {\bibinfo  {journal} {MNRAS}\ }\textbf {\bibinfo {volume} {242}},\
  \bibinfo {pages} {289} (\bibinfo {year} {1990})}\BibitemShut {NoStop}%
\bibitem [{\citenamefont {{Flanagan}}\ and\ \citenamefont
  {{Hughes}}(1998)}]{flanagan_1998}%
  \BibitemOpen
  \bibfield  {author} {\bibinfo {author} {\bibfnamefont {{\'E}.~{\'E}.}\
  \bibnamefont {{Flanagan}}}\ and\ \bibinfo {author} {\bibfnamefont {S.~A.}\
  \bibnamefont {{Hughes}}},\ }\href {https://doi.org/10.1103/PhysRevD.57.4535}
  {\bibfield  {journal} {\bibinfo  {journal} {\prd}\ }\textbf {\bibinfo
  {volume} {57}},\ \bibinfo {pages} {4535} (\bibinfo {year} {1998})},\ \Eprint
  {https://arxiv.org/abs/gr-qc/9701039} {arXiv:gr-qc/9701039 [gr-qc]}
  \BibitemShut {NoStop}%
\bibitem [{\citenamefont {{Ott}}\ \emph {et~al.}(2012)\citenamefont {{Ott}},
  \citenamefont {{Abdikamalov}}, \citenamefont {{O'Connor}}, \citenamefont
  {{Reisswig}}, \citenamefont {{Haas}}, \citenamefont {{Kalmus}}, \citenamefont
  {{Drasco}}, \citenamefont {{Burrows}},\ and\ \citenamefont
  {{Schnetter}}}]{ott_2012}%
  \BibitemOpen
  \bibfield  {author} {\bibinfo {author} {\bibfnamefont {C.~D.}\ \bibnamefont
  {{Ott}}}, \bibinfo {author} {\bibfnamefont {E.}~\bibnamefont
  {{Abdikamalov}}}, \bibinfo {author} {\bibfnamefont {E.}~\bibnamefont
  {{O'Connor}}}, \bibinfo {author} {\bibfnamefont {C.}~\bibnamefont
  {{Reisswig}}}, \bibinfo {author} {\bibfnamefont {R.}~\bibnamefont {{Haas}}},
  \bibinfo {author} {\bibfnamefont {P.}~\bibnamefont {{Kalmus}}}, \bibinfo
  {author} {\bibfnamefont {S.}~\bibnamefont {{Drasco}}}, \bibinfo {author}
  {\bibfnamefont {A.}~\bibnamefont {{Burrows}}},\ and\ \bibinfo {author}
  {\bibfnamefont {E.}~\bibnamefont {{Schnetter}}},\ }\href
  {https://doi.org/10.1103/PhysRevD.86.024026} {\bibfield  {journal} {\bibinfo
  {journal} {\prd}\ }\textbf {\bibinfo {volume} {86}},\ \bibinfo {eid} {024026}
  (\bibinfo {year} {2012})},\ \Eprint {https://arxiv.org/abs/1204.0512}
  {arXiv:1204.0512 [astro-ph.HE]} \BibitemShut {NoStop}%
\bibitem [{\citenamefont {{Choi}}\ \emph {et~al.}(2024)\citenamefont {{Choi}},
  \citenamefont {{Burrows}},\ and\ \citenamefont {{Vartanyan}}}]{choi_2024}%
  \BibitemOpen
  \bibfield  {author} {\bibinfo {author} {\bibfnamefont {L.}~\bibnamefont
  {{Choi}}}, \bibinfo {author} {\bibfnamefont {A.}~\bibnamefont {{Burrows}}},\
  and\ \bibinfo {author} {\bibfnamefont {D.}~\bibnamefont {{Vartanyan}}},\
  }\href {https://doi.org/10.3847/1538-4357/ad74f8} {\bibfield  {journal}
  {\bibinfo  {journal} {\apj}\ }\textbf {\bibinfo {volume} {975}},\ \bibinfo
  {eid} {12} (\bibinfo {year} {2024})},\ \Eprint
  {https://arxiv.org/abs/2408.01525} {arXiv:2408.01525 [astro-ph.HE]}
  \BibitemShut {NoStop}%
\bibitem [{\citenamefont {{De}}\ and\ \citenamefont
  {{Siegel}}(2021)}]{de_2021}%
  \BibitemOpen
  \bibfield  {author} {\bibinfo {author} {\bibfnamefont {S.}~\bibnamefont
  {{De}}}\ and\ \bibinfo {author} {\bibfnamefont {D.~M.}\ \bibnamefont
  {{Siegel}}},\ }\href {https://doi.org/10.3847/1538-4357/ac110b} {\bibfield
  {journal} {\bibinfo  {journal} {\apj}\ }\textbf {\bibinfo {volume} {921}},\
  \bibinfo {eid} {94} (\bibinfo {year} {2021})},\ \Eprint
  {https://arxiv.org/abs/2011.07176} {arXiv:2011.07176 [astro-ph.HE]}
  \BibitemShut {NoStop}%
\bibitem [{\citenamefont {{Balbus}}\ and\ \citenamefont
  {{Hawley}}(1998)}]{balbus_1998}%
  \BibitemOpen
  \bibfield  {author} {\bibinfo {author} {\bibfnamefont {S.~A.}\ \bibnamefont
  {{Balbus}}}\ and\ \bibinfo {author} {\bibfnamefont {J.~F.}\ \bibnamefont
  {{Hawley}}},\ }\href {https://doi.org/10.1103/RevModPhys.70.1} {\bibfield
  {journal} {\bibinfo  {journal} {Reviews of Modern Physics}\ }\textbf
  {\bibinfo {volume} {70}},\ \bibinfo {pages} {1} (\bibinfo {year}
  {1998})}\BibitemShut {NoStop}%
\bibitem [{\citenamefont {{Abbott}}\ \emph {et~al.}(2020)\citenamefont
  {{Abbott}} \emph {et~al.}}]{ligo_lrr_2020}%
  \BibitemOpen
  \bibfield  {author} {\bibinfo {author} {\bibfnamefont {B.~P.}\ \bibnamefont
  {{Abbott}}} \emph {et~al.},\ }\href
  {https://doi.org/10.1007/s41114-020-00026-9} {\bibfield  {journal} {\bibinfo
  {journal} {Living Reviews in Relativity}\ }\textbf {\bibinfo {volume} {23}},\
  \bibinfo {eid} {3} (\bibinfo {year} {2020})}\BibitemShut {NoStop}%
\bibitem [{\citenamefont {{Srivastava}}\ \emph {et~al.}(2022)\citenamefont
  {{Srivastava}}, \citenamefont {{Davis}}, \citenamefont {{Kuns}},
  \citenamefont {{Landry}}, \citenamefont {{Ballmer}}, \citenamefont {{Evans}},
  \citenamefont {{Hall}}, \citenamefont {{Read}},\ and\ \citenamefont
  {{Sathyaprakash}}}]{srivastava_2022}%
  \BibitemOpen
  \bibfield  {author} {\bibinfo {author} {\bibfnamefont {V.}~\bibnamefont
  {{Srivastava}}}, \bibinfo {author} {\bibfnamefont {D.}~\bibnamefont
  {{Davis}}}, \bibinfo {author} {\bibfnamefont {K.}~\bibnamefont {{Kuns}}},
  \bibinfo {author} {\bibfnamefont {P.}~\bibnamefont {{Landry}}}, \bibinfo
  {author} {\bibfnamefont {S.}~\bibnamefont {{Ballmer}}}, \bibinfo {author}
  {\bibfnamefont {M.}~\bibnamefont {{Evans}}}, \bibinfo {author} {\bibfnamefont
  {E.~D.}\ \bibnamefont {{Hall}}}, \bibinfo {author} {\bibfnamefont
  {J.}~\bibnamefont {{Read}}},\ and\ \bibinfo {author} {\bibfnamefont {B.~S.}\
  \bibnamefont {{Sathyaprakash}}},\ }\href
  {https://doi.org/10.3847/1538-4357/ac5f04} {\bibfield  {journal} {\bibinfo
  {journal} {\apj}\ }\textbf {\bibinfo {volume} {931}},\ \bibinfo {eid} {22}
  (\bibinfo {year} {2022})},\ \Eprint {https://arxiv.org/abs/2201.10668}
  {arXiv:2201.10668 [gr-qc]} \BibitemShut {NoStop}%
\bibitem [{\citenamefont {{Hild}}\ \emph {et~al.}(2011)\citenamefont {{Hild}}
  \emph {et~al.}}]{hild_2011}%
  \BibitemOpen
  \bibfield  {author} {\bibinfo {author} {\bibfnamefont {S.}~\bibnamefont
  {{Hild}}} \emph {et~al.},\ }\href
  {https://doi.org/10.1088/0264-9381/28/9/094013} {\bibfield  {journal}
  {\bibinfo  {journal} {Classical and Quantum Gravity}\ }\textbf {\bibinfo
  {volume} {28}},\ \bibinfo {eid} {094013} (\bibinfo {year} {2011})},\ \Eprint
  {https://arxiv.org/abs/1012.0908} {arXiv:1012.0908 [gr-qc]} \BibitemShut
  {NoStop}%
\bibitem [{\citenamefont {{Yagi}}\ and\ \citenamefont
  {{Seto}}(2017)}]{yagi_2017}%
  \BibitemOpen
  \bibfield  {author} {\bibinfo {author} {\bibfnamefont {K.}~\bibnamefont
  {{Yagi}}}\ and\ \bibinfo {author} {\bibfnamefont {N.}~\bibnamefont
  {{Seto}}},\ }\href {https://doi.org/10.1103/PhysRevD.95.109901} {\bibfield
  {journal} {\bibinfo  {journal} {\prd}\ }\textbf {\bibinfo {volume} {95}},\
  \bibinfo {eid} {109901} (\bibinfo {year} {2017})}\BibitemShut {NoStop}%
\bibitem [{\citenamefont {Yuan}\ \emph {et~al.}(2025)\citenamefont {Yuan},
  \citenamefont {Cardoso}, \citenamefont {Duque},\ and\ \citenamefont
  {Younsi}}]{yuan_2025}%
  \BibitemOpen
  \bibfield  {author} {\bibinfo {author} {\bibfnamefont {C.}~\bibnamefont
  {Yuan}}, \bibinfo {author} {\bibfnamefont {V.}~\bibnamefont {Cardoso}},
  \bibinfo {author} {\bibfnamefont {F.}~\bibnamefont {Duque}},\ and\ \bibinfo
  {author} {\bibfnamefont {Z.}~\bibnamefont {Younsi}},\ }\href
  {https://doi.org/10.1103/PhysRevD.111.063048} {\bibfield  {journal} {\bibinfo
   {journal} {Phys. Rev. D}\ }\textbf {\bibinfo {volume} {111}},\ \bibinfo
  {pages} {063048} (\bibinfo {year} {2025})},\ \Eprint
  {https://arxiv.org/abs/2502.07871} {arXiv:2502.07871 [gr-qc]} \BibitemShut
  {NoStop}%
\bibitem [{\citenamefont {{Fern{\'a}ndez}}\ and\ \citenamefont
  {{Thompson}}(2009)}]{FT09_sasi}%
  \BibitemOpen
  \bibfield  {author} {\bibinfo {author} {\bibfnamefont {R.}~\bibnamefont
  {{Fern{\'a}ndez}}}\ and\ \bibinfo {author} {\bibfnamefont {C.}~\bibnamefont
  {{Thompson}}},\ }\href {https://doi.org/10.1088/0004-637X/697/2/1827}
  {\bibfield  {journal} {\bibinfo  {journal} {\apj}\ }\textbf {\bibinfo
  {volume} {697}},\ \bibinfo {pages} {1827} (\bibinfo {year} {2009})},\ \Eprint
  {https://arxiv.org/abs/0811.2795} {arXiv:0811.2795 [astro-ph]} \BibitemShut
  {NoStop}%
\bibitem [{\citenamefont {{Graur}}\ \emph {et~al.}(2017)\citenamefont
  {{Graur}}, \citenamefont {{Bianco}}, \citenamefont {{Modjaz}}, \citenamefont
  {{Shivvers}}, \citenamefont {{Filippenko}}, \citenamefont {{Li}},\ and\
  \citenamefont {{Smith}}}]{graur_2017}%
  \BibitemOpen
  \bibfield  {author} {\bibinfo {author} {\bibfnamefont {O.}~\bibnamefont
  {{Graur}}}, \bibinfo {author} {\bibfnamefont {F.~B.}\ \bibnamefont
  {{Bianco}}}, \bibinfo {author} {\bibfnamefont {M.}~\bibnamefont {{Modjaz}}},
  \bibinfo {author} {\bibfnamefont {I.}~\bibnamefont {{Shivvers}}}, \bibinfo
  {author} {\bibfnamefont {A.~V.}\ \bibnamefont {{Filippenko}}}, \bibinfo
  {author} {\bibfnamefont {W.}~\bibnamefont {{Li}}},\ and\ \bibinfo {author}
  {\bibfnamefont {N.}~\bibnamefont {{Smith}}},\ }\href
  {https://doi.org/10.3847/1538-4357/aa5eb7} {\bibfield  {journal} {\bibinfo
  {journal} {\apj}\ }\textbf {\bibinfo {volume} {837}},\ \bibinfo {eid} {121}
  (\bibinfo {year} {2017})},\ \Eprint {https://arxiv.org/abs/1609.02923}
  {arXiv:1609.02923 [astro-ph.HE]} \BibitemShut {NoStop}%
\bibitem [{\citenamefont {{Abe}}\ \emph {et~al.}(2011)\citenamefont {{Abe}},
  \citenamefont {{Abe}}, \citenamefont {{Aihara}}, \citenamefont {{Fukuda}},
  \citenamefont {{Hayato}}, \citenamefont {{Huang}}, \citenamefont
  {{Ichikawa}}, \citenamefont {{Ikeda}}, \citenamefont {{Inoue}}, \citenamefont
  {{Ishino}}, \citenamefont {{Itow}}, \citenamefont {{Kajita}}, \citenamefont
  {{Kameda}}, \citenamefont {{Kishimoto}}, \citenamefont {{Koga}},
  \citenamefont {{Koshio}}, \citenamefont {{Lee}}, \citenamefont {{Minamino}},
  \citenamefont {{Miura}}, \citenamefont {{Moriyama}}, \citenamefont
  {{Nakahata}}, \citenamefont {{Nakamura}}, \citenamefont {{Nakaya}},
  \citenamefont {{Nakayama}}, \citenamefont {{Nishijima}}, \citenamefont
  {{Nishimura}}, \citenamefont {{Obayashi}}, \citenamefont {{Okumura}},
  \citenamefont {{Sakuda}}, \citenamefont {{Sekiya}}, \citenamefont
  {{Shiozawa}}, \citenamefont {{Suzuki}}, \citenamefont {{Suzuki}},
  \citenamefont {{Takeda}}, \citenamefont {{Takeuchi}}, \citenamefont
  {{Tanaka}}, \citenamefont {{Tasaka}}, \citenamefont {{Tomura}}, \citenamefont
  {{Vagins}}, \citenamefont {{Wang}},\ and\ \citenamefont
  {{Yokoyama}}}]{HyperK}%
  \BibitemOpen
  \bibfield  {author} {\bibinfo {author} {\bibfnamefont {K.}~\bibnamefont
  {{Abe}}}, \bibinfo {author} {\bibfnamefont {T.}~\bibnamefont {{Abe}}},
  \bibinfo {author} {\bibfnamefont {H.}~\bibnamefont {{Aihara}}}, \bibinfo
  {author} {\bibfnamefont {Y.}~\bibnamefont {{Fukuda}}}, \bibinfo {author}
  {\bibfnamefont {Y.}~\bibnamefont {{Hayato}}}, \bibinfo {author}
  {\bibfnamefont {K.}~\bibnamefont {{Huang}}}, \bibinfo {author} {\bibfnamefont
  {A.~K.}\ \bibnamefont {{Ichikawa}}}, \bibinfo {author} {\bibfnamefont
  {M.}~\bibnamefont {{Ikeda}}}, \bibinfo {author} {\bibfnamefont
  {K.}~\bibnamefont {{Inoue}}}, \bibinfo {author} {\bibfnamefont
  {H.}~\bibnamefont {{Ishino}}}, \bibinfo {author} {\bibfnamefont
  {Y.}~\bibnamefont {{Itow}}}, \bibinfo {author} {\bibfnamefont
  {T.}~\bibnamefont {{Kajita}}}, \bibinfo {author} {\bibfnamefont
  {J.}~\bibnamefont {{Kameda}}}, \bibinfo {author} {\bibfnamefont
  {Y.}~\bibnamefont {{Kishimoto}}}, \bibinfo {author} {\bibfnamefont
  {M.}~\bibnamefont {{Koga}}}, \bibinfo {author} {\bibfnamefont
  {Y.}~\bibnamefont {{Koshio}}}, \bibinfo {author} {\bibfnamefont {K.~P.}\
  \bibnamefont {{Lee}}}, \bibinfo {author} {\bibfnamefont {A.}~\bibnamefont
  {{Minamino}}}, \bibinfo {author} {\bibfnamefont {M.}~\bibnamefont {{Miura}}},
  \bibinfo {author} {\bibfnamefont {S.}~\bibnamefont {{Moriyama}}}, \bibinfo
  {author} {\bibfnamefont {M.}~\bibnamefont {{Nakahata}}}, \bibinfo {author}
  {\bibfnamefont {K.}~\bibnamefont {{Nakamura}}}, \bibinfo {author}
  {\bibfnamefont {T.}~\bibnamefont {{Nakaya}}}, \bibinfo {author}
  {\bibfnamefont {S.}~\bibnamefont {{Nakayama}}}, \bibinfo {author}
  {\bibfnamefont {K.}~\bibnamefont {{Nishijima}}}, \bibinfo {author}
  {\bibfnamefont {Y.}~\bibnamefont {{Nishimura}}}, \bibinfo {author}
  {\bibfnamefont {Y.}~\bibnamefont {{Obayashi}}}, \bibinfo {author}
  {\bibfnamefont {K.}~\bibnamefont {{Okumura}}}, \bibinfo {author}
  {\bibfnamefont {M.}~\bibnamefont {{Sakuda}}}, \bibinfo {author}
  {\bibfnamefont {H.}~\bibnamefont {{Sekiya}}}, \bibinfo {author}
  {\bibfnamefont {M.}~\bibnamefont {{Shiozawa}}}, \bibinfo {author}
  {\bibfnamefont {A.~T.}\ \bibnamefont {{Suzuki}}}, \bibinfo {author}
  {\bibfnamefont {Y.}~\bibnamefont {{Suzuki}}}, \bibinfo {author}
  {\bibfnamefont {A.}~\bibnamefont {{Takeda}}}, \bibinfo {author}
  {\bibfnamefont {Y.}~\bibnamefont {{Takeuchi}}}, \bibinfo {author}
  {\bibfnamefont {H.~K.~M.}\ \bibnamefont {{Tanaka}}}, \bibinfo {author}
  {\bibfnamefont {S.}~\bibnamefont {{Tasaka}}}, \bibinfo {author}
  {\bibfnamefont {T.}~\bibnamefont {{Tomura}}}, \bibinfo {author}
  {\bibfnamefont {M.~R.}\ \bibnamefont {{Vagins}}}, \bibinfo {author}
  {\bibfnamefont {J.}~\bibnamefont {{Wang}}},\ and\ \bibinfo {author}
  {\bibfnamefont {M.}~\bibnamefont {{Yokoyama}}},\ }\href
  {https://doi.org/10.48550/arXiv.1109.3262} {\bibfield  {journal} {\bibinfo
  {journal} {arXiv e-prints}\ ,\ \bibinfo {eid} {arXiv:1109.3262}} (\bibinfo
  {year} {2011})},\ \Eprint {https://arxiv.org/abs/1109.3262} {arXiv:1109.3262
  [hep-ex]} \BibitemShut {NoStop}%
\bibitem [{\citenamefont {An}\ \emph {et~al.}(2016)\citenamefont {An} \emph
  {et~al.}}]{JUNO:2015zny}%
  \BibitemOpen
  \bibfield  {author} {\bibinfo {author} {\bibfnamefont {F.}~\bibnamefont {An}}
  \emph {et~al.} (\bibinfo {collaboration} {JUNO}),\ }\href
  {https://doi.org/10.1088/0954-3899/43/3/030401} {\bibfield  {journal}
  {\bibinfo  {journal} {J. Phys. G}\ }\textbf {\bibinfo {volume} {43}},\
  \bibinfo {pages} {030401} (\bibinfo {year} {2016})},\ \Eprint
  {https://arxiv.org/abs/1507.05613} {arXiv:1507.05613 [physics.ins-det]}
  \BibitemShut {NoStop}%
\bibitem [{\citenamefont {{Tamborra}}\ \emph
  {et~al.}(2014{\natexlab{b}})\citenamefont {{Tamborra}}, \citenamefont
  {{Hanke}}, \citenamefont {{Janka}}, \citenamefont {{M{\"u}ller}},
  \citenamefont {{Raffelt}},\ and\ \citenamefont {{Marek}}}]{tamborra_2014b}%
  \BibitemOpen
  \bibfield  {author} {\bibinfo {author} {\bibfnamefont {I.}~\bibnamefont
  {{Tamborra}}}, \bibinfo {author} {\bibfnamefont {F.}~\bibnamefont {{Hanke}}},
  \bibinfo {author} {\bibfnamefont {H.-T.}\ \bibnamefont {{Janka}}}, \bibinfo
  {author} {\bibfnamefont {B.}~\bibnamefont {{M{\"u}ller}}}, \bibinfo {author}
  {\bibfnamefont {G.~G.}\ \bibnamefont {{Raffelt}}},\ and\ \bibinfo {author}
  {\bibfnamefont {A.}~\bibnamefont {{Marek}}},\ }\href
  {https://doi.org/10.1088/0004-637X/792/2/96} {\bibfield  {journal} {\bibinfo
  {journal} {\apj}\ }\textbf {\bibinfo {volume} {792}},\ \bibinfo {eid} {96}
  (\bibinfo {year} {2014}{\natexlab{b}})},\ \Eprint
  {https://arxiv.org/abs/1402.5418} {arXiv:1402.5418 [astro-ph.SR]}
  \BibitemShut {NoStop}%
\bibitem [{\citenamefont {{Foucart}}\ \emph {et~al.}(2023)\citenamefont
  {{Foucart}}, \citenamefont {{Duez}}, \citenamefont {{Haas}}, \citenamefont
  {{Kidder}}, \citenamefont {{Pfeiffer}}, \citenamefont {{Scheel}},\ and\
  \citenamefont {{Spira-Savett}}}]{foucart_2023}%
  \BibitemOpen
  \bibfield  {author} {\bibinfo {author} {\bibfnamefont {F.}~\bibnamefont
  {{Foucart}}}, \bibinfo {author} {\bibfnamefont {M.~D.}\ \bibnamefont
  {{Duez}}}, \bibinfo {author} {\bibfnamefont {R.}~\bibnamefont {{Haas}}},
  \bibinfo {author} {\bibfnamefont {L.~E.}\ \bibnamefont {{Kidder}}}, \bibinfo
  {author} {\bibfnamefont {H.~P.}\ \bibnamefont {{Pfeiffer}}}, \bibinfo
  {author} {\bibfnamefont {M.~A.}\ \bibnamefont {{Scheel}}},\ and\ \bibinfo
  {author} {\bibfnamefont {E.}~\bibnamefont {{Spira-Savett}}},\ }\href
  {https://doi.org/10.1103/PhysRevD.107.103055} {\bibfield  {journal} {\bibinfo
   {journal} {\prd}\ }\textbf {\bibinfo {volume} {107}},\ \bibinfo {eid}
  {103055} (\bibinfo {year} {2023})},\ \Eprint
  {https://arxiv.org/abs/2210.05670} {arXiv:2210.05670 [astro-ph.HE]}
  \BibitemShut {NoStop}%
\bibitem [{\citenamefont {Childs}\ \emph {et~al.}(2012)\citenamefont {Childs}
  \emph {et~al.}}]{VisIt}%
  \BibitemOpen
  \bibfield  {author} {\bibinfo {author} {\bibfnamefont {H.}~\bibnamefont
  {Childs}} \emph {et~al.},\ }in\ \href
  {https://escholarship.org/uc/item/69r5m58v} {\emph {\bibinfo {booktitle}
  {{High Performance Visualization--Enabling Extreme-Scale Scientific
  Insight}}}}\ (\bibinfo  {publisher} {eScholarship, University of
  California},\ \bibinfo {year} {2012})\ pp.\ \bibinfo {pages}
  {357--372}\BibitemShut {NoStop}%
\bibitem [{\citenamefont {Hunter}(2007)}]{hunter2007}%
  \BibitemOpen
  \bibfield  {author} {\bibinfo {author} {\bibfnamefont {J.~D.}\ \bibnamefont
  {Hunter}},\ }\href {https://doi.org/10.1109/MCSE.2007.55} {\bibfield
  {journal} {\bibinfo  {journal} {Computing In Science \& Engineering}\
  }\textbf {\bibinfo {volume} {9}},\ \bibinfo {pages} {90} (\bibinfo {year}
  {2007})}\BibitemShut {NoStop}%
\bibitem [{\citenamefont {{Loken}}\ \emph {et~al.}(2010)\citenamefont
  {{Loken}}, \citenamefont {{Gruner}}, \citenamefont {{Groer}}, \citenamefont
  {{Peltier}}, \citenamefont {{Bunn}}, \citenamefont {{Craig}}, \citenamefont
  {{Henriques}}, \citenamefont {{Dempsey}}, \citenamefont {{Yu}}, \citenamefont
  {{Chen}}, \citenamefont {{Dursi}}, \citenamefont {{Chong}}, \citenamefont
  {{Northrup}}, \citenamefont {{Pinto}}, \citenamefont {{Knecht}},\ and\
  \citenamefont {{Van Zon}}}]{SciNet}%
  \BibitemOpen
  \bibfield  {author} {\bibinfo {author} {\bibfnamefont {C.}~\bibnamefont
  {{Loken}}}, \bibinfo {author} {\bibfnamefont {D.}~\bibnamefont {{Gruner}}},
  \bibinfo {author} {\bibfnamefont {L.}~\bibnamefont {{Groer}}}, \bibinfo
  {author} {\bibfnamefont {R.}~\bibnamefont {{Peltier}}}, \bibinfo {author}
  {\bibfnamefont {N.}~\bibnamefont {{Bunn}}}, \bibinfo {author} {\bibfnamefont
  {M.}~\bibnamefont {{Craig}}}, \bibinfo {author} {\bibfnamefont
  {T.}~\bibnamefont {{Henriques}}}, \bibinfo {author} {\bibfnamefont
  {J.}~\bibnamefont {{Dempsey}}}, \bibinfo {author} {\bibfnamefont {C.-H.}\
  \bibnamefont {{Yu}}}, \bibinfo {author} {\bibfnamefont {J.}~\bibnamefont
  {{Chen}}}, \bibinfo {author} {\bibfnamefont {L.~J.}\ \bibnamefont {{Dursi}}},
  \bibinfo {author} {\bibfnamefont {J.}~\bibnamefont {{Chong}}}, \bibinfo
  {author} {\bibfnamefont {S.}~\bibnamefont {{Northrup}}}, \bibinfo {author}
  {\bibfnamefont {J.}~\bibnamefont {{Pinto}}}, \bibinfo {author} {\bibfnamefont
  {N.}~\bibnamefont {{Knecht}}},\ and\ \bibinfo {author} {\bibfnamefont
  {R.}~\bibnamefont {{Van Zon}}},\ }in\ \href
  {https://doi.org/10.1088/1742-6596/256/1/012026} {\emph {\bibinfo {booktitle}
  {Journal of Physics Conference Series}}},\ \bibinfo {series} {Journal of
  Physics Conference Series}, Vol.\ \bibinfo {volume} {256}\ (\bibinfo {year}
  {2010})\ p.\ \bibinfo {pages} {012026}\BibitemShut {NoStop}%
\bibitem [{\citenamefont {Ponce}\ \emph {et~al.}(2019)\citenamefont {Ponce},
  \citenamefont {van Zon}, \citenamefont {Northrup}, \citenamefont {Gruner},
  \citenamefont {Chen}, \citenamefont {Ertinaz}, \citenamefont {Fedoseev},
  \citenamefont {Groer}, \citenamefont {Mao}, \citenamefont {Mundim},
  \citenamefont {Nolta}, \citenamefont {Pinto}, \citenamefont {Saldarriaga},
  \citenamefont {Slavnic}, \citenamefont {Spence}, \citenamefont {Yu},\ and\
  \citenamefont {Peltier}}]{Niagara}%
  \BibitemOpen
  \bibfield  {author} {\bibinfo {author} {\bibfnamefont {M.}~\bibnamefont
  {Ponce}}, \bibinfo {author} {\bibfnamefont {R.}~\bibnamefont {van Zon}},
  \bibinfo {author} {\bibfnamefont {S.}~\bibnamefont {Northrup}}, \bibinfo
  {author} {\bibfnamefont {D.}~\bibnamefont {Gruner}}, \bibinfo {author}
  {\bibfnamefont {J.}~\bibnamefont {Chen}}, \bibinfo {author} {\bibfnamefont
  {F.}~\bibnamefont {Ertinaz}}, \bibinfo {author} {\bibfnamefont
  {A.}~\bibnamefont {Fedoseev}}, \bibinfo {author} {\bibfnamefont
  {L.}~\bibnamefont {Groer}}, \bibinfo {author} {\bibfnamefont
  {F.}~\bibnamefont {Mao}}, \bibinfo {author} {\bibfnamefont {B.~C.}\
  \bibnamefont {Mundim}}, \bibinfo {author} {\bibfnamefont {M.}~\bibnamefont
  {Nolta}}, \bibinfo {author} {\bibfnamefont {J.}~\bibnamefont {Pinto}},
  \bibinfo {author} {\bibfnamefont {M.}~\bibnamefont {Saldarriaga}}, \bibinfo
  {author} {\bibfnamefont {V.}~\bibnamefont {Slavnic}}, \bibinfo {author}
  {\bibfnamefont {E.}~\bibnamefont {Spence}}, \bibinfo {author} {\bibfnamefont
  {C.-H.}\ \bibnamefont {Yu}},\ and\ \bibinfo {author} {\bibfnamefont {W.~R.}\
  \bibnamefont {Peltier}},\ }in\ \href
  {https://doi.org/10.1145/3332186.3332195} {\emph {\bibinfo {booktitle}
  {Proceedings of the Practice and Experience in Advanced Research Computing on
  Rise of the Machines (Learning)}}},\ \bibinfo {series and number} {PEARC
  '19}\ (\bibinfo  {publisher} {Association for Computing Machinery},\ \bibinfo
  {address} {New York, NY, USA},\ \bibinfo {year} {2019})\BibitemShut {NoStop}%
\bibitem [{\citenamefont {{Press}}\ \emph {et~al.}(1992)\citenamefont
  {{Press}}, \citenamefont {{Teukolsky}}, \citenamefont {{Vetterling}},\ and\
  \citenamefont {{Flannery}}}]{NR92}%
  \BibitemOpen
  \bibfield  {author} {\bibinfo {author} {\bibfnamefont {W.~H.}\ \bibnamefont
  {{Press}}}, \bibinfo {author} {\bibfnamefont {S.~A.}\ \bibnamefont
  {{Teukolsky}}}, \bibinfo {author} {\bibfnamefont {W.~T.}\ \bibnamefont
  {{Vetterling}}},\ and\ \bibinfo {author} {\bibfnamefont {B.~P.}\ \bibnamefont
  {{Flannery}}},\ }\href@noop {} {\emph {\bibinfo {title} {{Numerical recipes
  in C. The art of scientific computing}}}}\ (\bibinfo {year}
  {1992})\BibitemShut {NoStop}%
\end{thebibliography}%

\end{document}